\pdfoutput=1
\newcommand*{\ATLASLATEXPATH}{}
\documentclass[cernpreprint, texlive=2016, UKenglish]{\ATLASLATEXPATH atlasdoc}
 
\usepackage{\ATLASLATEXPATH atlaspackage}
\usepackage[backref=false]{\ATLASLATEXPATH atlasbiblatex}
 
\usepackage[jetetmiss=true]{\ATLASLATEXPATH atlasphysics}
 
\graphicspath{{logos/}{figures/}}
 
\usepackage{ANA-TOPQ-2020-03-PAPER-defs}

\AtlasTitle{Measurements of inclusive and differential cross-sections of combined $t\bar{t}\gamma$ and $tW\gamma$ production in the $e\mkern-2mu\mu$ channel at 13\,TeV with the ATLAS detector}
 
\AtlasAbstract{
Inclusive and differential cross-sections for the production of top quarks in association with a photon are measured with proton--proton collision data corresponding to an integrated luminosity of 139 fb$^{-1}$.
The data were collected by the ATLAS detector at the LHC during Run 2 between 2015 and 2018 at a centre-of-mass energy of 13~TeV.
The measurements are performed in a fiducial volume defined at parton level.
Events with exactly one photon, one electron and one muon of opposite sign, and at least two jets, of which at least one is $b$-tagged, are selected.
The fiducial cross-section is measured to be $39.6\,^{+2.7}_{-2.3}\,\textrm{fb}$.
Differential cross-sections as functions of several observables are compared with state-of-the-art Monte Carlo simulations and next-to-leading-order theoretical calculations.
These include cross-sections as functions of photon kinematic variables, angular variables related to the photon and the leptons, and
angular separations between the two leptons in the event. All measurements are in agreement with the predictions from the Standard Model.
 
}
 
\author{The ATLAS Collaboration}
 
\AtlasRefCode{TOPQ-2020-03}

\AtlasJournal{JHEP}
 
\PreprintIdNumber{CERN-EP-2020-100}
 
\AtlasJournalRef{JHEP 09 (2020) 049}
\AtlasDOI{10.1007/JHEP09(2020)049}
\hypersetup{pdftitle={ATLAS document},pdfauthor={The ATLAS Collaboration}}
 
\begin{document}
 
\maketitle

\section{Introduction}
 
Precise measurements of top-quark production and decay properties provide crucial information for testing the predictions of the Standard Model (SM) and its possible extensions. In particular, the study of the associated production of a top-quark pair (\ttbar) with a high-energy photon probes the $t\gamma$ electroweak coupling. Furthermore, measurements of the inclusive and differential cross-sections of this process are of particular interest because these topologies are sensitive, for instance, to new physics through anomalous dipole moments of the top quark~\cite{Baur:2004uw, Bouzas:2012av, Schulze:2016qas} and in the context of effective field theories~\cite{Bylund:2016phk}.
 
First evidence for the production of \ttbar in association with a photon (\ttg) was reported by the CDF Collaboration~\cite{Aaltonen:2011sp}, while the observation of the \ttg process was established by the ATLAS Collaboration in proton--proton ($pp$) collisions at $\sqrt{s}=7~\TeV$~\cite{TOPQ-2012-07}. Both the ATLAS and CMS Collaborations measured the \ttg cross-section at $\sqrt{s}=8~\TeV$~\cite{TOPQ-2015-21,CMS-TOP-14-008}. First measurements of the inclusive and differential cross-sections at $\sqrt{s}=13~\TeV$ were performed by the ATLAS Collaboration~\cite{TOPQ-2017-14}.
 
This \paperornote presents a measurement of the fiducial inclusive and differential combined \ttg+~\tWg production cross-sections in the final state with one electron and one muon, referred to as the \chemu channel. Events where the electrons and muons arise from the leptonic decays of $\tau$-leptons are considered as background. The measurement is performed using the full data set recorded at the LHC between 2015 and 2018 at a centre-of-mass energy of $\sqrt{s}=13~\TeV$ and corresponding to an integrated luminosity of 139~\ifb. The fiducial inclusive cross-section is measured using a profile likelihood fit to the distribution of \ST, defined as the scalar sum of all transverse momenta in the event, including leptons, photons, jets and missing transverse momentum. The differential cross-sections, absolute and normalised to unity, are measured in the same fiducial region as the inclusive cross-section, as functions of photon kinematic variables, angular variables related to the photon and the leptons, and angular separations between the two leptons in the event.
 
Compared to the previous \ttg ATLAS analysis with 13~\TeV\ data~\cite{TOPQ-2017-14}, only the $e\mu$ channel is considered since it provides a clean final state with a small background contribution and, thus, no multivariate analysis techniques are needed to separate signal and background processes. Additionally, the cross-sections are measured at parton level rather than at particle level to allow comparison with the theory calculation in Refs.~\cite{Bevilacqua:2018woc,Bevilacqua:2018dny}. The calculation constitutes the first full computation for \ttbar production with a hard final-state photon in hadronic collisions at next-to-leading order (NLO) in quantum chromodynamics (QCD), $pp\rightarrow bWbW\gamma$, including all resonant and non-resonant diagrams, interferences, and off-shell effects of the top quarks and the $W$ bosons.
Therefore, in this \paperornote the combined cross-section of resonant \ttg and non-resonant \tWg production is measured, referred to as signal in the following.
Example Feynman diagrams at leading order in QCD for \ttg and \tWg production are shown in Figure~\ref{fig:diagrams}.
 
The \paperornote is organised as follows. The ATLAS detector is briefly introduced in Section~\ref{sec:detector}. Details of the event-simulation generators and their theoretical predictions are given in Section~\ref{sec:modelling}. The event selection and the analysis strategy are presented in Sections~\ref{sec:selection} and~\ref{sec:strategy}. The systematic uncertainties are described in Section~\ref{sec:systematics}. The results for the fiducial inclusive and differential cross-sections are presented in Sections~\ref{sec:inclusive} and~\ref{sec:differential}, respectively. Finally, a summary is given in Section~\ref{sec:conclusion}.
 
\begin{figure}
\centering
\includegraphics[width=0.35\linewidth]{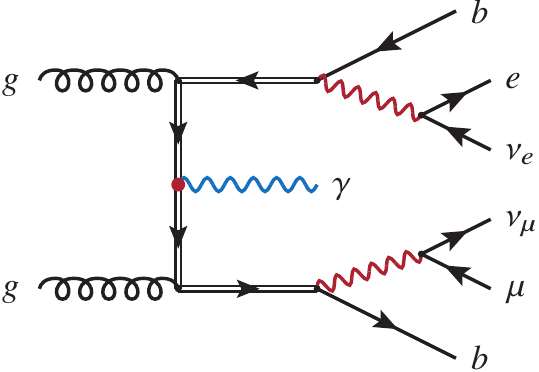}
\hspace{1cm}
\includegraphics[width=0.4\linewidth]{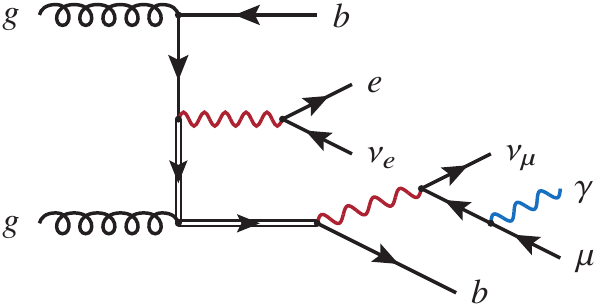}
\caption{
Example Feynman diagrams at leading order for \ttg (left) and \tWg production (right) in the $e\mu$ channel.
The top-quark mass resonances are marked with double-lined arrows, while $W$ bosons are marked in red.
}
\label{fig:diagrams}
\end{figure}

\section{ATLAS detector}
\label{sec:detector}
 
\newcommand{\AtlasCoordFootnote}{
ATLAS uses a right-handed coordinate system with its origin at the nominal interaction point (IP) in the centre
of the detector and the $z$-axis along the beam pipe. The $x$-axis points from the IP to the centre of the
LHC ring, and the $y$-axis points upwards. Cylindrical coordinates $(r,\phi)$ are used in the transverse plane,
$\phi$ being the azimuthal angle around the $z$-axis. The pseudorapidity is defined in terms of the polar angle
$\theta$ as $\eta = -\ln \tan(\theta/2)$. Angular distance is measured in units of $\Delta R \equiv \sqrt{(\Delta\eta)^{2} + (\Delta\phi)^{2}}$.}
 
ATLAS~\cite{PERF-2007-01,ATLAS-TDR-19,Abbott:2018ikt} is a multipurpose detector with a
forward--backward symmetric cylindrical geometry with respect to the LHC beam axis.\footnote{\AtlasCoordFootnote}
The innermost layers consist of tracking detectors in the pseudorapidity range
$|\eta|<2.5$. This inner detector (ID) is surrounded by a thin superconducting solenoid
that provides a \SI{2}{\tesla} axial magnetic field. It is enclosed by the electromagnetic and
hadronic calorimeters, which cover $|\eta|<4.9$. The outermost layers of ATLAS consist of an external
muon spectrometer within  $|\eta|<2.7$, incorporating three large toroidal magnetic assemblies with eight coils each.
The field integral of the toroids ranges between 2.0 and 6.0$\,$Tm for most of the acceptance.
The muon spectrometer includes precision tracking chambers and fast detectors for triggering.
A two-level trigger system~\cite{TRIG-2016-01} reduces the recorded event rate to an average of \SI{1}{\kilo\hertz}.

\section{Signal and background modelling}
\label{sec:modelling}
 
The estimation of signal and background contributions relies on the modelling of these processes with simulated events produced with Monte Carlo (MC) event generators. The response of the ATLAS detector was simulated~\cite{SOFT-2010-01} with \textsc{Geant4}~\cite{Agostinelli:2002hh}.
For some of the estimates of modelling uncertainties, the fast-simulation package \textsc{AtlFast-II} was used instead of the full detector simulation. Additional \pp~interactions (pile-up) were generated with \PYTHIA~8~\cite{Sjostrand:2006za:mod,Sjostrand:2014zea} using a set of tuned parameters called the A3 tune~\cite{ATL-PHYS-PUB-2016-017} and the \nnpdflo parton distribution function (PDF) set~\cite{Pumplin:2002vw}. Corrections to the pile-up profile, selection efficiencies, energy scales and resolutions derived from dedicated data samples are applied to the MC simulation to improve agreement with data.
 
This analysis uses both \emph{inclusive} samples, in which processes were generated at matrix-element (ME) level without explicitly including a photon in the final state, and \emph{dedicated} samples for certain processes, where photons were included in the ME-level generation step.
Dedicated samples with a photon in the ME were generated for the \tty and \Wty final states, as well as for $V\gamma$ processes with additional jets.
Here, $V$ denotes either a $W$ or a $Z$ boson.
Although no photons were generated at ME level in the inclusive samples, initial- and final-state radiation of photons is accounted for by the showering algorithm. Combining inclusive and dedicated samples for the modelling of processes might result in double-counting photon radiation in certain phase-space regions. As a consequence, a procedure to remove overlaps between the inclusive and dedicated samples was performed.
Photon radiation simulated at ME level in dedicated samples achieves higher accuracy than the photon radiation in the showering algorithm.
On the other hand, kinematic requirements are applied to the kinematic properties of the photons at ME level in the dedicated samples.
In the overlap-removal procedure, all events from the dedicated samples are kept while events from the inclusive samples are discarded if they contain a parton-level photon that fulfils the dedicated samples' kinematic requirements of $\pT(\gamma)>\SI{15}{\gev}$ and $\Delta R(\gamma, \ell) > 0.2$, where $\pT(\gamma)$ is the photon's transverse momentum and $\Delta R(\gamma, \ell)$ is the angular distance between the photon and any charged lepton.
 
The dedicated sample for the \tty signal process was simulated using the \madgraph generator (v2.3.3)~\cite{Alwall:2014hca} and the \nnpdflo PDF set at leading order (LO) in QCD. The events were generated as a doubly resonant $2\to7$~process, e.g.\ as $pp \to b\ell\nu b\ell\nu\gamma$, thus, diagrams where the photon is radiated from the initial state (in the case of quark--antiquark annihilation), intermediate top quarks, the \bquarks, and the intermediate $W$~bosons, as well as the decay products of the $W$ bosons, are included. To prevent divergences, the photon was required to have $\pT > 15~\GeV$ and $|\eta| < 5.0$ and the leptons to satisfy $|\eta| < 5.0$. The $\Delta R$ between the photon and any of the charged particles among the seven final-state particles were required to be greater than 0.2.
The top-quark mass in this and all other samples was set to \SI{172.5}{\gev}. The renormalisation and the factorisation scales were set to 0.5$\times \sum_i \sqrt{m^2_i+p^2_{\text{T},i}}$, where the sum runs over all the particles generated from the ME calculation. The event generation was interfaced to \PYTHIA~8 (v8.212) using the A14 tune~\cite{ATL-PHYS-PUB-2014-021} to model parton showers, hadronisation, fragmentation and the underlying event. Heavy-flavour hadron decays were modelled with \evtgen~\cite{Lange:2001uf}; this program was used for all samples, except for those generated using the \sherpa MC program~\cite{Gleisberg:2008ta:mod,Hoeche:2009rj}.
In the latter case, heavy-flavour decays were modelled directly with \sherpa.
 
Two dedicated samples for the \tWy process were generated with the \madgraph generator as well.
The first one was produced at LO in the five-flavour scheme for the $2\rightarrow3$ process (e.g.\ $pp \to tW\gamma$) assuming a stable top quark. The second set of events was generated at LO as a $2 \to 6$ process (e.g.\ $pp \to b\ell\nu\ell\nu\gamma$) in the five-flavour scheme, where the photon is radiated from any other charged final-state particle. In the five-flavour scheme, the \bquarks are treated as massless and the LO representation of the process includes a \bquark in the initial state. The two sets of events are complementary and, once combined, provide a full simulation of the \Wty process.
Both samples make use  of the \nnpdflo PDF set and were interfaced to \pythia~8 (v8.212) for parton showering using the A14 tune.
The photon was also required to have $\pT > 15~\GeV$ and $|\eta| < 5.0$ and to be separated by $\Delta R>0.2$ from any parton.
Although possible interference effects between \tty and \tWy are still missing in the simulated LO samples, the \tWy process is treated as part of the signal in this analysis.

Events with $W\gamma$~and $Z\gamma$~final states (with additional jets) were simulated as dedicated samples.
The $W\gamma$ processes were simulated with \sherpa 2.2.2  at NLO accuracy in QCD using the \nnpdfnnlo PDF set, whereas $Z\gamma$ events were generated with \sherpa 2.2.4 at LO in QCD with the same PDF set. The samples are normalised to the cross-sections given by the corresponding MC simulation. The \sherpa generator performs all steps of the event generation, from the hard process to the observable particles.
All samples were matched and merged by the \sherpa-internal parton showering based on Catani--Seymour dipoles~\cite{Gleisberg:2008fv,Schumann:2007mg} using the MEPS@NLO prescription~\cite{Hoeche:2011fd,Catani:2001cc,Hoeche:2012yf:mod}.
Virtual corrections for the NLO accuracy in QCD in the matrix element were provided by the OpenLoops library~\cite{Cascioli:2011va:mod,Denner:2016kdg}.

Inclusive \ttbar production processes were simulated at matrix-element level at NLO accuracy in QCD using \powhegbox{}~\textsc{v2}~\cite{Nason:2004rx,Frixione:2007vw,Alioli:2010xd}. The calculation used the \nnpdfnlo PDF set~\cite{Ball:2014uwa}. The parton shower was  generated with \PYTHIA~8 (v8.230), for which the A14 tune~\cite{ATL-PHYS-PUB-2012-003} was used.
The \ttbar events are normalised to a cross-section value calculated with the \textsc{Top++2.0} program at next-to-next-to-leading order (NNLO) in perturbative QCD, including soft-gluon resummation to next-to-next-to-leading-logarithm order (see Ref.~\cite{Czakon:2011xx:mod} and references therein).
 
Events with inclusive $W$- and $Z$-boson production in association with additional jets were simulated with \sherpa 2.2.1~\cite{Gleisberg:2008ta:mod,Hoeche:2009rj} at NLO in QCD.
The \nnpdfnlo PDF set was used in conjunction with a dedicated tune provided by the \sherpa authors.
The samples are normalised to the NNLO cross-section in QCD~\cite{STDM-2015-03}.
 
Events with two directly produced vector bosons, i.e.\ $\mathit{WW}$, $\mathit{WZ}$ and $\mathit{ZZ}$, were generated with \sherpa versions 2.2.2 (purely leptonic decays) and 2.2.1 (all others) at LO in QCD.
The \nnpdfnnlo PDF set was used in conjunction with a dedicated tune provided by the \sherpa authors.
The samples are normalised to NLO accuracy cross-sections in QCD~\cite{Campbell:1999ah}.
 
Events with a \ttbar pair and an associated $W$~or $Z$~boson ($t\bar{t}V$) were simulated at NLO at the ME level with \madgraph using the \nnpdfnlo PDF set.
The ME generator was interfaced to \pythia{}~8 (v8.210), for which the A14 tune was used in conjunction with the \nnpdflo PDF set.
The samples are normalised to NLO in QCD and electroweak theory~\cite{deFlorian:2016spz}.
 
The background processes are sorted into three categories based on the origin of the reconstructed photon required in the event selection.
The three are estimated from MC simulation by categorising events from all considered samples that are not classified as signal events. The MC simulations for all categories include processes without prompt photons such as \ttbar, $W$+jets, $Z$+jets, diboson and \ttV production, as well as background processes with an additional prompt photon.
The first category is labelled \emph{\cathfake} and contains any type of hadronic fakes that mimic a photon signature in the detector.
This category includes not only photon signatures faked by hadronic energy depositions in the electromagnetic calorimeter, but also hadron decays involving photons, for example $\pi^0 \to \gamma\gamma$ decays.
It also includes processes with a prompt photon, where the prompt photon is not reconstructed in the detector or does not pass the selection requirements, but a \hfake does.
Studies performed with data-driven techniques following the approach described in Ref.~\cite{TOPQ-2017-14} show that possible data-driven corrections have a negligible effect on the distribution shapes of relevant observables. Possible differences in the total expected number of events are covered by a normalisation uncertainty as described in Section~\ref{sec:systematics}.
The second category is labelled \emph{e-fake} and contains processes with an electron mimicking a photon signature in the calorimeter.
Similarly to the h-fake category, this category includes contributions from processes without a prompt photon but with an \efake, as well as processes with a prompt photon in the simulation but an \efake in the reconstruction.
This category represents a minor background contribution.
The third category is called \emph{\catprompty background} and contains any type of background process with a prompt photon. The background contribution from \ttbar production with a photon produced in an additional $pp$ interaction in the same bunch crossing was found to be negligible.
This was estimated by comparing the significance of the distance in $z$ between the photon's origin and the primary vertex in data and simulation.
 
The \tty and \tWy events where one or both $W$ bosons decay into $\tau$-leptons, which then subsequently decay into $e$ or $\mu$, are categorised as \othertty, and not as \chemu signal, following the definition of signal events in the theory calculation in Refs.~\cite{Bevilacqua:2018woc,Bevilacqua:2018dny}. Single-lepton events, where a second lepton is faked by hadronic energy depositions, are also included in the category \othertty. The contribution of \tty single-lepton events was found to be negligible in the \chemu final state in the previous measurement~\cite{TOPQ-2017-14} and it is therefore estimated from the MC simulation.

\section{Event selection}
\label{sec:selection}
 
The data set used in this analysis corresponds to the 139~\ifb{} of integrated luminosity collected with the ATLAS detector during the Run~2 period. Each event in data and simulation is required to have at least one reconstructed primary vertex with at least two associated reconstructed tracks.
Furthermore, only events where at least one of the single-electron~\cite{Aad:2019wsl} or single-muon~\cite{Aad:2020uyd} triggers was fired are selected.
 
The main physics objects considered in this analysis are electrons, muons, photons, jets, \bjets and missing transverse momentum.
Electrons are reconstructed from energy deposits in the electromagnetic calorimeter associated with reconstructed tracks in the ID system.
They are identified with a combined likelihood technique~\cite{EGAM-2018-01} using a `tight' working point, and are required to be isolated based on calorimeter and tracking quantities.
The $\pT$- and $\eta$-dependent isolation criteria yield an efficiency of 90\% for electrons with $\pT = 25$~GeV and 99\% for those with $\pT = 60$~GeV.
The origin of the electron track has to be compatible with the primary vertex.
Electrons are calibrated with the method described in Ref.~\cite{EGAM-2018-01}.
They are selected if they fulfil $\pt>\SI{25}{\GeV}$ and $|\eta_{\mathrm{clus}}| < 2.47$, excluding the calorimeter barrel/endcap transition region $1.37 < |\eta_{\mathrm{clus}}| < 1.52$.
\footnote{$\eta_{\mathrm{clus}}$ denotes the pseudorapidity of the calorimeter cell cluster associated with the electron.}
 
Muons are reconstructed with an algorithm that combines the track segments in the various layers of the muon spectrometer and the tracks in the ID system.
The reconstruction, identification and calibration methods are described in Ref.~\cite{PERF-2015-10}.
Muons are required to be isolated according to track- and calorimeter-based criteria similar to those applied to electrons.
Only muons with calibrated $\pT > \SI{25}{\GeV}$ and $|\eta| < 2.5$ and passing `medium' quality requirements are considered.
The muon track is also required to originate from the primary collision vertex.
 
Photons are reconstructed from energy deposits in the central region of the electromagnetic calorimeters.
If the cluster considered is not matched to any reconstructed track in the ID system, the photon candidate is classified as unconverted.
If the cluster is matched with one or two reconstructed tracks that are consistent with originating from a photon conversion and if, in addition, a conversion
vertex can be found, the photon candidate is classified as converted.
Both kinds of photons are considered in this analysis.
Photons are reconstructed and identified as described in Ref.~\cite{PERF-2017-02} and their energies are calibrated with the method described in Ref.~\cite{PERF-2017-03}.
They are subject to a tight isolation requirement defined as $\left. \ET^{\mathrm{iso}} \right|_{\Delta R < 0.4} < 0.022 \cdot \ET(\gamma) + \SI{2.45}{\GeV}$ in conjunction with $\left. \pT^{\mathrm{iso}} \right|_{\Delta R < 0.2} < 0.05 \cdot \ET(\gamma)$, where $\ET^{\mathrm{iso}}$ refers to the calorimeter isolation within $\Delta R < 0.4$ around the direction of the photon candidate and $\pT^{\mathrm{iso}}$ is the track isolation within $\Delta R < 0.2$~\cite{PERF-2017-02}.
Only photons with calibrated $\ET > \SI{20}{\GeV}$ and $|\eta_{\mathrm{clus}}| < 2.37$, excluding the calorimeter transition region $1.37 < |\eta_{\mathrm{clus}}| < 1.52$, are considered.
 
Jets are reconstructed using the anti-$k_{t}$ algorithm~\cite{Cacciari:2008gp} in the \Fastjet implementation~\cite{Fastjet} with a distance parameter $R=0.4$.
They are reconstructed from topological clusters of cells in the calorimeter~\cite{ATL-PHYS-PUB-2015-036}.
The jet energy scale and jet energy resolution are calibrated using information from both simulation and data~\cite{PERF-2016-04}.
The jets are required to have $\pT > \SI{25}{\GeV}$ and $|\eta| < 2.5$.
Jets with a large contribution from pile-up vertices are identified with the \emph{Jet Vertex Tagger}~\cite{ATLAS-CONF-2014-018} and rejected.
 
The \btag algorithm (MV2c10) applied to the selected jets to identify those from \bquark hadronisation~\cite{ATL-PHYS-PUB-2017-013} labelled as \bjets is based on a boosted decision tree combining information from other algorithms using track impact parameters and secondary vertices, and a multi-vertex reconstruction algorithm.
A working point with a selection efficiency of 85\% on simulated \ttbar events is used, corresponding to rejection factors of 3.1 and 35 for jets initiated by charm quarks and light-flavour partons, respectively. The flavour-tagging efficiency for \bjets, as well as for $c$-jets and light-flavour jets, is calibrated as described in Ref.~\cite{PERF-2016-05}.
 
The reconstructed missing transverse momentum \met~\cite{ATLAS-CONF-2018-023,PERF-2016-07} is computed as the negative vector sum over all reconstructed, fully calibrated physics objects, including photons, and the remaining unclustered energy, also called the \emph{soft term}.
The soft term is estimated from low-\pt tracks associated with the primary vertex but not with any reconstructed object.
 
An overlap-removal procedure is applied to avoid the reconstruction of the same energy clusters or tracks as different objects. First, electron candidates sharing their track with a muon candidate are removed and jets within a $\Delta R = 0.2$ cone around any remaining electron are excluded. Secondly, electrons within a $\Delta R = 0.4$ cone around any remaining jet are removed. If the distance between a jet and any muon candidate is $\Delta R < 0.4$, the muon candidate is discarded if the jet has more than two associated tracks, otherwise the jet is removed. Finally, photons within a $\Delta R = 0.4$ cone around any remaining electron or muon are removed and then jets within a $\Delta R = 0.4$ cone around any remaining photon are excluded.
 
The selected events must have exactly one electron and exactly one muon, each with $\pT > \SI{25}{\GeV}$.
At least one of these leptons has to be matched to a fired single-lepton trigger.
Since the \pt threshold of the single-lepton triggers was increased over the different data-taking periods due to increased collisions rates,
the offline \pt thresholds for these electrons and muons that are matched to a fired single-lepton trigger are chosen to be $25~\GeV$ in 2015, $27~\GeV$ in 2016, and $28~\GeV$ in 2017 and 2018
in order to lie above the trigger thresholds.
Electrons and muons must have opposite-sign charges and the \chemu invariant mass is required to be higher than $15~\GeV$. The event is required to have at least two jets and at least one of the jets must be \btagged.
In addition, all events must contain exactly one reconstructed photon fulfilling the condition that $\Delta R$ between the selected photon and any of the leptons is greater than 0.4.
 
The observed event yields after selection are listed in Table~\ref{tab:prefityields} for the different signal and background categories described in Section~\ref{sec:modelling}.
The LO cross-section of the MC samples underestimates the expected number of signal events; therefore, for illustration purposes the combination of all \tty and \Wty categories is normalised to match the event yields in data. Correction factors for detector effects (described in Section~\ref{sec:systematics}) are applied, when needed, to improve the description of the data by the simulation.
 
\begin{table}
\centering
\caption{Event yields before the profile likelihood fit of the signal and background processes to data after the full selection. All categories are estimated from MC simulation and include correction factors for detector effects as described in Section~\ref{sec:systematics}.
The combination of all \tty and \Wty categories is scaled to match the event yields in data.
The quoted uncertainties correspond to the total statistical and systematic uncertainties (cf.\ Section~\ref{sec:systematics}) added in quadrature.}
\label{tab:prefityields}
\sisetup{table-format=4.0(3)}
\begin{tabular}{
l
S[table-format=4.0] @{${}\pm{}$} S[table-format=3.0]
}
\toprule
& \multicolumn{2}{c}{Events} \\
\midrule
\tty $e\mu$             & 2391 & 130 \\
\Wty $e\mu$             & 156  & 15 \\
\othertty               & 279  & 15 \\
h-fake                  & 78   & 40 \\
e-fake                  & 23   & 12 \\
Prompt $\gamma$ bkg.    & 87   & 40 \\
\midrule
Total                   & 3014 & 160 \\
\midrule
Data                    & \multicolumn{2}{l}{3014} \\
\bottomrule
\end{tabular}
\end{table}
 
The modelling of signal and background processes is inspected through the comparison of distributions. A selection of these distributions showing a comparison between the MC simulation before the profile likelihood fit and data is presented in Figure~\ref{fig:controlplots_with_syst_1}. The combination of all \tty and \Wty categories is normalised to match the event yields in data as done in Table~\ref{tab:prefityields} to allow a comparison of the shapes of the kinematic variables.
All systematic uncertainties that are introduced in Section~\ref{sec:systematics} are included in these distributions and their sum in quadrature, which assumes they are fully uncorrelated, is illustrated by the shaded error bands.
 
\begin{figure}
\centering
\includegraphics[width=0.33\textwidth]{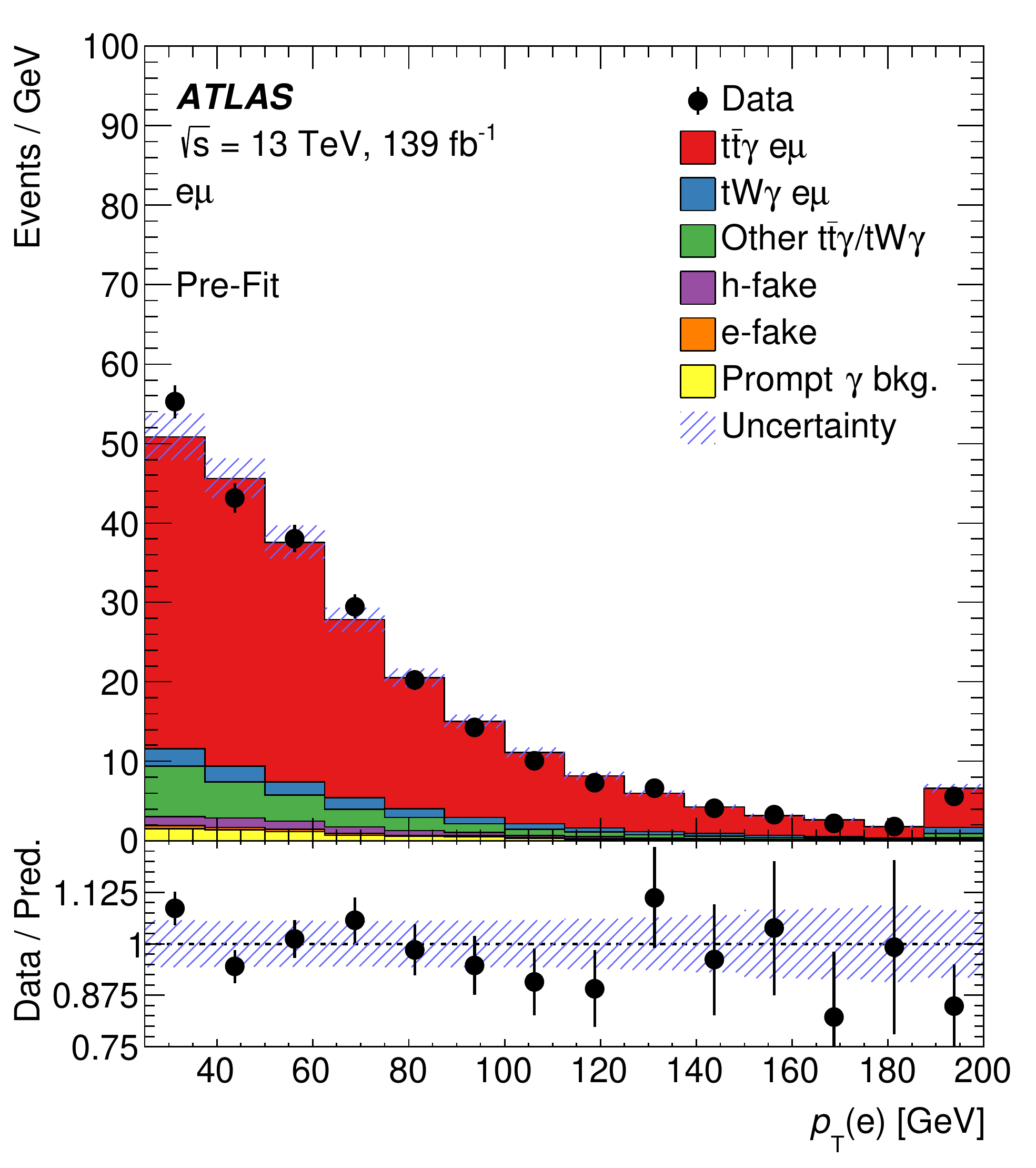}\includegraphics[width=0.33\textwidth]{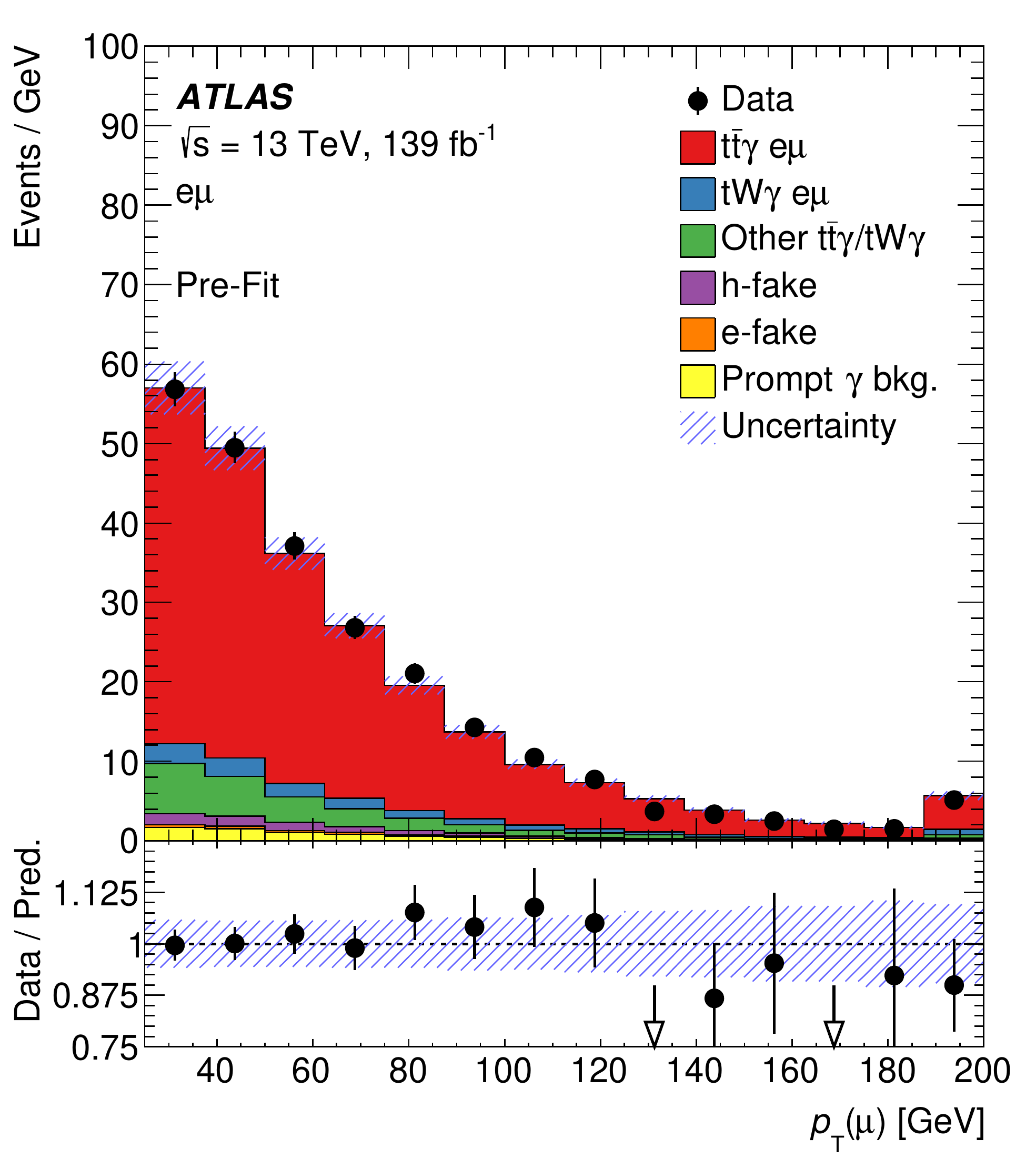}\includegraphics[width=0.33\textwidth]{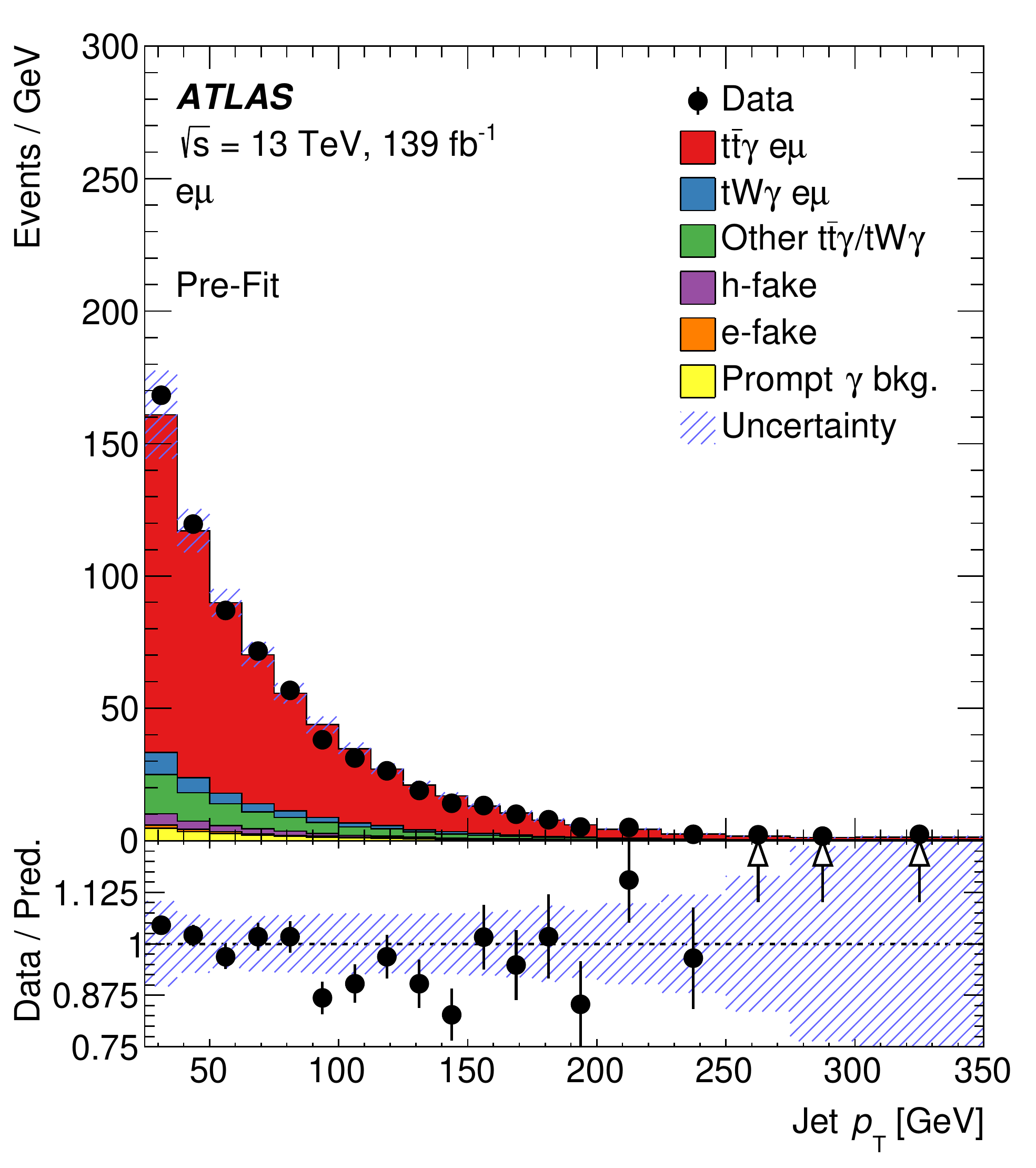}\\
\includegraphics[width=0.33\textwidth]{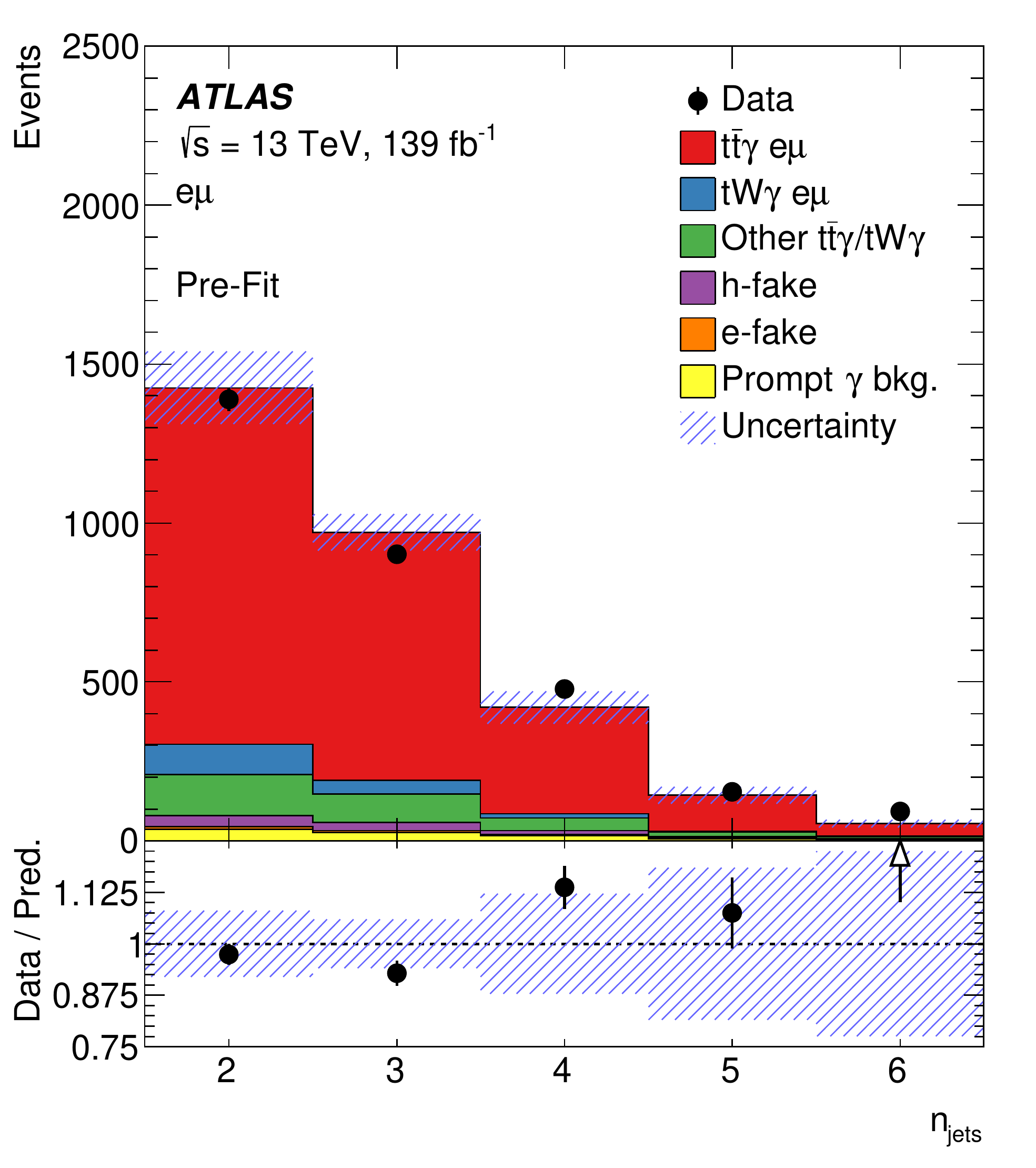}\includegraphics[width=0.33\textwidth]{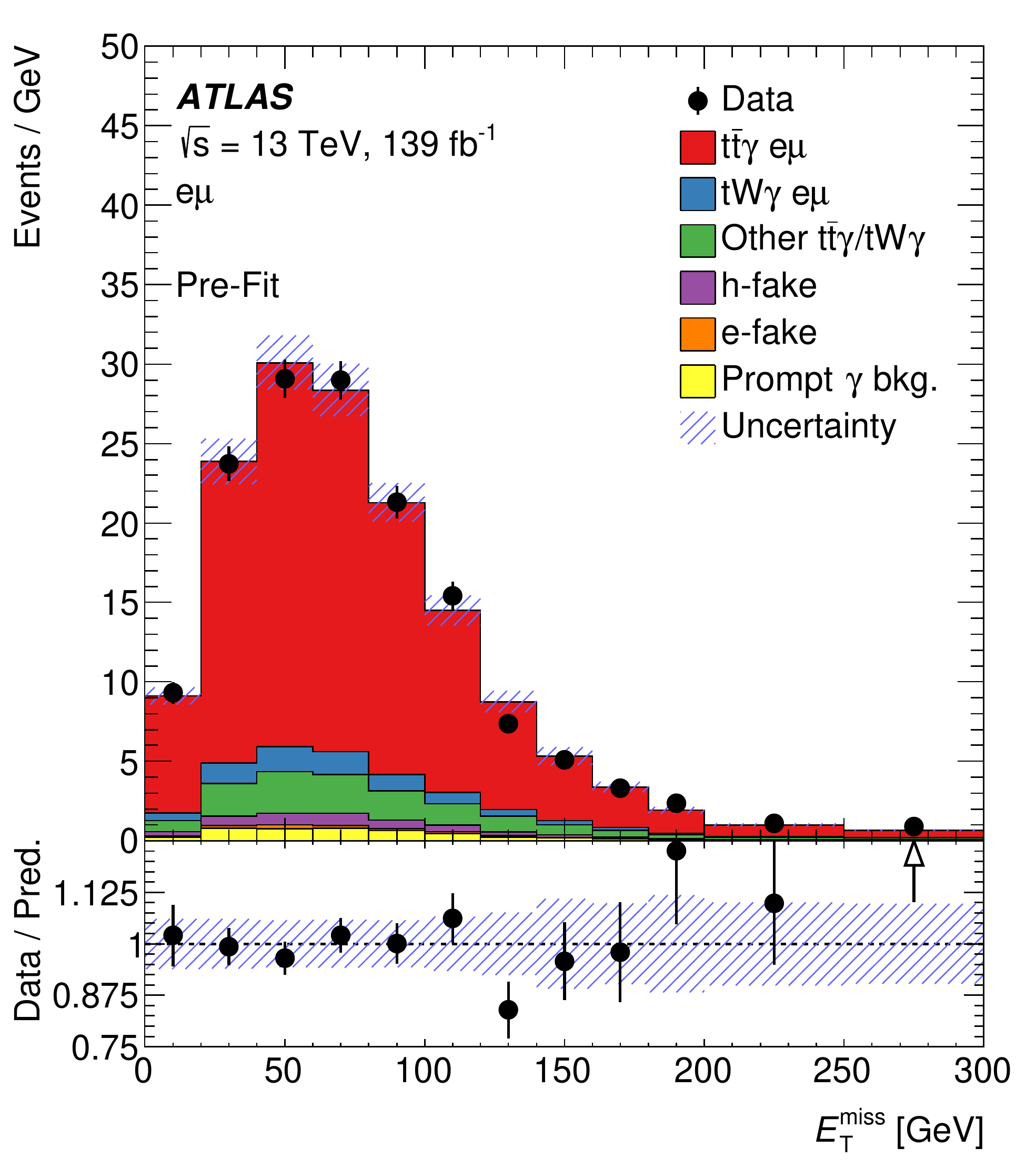}\includegraphics[width=0.33\textwidth]{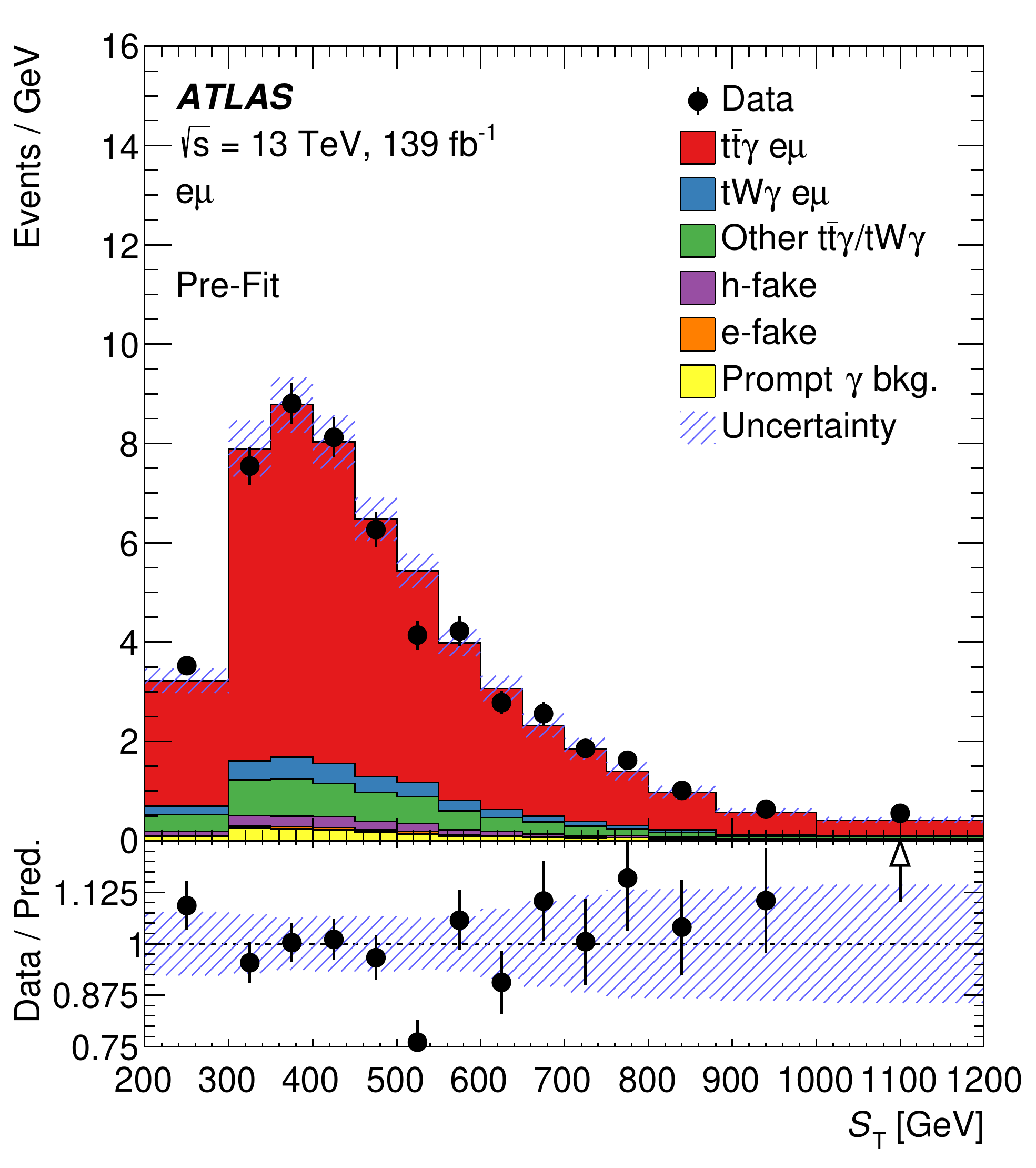}
\caption{Distributions of the transverse momentum of the electron, the muon and all jets (top row), and the number of jets, \met and \ST (bottom row) after event selection and before the profile likelihood fit. The combination of all \tty and \Wty categories is scaled to match the event yields in data. The shaded bands correspond to the statistical and systematic uncertainties (cf.\ Section~\ref{sec:systematics}) added in quadrature. Overflow events are included in the last bin of each distribution. In the case of the \ST distribution, the underflow events are included in the first bin. The lower part of each plot shows the ratio of the data to the prediction.}
\label{fig:controlplots_with_syst_1}
\end{figure}

\section{Analysis strategy}
\label{sec:strategy}
 
The inclusive and differential cross-sections are measured in the fiducial region described in Section~\ref{sec:fiducial} and the same sources of background contributions and systematic uncertainties are considered.
In the fiducial inclusive cross-section the \ST distribution is fitted and the post-fit background yields and systematic uncertainties are used to extract the signal cross-section, while no fit is performed for the determination of the differential cross-sections.
 
\subsection{Fiducial region definition}
\label{sec:fiducial}
 
The cross-sections are reported at parton level in a fiducial region, defined by the kinematic properties of the signal process, in which all selected final-state objects are produced within the detector acceptance.
This is done in a way that mimics the event selection as defined in the theoretical calculation. Objects at parton level are taken from the MC simulation history.
Photons and leptons are selected as stable particles after final-state radiation. The leptons ($\ell = e, \mu$) must originate from $W$-boson decays and they are dressed with nearby photons within a cone of size of $\Delta R=0.1$ around them and must have $\pT > 25~\GeV$ and $|\eta|<$~2.5. Only events with exactly one electron and one muon are considered. Events with leptons originating from an intermediate $\tau$-lepton in the top-quark decay chain are not considered.
The $b$-jets at parton level in the calculation from Refs.~\cite{Bevilacqua:2018woc,Bevilacqua:2018dny} are jets clustered with the anti-$k_{t}$ algorithm with a distance parameter of $R=0.4$. Since showering and hadronisation effects are not considered in this calculation, the jets correspond to the $b$-quarks from the top-quark decay (with an additional parton in the cases where the NLO real emission leads to a parton close by a $b$-quark). To mimic this definition in the LO MC simulation, parton-level $b$-jets are defined as follows. The anti-$k_t$ algorithm with a distance parameter $R = 0.4$ is applied to all partons that are radiated from the two $b$-quarks (including the $b$-quarks themselves) and from the two initial partons. The jets that include a $b$-quark from the decay of a top quark are selected as $b$-jets. The event is kept if there are two $b$-jets satisfying $\pt > 25~\GeV$ and $|\eta| < 2.5$. Exactly one photon with $\et > 20~\GeV$ and $|\eta| < 2.37$ is required. Photons are required to be isolated from nearby jets by imposing a modified cone approach as described in Ref.~\cite{Frixione:1998jh}, as it is also done in the theory calculation in Refs.~\cite{Bevilacqua:2018woc,Bevilacqua:2018dny}, to ensure soft and collinear safety.
The event is dropped if any of the following requirements is not fulfilled: $\Delta R(\gamma, \ell) > 0.4$, $\Delta R(e,\mu) > 0.4$, $\Delta R(b,b)>0.4$ or $\Delta R(\ell,b)>0.4$.

\subsection{Fiducial inclusive cross-section}
The fiducial inclusive cross-section is extracted using a binned profile likelihood fit to the full \ST distribution. The distribution of \ST provides good separation between signal and background and was found to be less sensitive to systematic uncertainties than other distributions considered, such as the jet multiplicity or the \pt of individual jets.
The expected signal and background distributions are modelled in the fit using template distributions taken from the simulated samples.
The parameter of interest, the fiducial cross-section $\sigma_{\text{fid}}$, is related to the number of signal events in bin $i$ of the \ST distribution as:
\begin{equation*}
N_i^s = L \times \sigma_{\text{fid}} \times C \times f_i^{\ST}.
\end{equation*}
The term $L$ is the integrated luminosity, $f_i^{\ST}$ is
the fraction of generated signal events falling into bin $i$ of the \ST distribution after fiducial requirements are applied,
and $C$ is the correction factor for the signal efficiency~$\epsilon$ and for migration into the fiducial region~$f_{\textrm{out}}$, defined as follows:
\begin{align*}
f_{\textrm{out}} = \frac{N_{\textrm{reco}}^{\textrm{non-fid}}}{N_{\textrm{reco}}}\,, \quad\quad
\epsilon = \frac{N_{\textrm{reco}}^{\textrm{fid}}}{N_{\textrm{MC}}^{\textrm{fid}}} \quad\quad
\Rightarrow C = \frac{\epsilon}{1 - f_{\textrm{out}}} = \frac{ N_{\textrm{reco}} }{ N_{\textrm{MC}}^{\textrm{fid}}}\,,
\end{align*}
where $N_{\textrm{reco}}$ is the number of simulated signal events passing the event selection described in Section~\ref{sec:selection}, $N_{\textrm{MC}}^{\textrm{fid}}$ is the corresponding number of signal events generated in the fiducial region defined in Section~\ref{sec:fiducial}, and $N_{\textrm{reco}}^{\textrm{fid}}$ and $N_{\textrm{reco}}^{\textrm{non-fid}}$ are the numbers of signal events that pass the event selection and are generated within and outside the fiducial region, respectively.
The efficiency and outside migration are obtained from simulated \tty and \tWy events.
The correction factor is estimated from the signal simulation to be $C=0.462\pm0.002$ (statistical uncertainty only).

The likelihood function $\mathcal{L}$, based on Poisson statistics, is given by:
\begin{equation*}
\mathcal{L} = \prod_i P \left(N_i^{\text{obs}} | N_i^s(\vec{\theta}) + \sum_b N_i^b(\vec{\theta})\right) \times \prod_t G(0|\theta_t,1),
\end{equation*}
where $N_i^{\text{obs}}$, $N_i^s$, and $N_i^b$ are the observed number of events in data, the predicted number of signal events, and the estimated number of background events in bin $i$ of the \ST distribution, respectively. The rates of those \tty and \tWy events not counted as part of the signal and categorised as \othertty are scaled with the same parameter as the signal events in the fit,
i.e.\ no independent production cross-section is assumed for these parts of the simulated  \tty/\tWy process.
The vector $\vec{\theta}$, of components $\theta_t$, represents the nuisance parameters that describe the sources of systematic uncertainties. Each nuisance parameter~$\theta_t$ is constrained by a Gaussian distribution, $G(0|\theta_t,1)$. The width of the Gaussian function corresponds to a change of $\pm$1 standard deviation of the corresponding quantity in the likelihood. For systematic uncertainties related to the finite number of simulated MC events, the Gaussian terms in the likelihood are replaced by Poisson terms. The cross-section is measured by profiling the nuisance parameters and minimising $-2\ln\mathcal{L}$~\cite{Cowan:2010js:fixed}.
 
\subsection{Absolute and normalised differential cross-sections}
The measurements of the absolute and normalised differential cross-sections are performed as functions of the \pT and $|\eta|$ of the photon, and of angular variables between the photon and the leptons: $\Delta R$ between the photon and the closest lepton $\Delta R (\gamma,\ell)_{\mathrm{min}}$, as well as $\Delta\phi(\ell, \ell)$ and $|\Delta\eta(\ell,\ell)|$ between the two leptons. The kinematic properties of the photon are sensitive to the $t \gamma$ coupling.
In particular, $\Delta R (\gamma,\ell)_{\mathrm{min}}$ is related to the angle between the top quark and the radiated photon, which could give insight into the structure of this coupling.
The distributions of $\Delta \phi(\ell,\ell)$ and $|\Delta\eta(\ell,\ell)|$  are sensitive to the \ttbar spin correlation. The corresponding distributions in data and SM simulations are compared in Figure~\ref{fig:controlplots_unfolding}. The simulation describes reasonably well the data within the uncertainties although it favours smaller $\Delta R (\gamma,\ell)_{\mathrm{min}}$ and larger $\Delta \phi(\ell,\ell)$ values than the observed ones.
 
\begin{figure}[!htbp]
\centering
\includegraphics[width=0.33\linewidth]{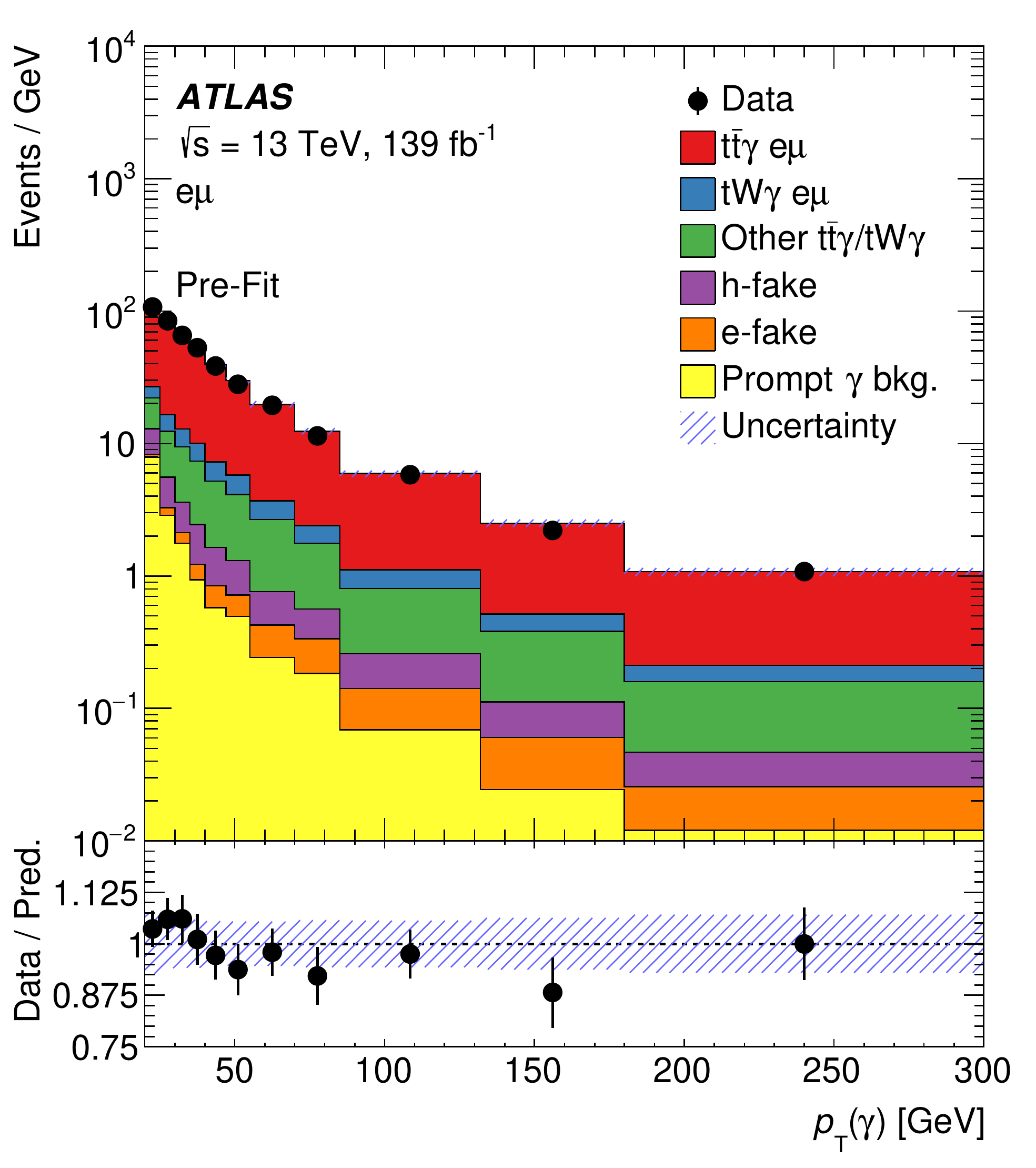}\includegraphics[width=0.33\linewidth]{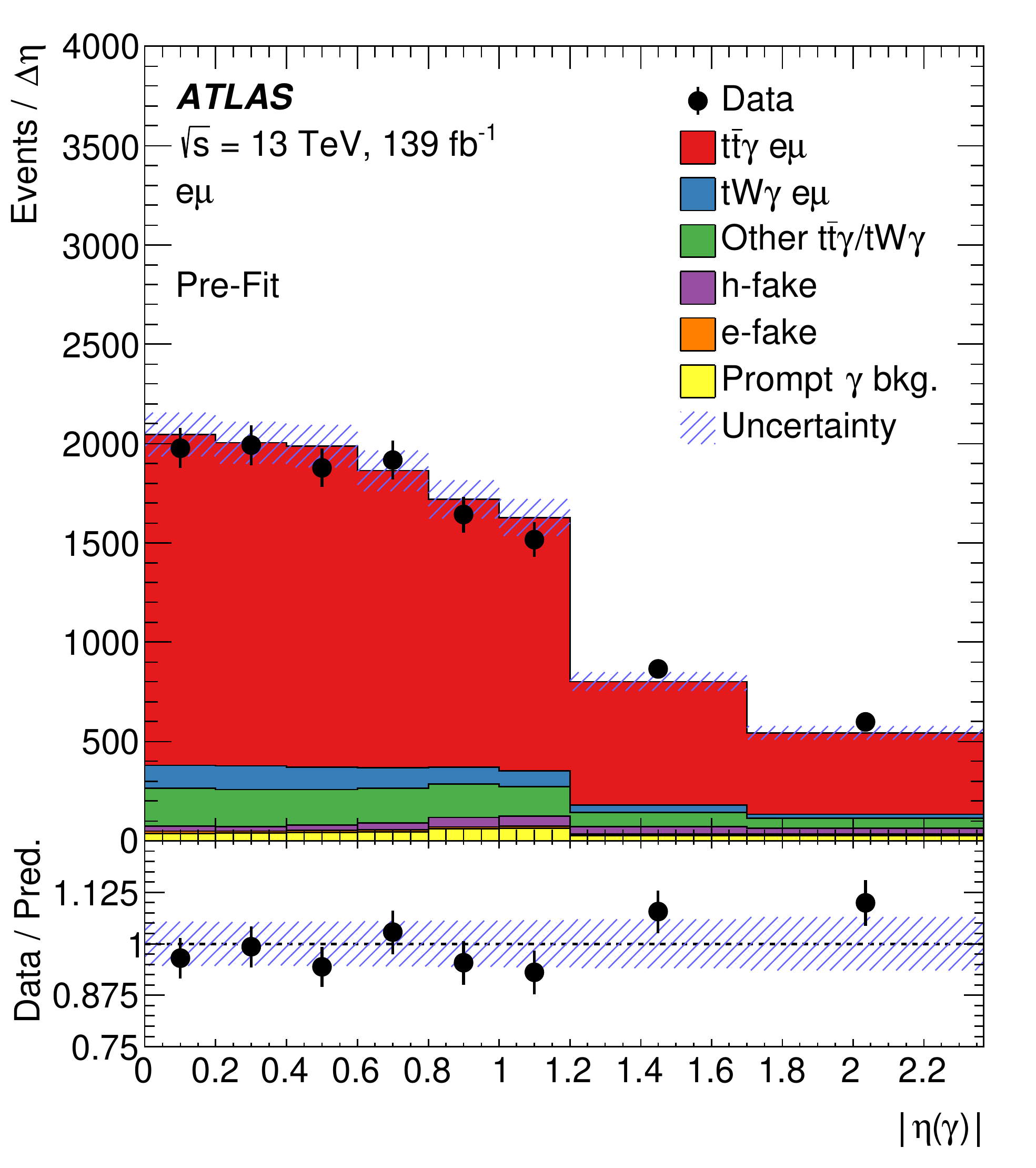}\\
\includegraphics[width=0.33\linewidth]{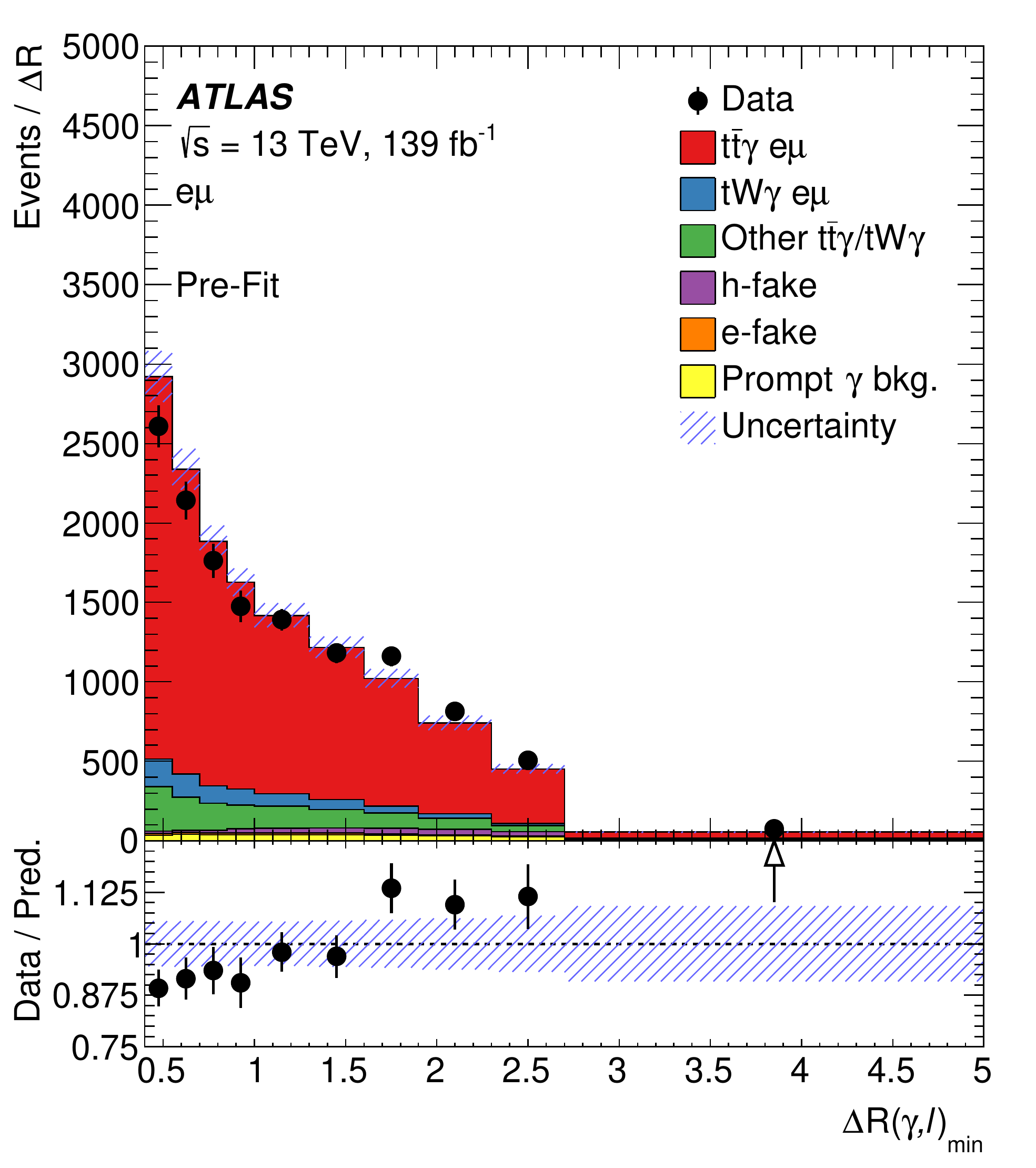}\includegraphics[width=0.33\linewidth]{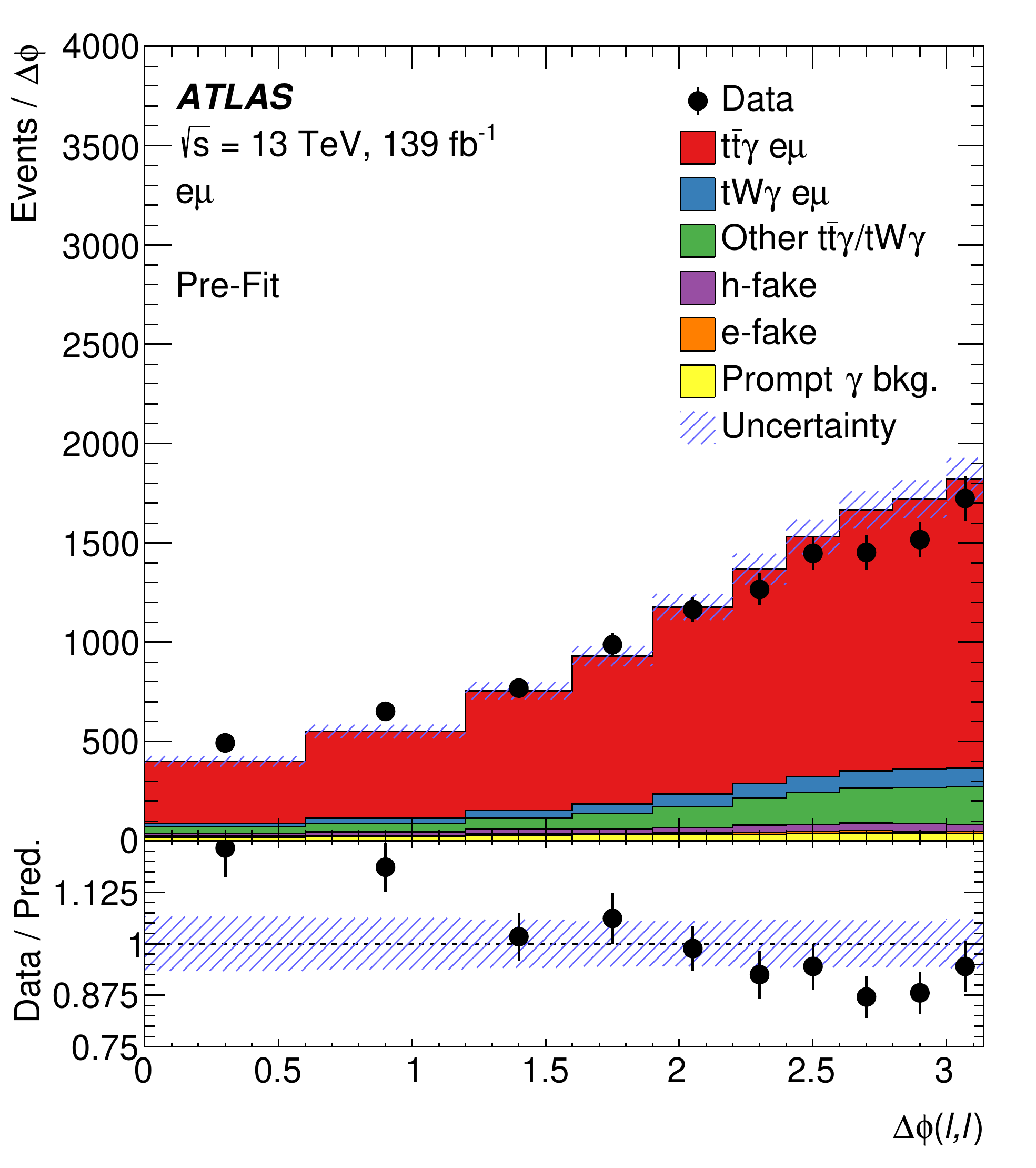}\includegraphics[width=0.33\linewidth]{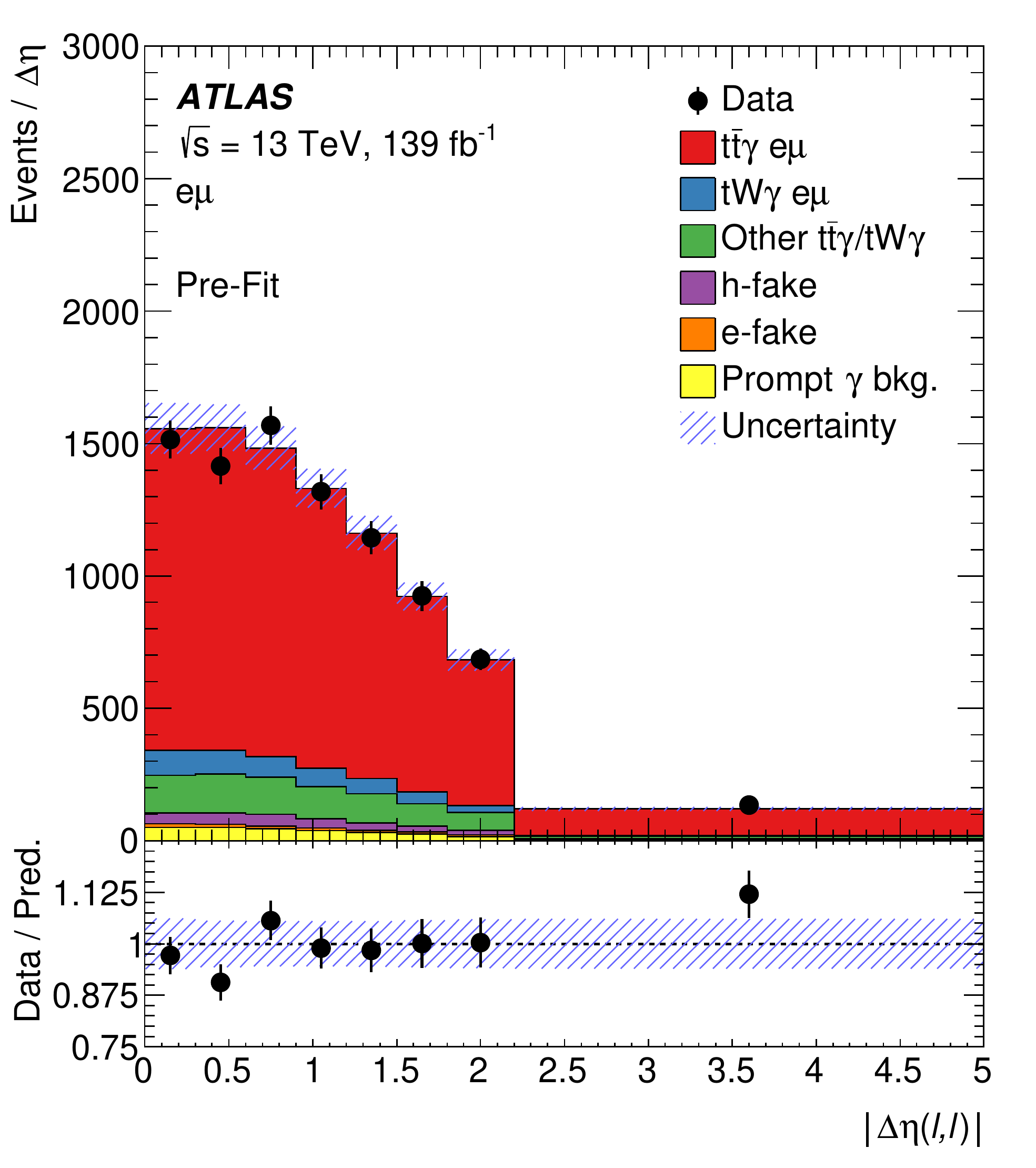}
\caption [] {Distributions of the photon \pT and $|\eta|$ in the top row, and $\Delta R (\gamma,\ell)_{\mathrm{min}}$,  $\Delta\phi(\ell,\ell)$ and $|\Delta\eta(\ell, \ell)|$ in the bottom row after event selection and before the profile likelihood fit. The combination of all \tty and \Wty categories is scaled to match the event yields in data. The shaded bands correspond to the statistical and systematic uncertainties (cf.\ Section~\ref{sec:systematics}) added in quadrature. When overflow events are present, they are included in the last bin of the distribution. The lower part of each plot shows the ratio of the data to the prediction.}
\label{fig:controlplots_unfolding}
\end{figure}
 
The data are corrected for detector resolution and acceptance effects to parton level in the fiducial phase space using an iterative matrix unfolding that uses Bayes' theorem~\cite{DAgostini:1994fjx} implemented in the \textsc{RooUnfold} package~\cite{Adye:2011gm}. The differential cross-section is defined as:
\begin{equation*}
\frac{\mathrm{d}\sigma}{\mathrm{d}X_k} = \frac{1}{L \times \Delta X_k \times \epsilon_k} \times \sum_j M_{jk}^{-1} \times (N^{\mathrm{obs}}_{j} - N^{b}_{j})\times f_{e\mu,j}\times (1-f_{\mathrm{out},j})\,.
\end{equation*}
The indices $j$ and $k$ represent the bin indices of the observable $X$ at detector and parton levels, respectively. The variable $N^{\textrm{obs}}_{j}$ is the number of observed events, and $N^{b}_{j}$ is the number of estimated non-\ttg/\tWy background events (pre-fit) in bin $j$ at detector level. The contribution from the \othertty category is taken into account by correcting the remaining number of observed events by the signal fraction, $f_{e\mu,j}$, defined as the ratio of the number of selected \ttg and \tWy \chemu events to the total number of selected \ttg and \tWy events, as determined from simulation. This avoids the dependence on the signal cross-section used for the normalisation. The efficiency $\epsilon_k$ is the fraction of signal events generated at parton level in bin $k$ of the fiducial region that are reconstructed and selected at detector level. The total integrated luminosity is denoted by $L$,  and $\Delta X_k$  represents the bin width. The migration matrix $M_{kj}$ describes the detector response and expresses the probability for an event in bin $k$ at parton level to be reconstructed in bin $j$ at detector level, calculated from events passing both the fiducial-region selection and the event selection.
The outside-migration fraction $f_{\mathrm{out},j}$ is the fraction of signal events generated outside the fiducial region but reconstructed and selected in bin $j$ at detector level. The normalised differential cross-section is derived by dividing the absolute result by the total cross-section, obtained by integrating over all bins of the observable.
 
The signal MC samples are used to determine $\epsilon_k$, $f_{\mathrm{out},j}$, and $M_{kj}$. The unfolding method relies on the Bayesian probability formula, starting from a given prior of the parton-level distribution and iteratively updating it with the posterior distribution. The binning choices of the unfolded observables take into account the detector resolution and the expected statistical uncertainty.
The bin width has to be larger than twice the resolution, and the statistical uncertainty is required to be around or below 10\% across all bins, with the latter being the limiting factor in most of the cases.
The resolution of the lepton and photon momenta is very high and, therefore, the fraction of events migrating from one bin to another is small. In all bins, the purity, defined as the fraction of reconstructed events that originate from the same bin at parton level, is larger than 80\%, and it is above 90\% for all observables except for the \pt of the photon. The number of iterations chosen is two, which provides good convergence of the unfolding distribution and a statistically stable result. For illustration purposes, the migration matrix is presented in the left panel of Figure~\ref{fig:effmigsl_pt}, while the right panel shows the efficiency, outside-migration fraction and the resulting $C$ correction factor obtained for the distribution of the photon \pt. The performance of the unfolding procedure is tested for possible biases from the choice of input model. It was verified that when reweighting the shape of the signal simulation by up to 50\% bin-by-bin with respect to the nominal shape, the unfolding procedure based on the nominal response matrix reproduces the altered shapes.
 
\begin{figure}[!htbp]
\centering
\includegraphics[width=0.5\linewidth]{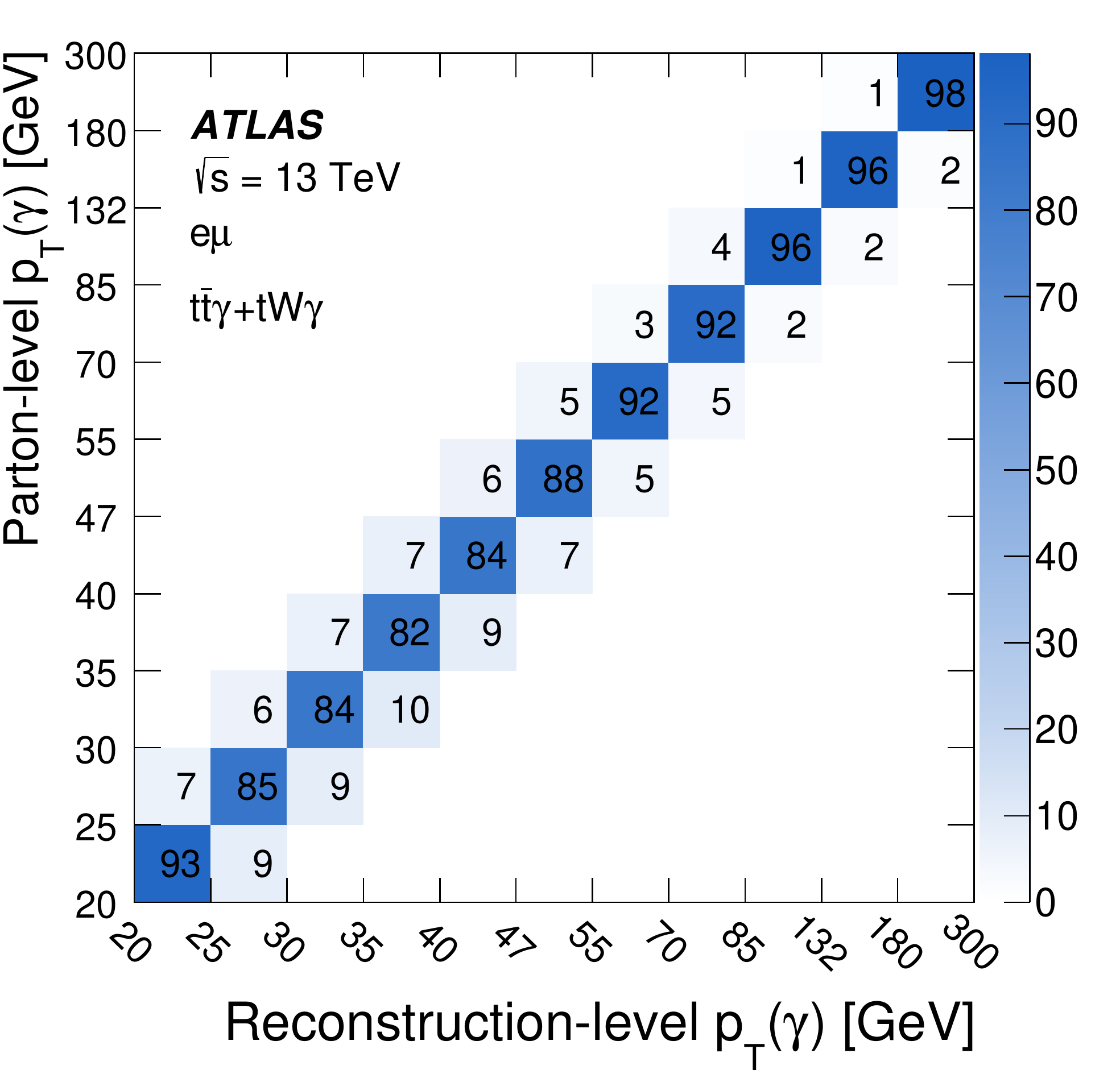}\includegraphics[width=0.5\linewidth]{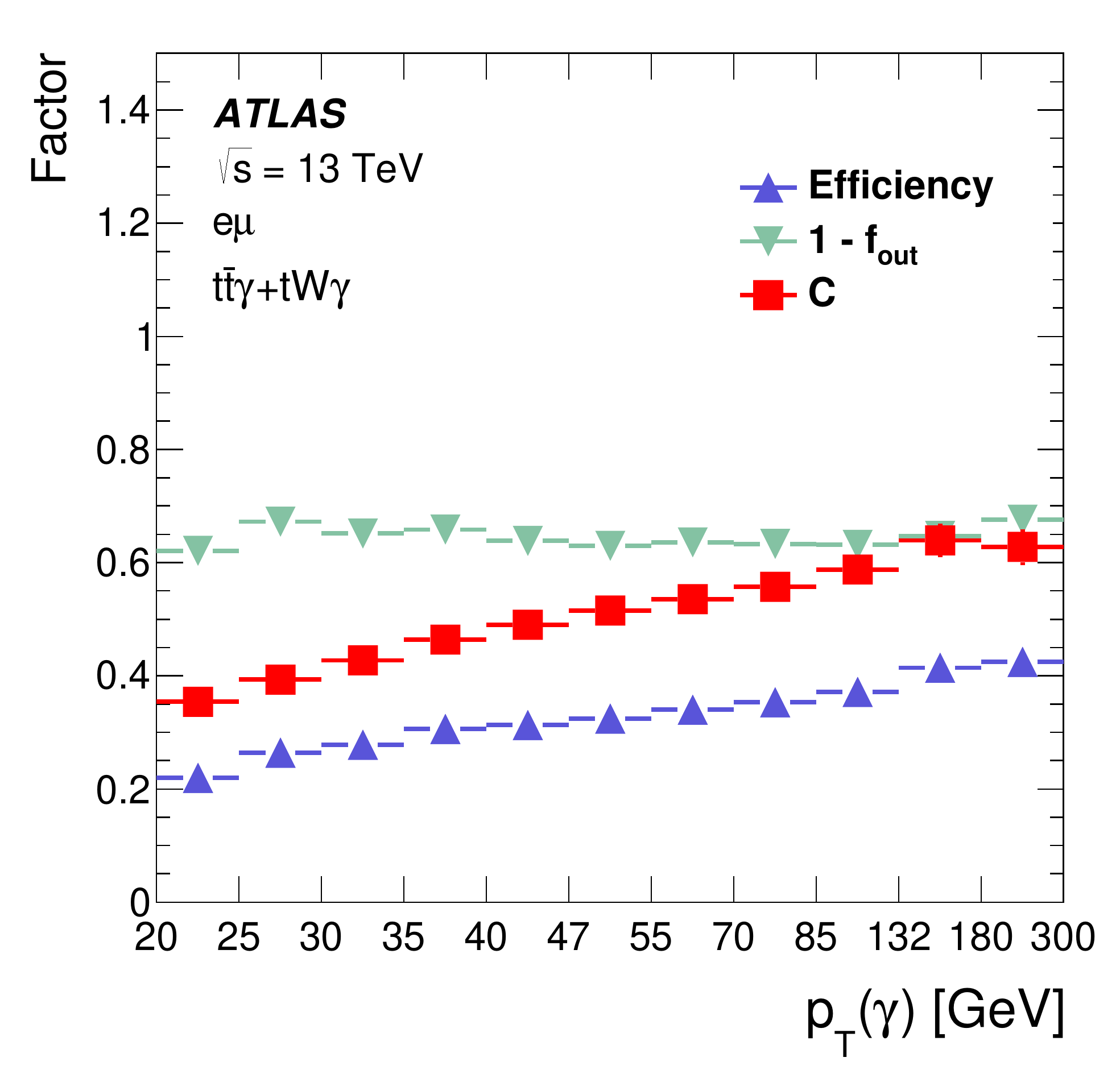}
\caption [] {Left: migration matrix relating the photon \pT at the reconstruction and parton levels in the fiducial phase space, normalised by column and shown as percentages. Right: signal reconstruction and selection efficiency ($\epsilon$), ($1-f_{\mathrm{out}}$) fraction and resulting $C$ correction factor as a function of the photon \pT.}
\label{fig:effmigsl_pt}
\end{figure}

\section{Systematic uncertainties}
\label{sec:systematics}
 
Various systematic uncertainties arising from detector effects are considered, along with theoretical uncertainties.
Signal and background predictions are both subject to these uncertainties.
 
\subsection{Experimental uncertainties}
 
Experimental systematic uncertainties affect the normalisation and shape of the distributions of the simulated signal and background samples. These include reconstruction and identification efficiency uncertainties, as well as uncertainties in the energy and momentum scale and resolution for the reconstructed physics objects in the analysis, including leptons, photons, jets and \met. In addition, uncertainties in the flavour-tagging of jets, the jet vertex tagger (JVT) discriminant, the integrated luminosity value and the pile-up simulation are considered.
 
The photon identification and isolation efficiencies as well as the efficiencies of the lepton reconstruction, identification, isolation, and trigger in the MC samples are all corrected using scale factors to match the corresponding values in data. Similarly, corrections to the lepton and photon momentum scale and resolution are applied in simulation~\cite{PERF-2017-03,PERF-2015-10}. All these corrections, which are \pT and $\eta$ dependent, are varied within their uncertainties.
 
The jet energy scale (JES) uncertainty is derived using a combination of simulations, test-beam data and \textit{in situ} measurements~\cite{PERF-2016-04}. Additional contributions from jet-flavour composition, $\eta$-intercalibration, punch-through, single-particle response, calorimeter response to different jet flavours, and pile-up are taken into account, resulting in 30 uncorrelated JES uncertainty subcomponents, of which 29 are non-zero in a given event depending on the type of simulation used. The most relevant JES uncertainties are related to the pile-up correction (\textit{JES pile-up correction}) and modelling aspects of the in situ calibration (\textit{JES in situ calibration}). The jet energy resolution (JER) in simulation is smeared by the measured JER uncertainty~\cite{ATL-PHYS-PUB-2015-015} split into eight uncorrelated sources. The uncertainty associated with the JVT discriminant is obtained by varying the efficiency correction factors (labelled \textit{jet vertex tagging} in the results, cf.\ Figure~\ref{fig:ranking}).
 
The uncertainties related to the $b$-jet tagging calibration are determined separately for $b$-jets, $c$-jets and light-flavour jets~\cite{Aaboud:2018xwy,ATLAS-CONF-2018-001,ATLAS-CONF-2018-006}.
For each jet category, the uncertainties are decomposed into several uncorrelated components.
The corrections are varied by their measured uncertainties.
 
The uncertainties associated with energy scales and resolutions of photons, leptons and jets are propagated to the \MET. Additional uncertainties originate from the modelling of its soft term~\cite{ATLAS-CONF-2013-082}.
 
The uncertainty in the combined 2015--2018 integrated luminosity is 1.7\%~\cite{ATLAS-CONF-2019-021}, obtained using the LUCID-2 detector \cite{LUCID2} for the primary luminosity measurements.
 
The uncertainty associated with the modelling of pile-up in the simulation is assessed by varying the pile-up reweighting in the simulation within its uncertainties.

\subsection{Signal and background modelling uncertainties}
\label{sec:modellingsyst}
 
The \ttg signal modelling uncertainties include the uncertainties owing to the choice of QCD scales, parton shower, amount of initial-state radiation (ISR), and PDF set. The effect of the QCD scale uncertainty is evaluated by varying the renormalisation and factorisation scales separately up and down by a factor of two from their nominal chosen values. The uncertainty from the parton shower and hadronisation (\textit{\tty PS model}) is estimated by comparing the \tty nominal samples, produced with \madgraph + \pythia~8, with an alternative sample interfaced to \herwig~7~\cite{Bahr:2008pv,Bellm:2015jjp}. The ISR uncertainty (\textit{\tty ISR}) is studied by comparing the nominal \madgraph + \pythia~8 sample with the results of varying the A14 tune parameter for radiation~\cite{ATL-PHYS-PUB-2014-021}.
The PDF uncertainty (\textit{\tty PDF}) is evaluated using the standard deviation in each bin of the respective distribution formed by the set of 100 replicas of the NNPDF set~\cite{Pumplin:2002vw}.
 
For the \Wty process the uncertainties due to the choice of renormalisation and factorisation scales are also estimated by varying them up and down separately by a factor of two relative to the nominal sample value.
A systematic uncertainty from the parton shower and hadronisation model is considered by comparing \PYTHIA~8 and \herwig~7 both interfaced to \madgraph.
The \Wty modelling uncertainties are treated as uncorrelated with the \ttg signal modelling uncertainties.
 
The \Wty process was generated in the five-flavour scheme at leading order in QCD and one of the two \bquarks is not included in the matrix-element generation step.
This \bquark, expected to be produced in the initial state through the PDF, is only found in a fraction of the events at parton level in the MC simulation.
The fractions of generated \Wty events without a second \bquark were found to be around 30\% and 50\% for the MC samples interfaced with \herwig and \pythia, respectively.
Therefore, an additional uncertainty associated with this possibly lost \bquark is assigned (\textit{\Wty parton definition}) as follows.
Relative to the nominal \Wty simulation, the parton-level event yields are doubled, assuming all $b$-jets are found, while the number of reconstructed events is kept constant. This leads to a variation of the correction factor~$C$ of 2.8\%.
 
Several uncertainties in the modelling of \ttbar processes, which give a dominant contribution to the \cathfake and \catprompty background categories, are considered as shape-only uncertainties.
The uncertainties associated with the parton shower and hadronisation are estimated by comparing the nominal simulation with alternative showering by \herwig~7.
Uncertainties in the modelling of final-state radiation are estimated by evaluating the effects of varying four different parameters in the \powheg + \pythia~8 generator set-up described in the following.
Uncertainties due to the renormalisation and factorisation scales are estimated by varying them up and down independently by a factor of two relative to the default scale choice.
These scale variations are implemented with corresponding weights which are available as part of the nominal MC sample.
Uncertainties due to the value of \alphas used in the ISR parton shower modelling are estimated by comparing the nominal \powheg + \pythia~8 simulation with alternative samples that correspond to higher and lower radiation parameter settings in the A14 tune, controlled by the \textit{var3c} parameter in \pythia~8.
This parameter is varied within its uncertainties corresponding to variations of $\alpha_S(m_Z)$ between 0.115 and 0.140.
An additional ISR uncertainty is obtained by comparing the nominal sample with an additional one where the $h_{\mathrm{damp}}$ parameter, which controls the \pt of the first additional emission, is varied by a factor of two
as supported by measurements reported in Ref.~\cite{ATL-PHYS-PUB-2018-009}.
 
In addition to those background modelling uncertainties, global normalisation uncertainties of 50\% are assigned to the following three categories: \hfakes, \efakes and \catprompty background~\cite{TOPQ-2017-14} (\textit{h-fakes}, \textit{e-fakes}, and \textit{\catprompty normalisation}).

\subsection{Treatment of the systematic uncertainties in the measurements}
\label{sec:systtreatment}
 
As stated in Section~\ref{sec:strategy}, the impact of systematic uncertainties on the fiducial inclusive cross-section measurement is taken into account via nuisance parameters in the likelihood function.
The nuisance parameters $\vec{\theta}$ are profiled in the maximum-likelihood fit.
Variations of the nuisance parameters can affect the rate of events as well as the shape of the \ST distribution.
In the case of signal modelling uncertainties, the rate uncertainty is composed of variations of the efficiency $\epsilon$ and the fraction $f_{\mathrm{out}}$.
All MC samples used to evaluate signal modelling uncertainties are scaled to the same  number of events in the fiducial phase space, $N_{\textrm{MC}}^{\textrm{fid}}$.
The only uncertainty that is not included as a nuisance parameter in the profile likelihood fit is the uncertainty from the \Wty parton definition.
This uncertainty does not affect the number of reconstructed events in the corresponding template in the profile likelihood fit.
It comprises only an uncertainty in the number of generated events in the fiducial phase space.
Thus, the \Wty parton definition uncertainty is added in quadrature to the post-fit uncertainty of the profile likelihood fit.
 
To reduce the sensitivity to statistical fluctuations due to the limited number of events in the MC samples used in systematic variations, \emph{smoothing} techniques are applied to the MC templates used to evaluate the signal and background modelling systematic uncertainties in the template fit. Additionally, the systematic uncertainties are symmetrised, taking the average of the up- and down-variation as the uncertainty. In the cases where both variations have the same sign or only one variation is available (e.g.\ the uncertainty from the parton shower and hadronisation signal modelling) the largest variation or the available one, respectively, is taken as both the up- and down-variations for the corresponding source.
The ISR uncertainty suffers from statistical fluctuations in the available \tty MC samples, so a more conservative approach is chosen for the symmetrisation.
In this case, the largest of the two variations is taken and mirrored around the nominal prediction.
 
In the case of the differential cross-section measurements, each systematic uncertainty is determined individually in each bin of the measurement by varying the corresponding efficiency, resolution, and model parameter within its uncertainty. The same symmetrisation approach described for the fiducial inclusive cross-section is used for this measurement. For each variation, the measured differential cross-section is recalculated and the deviation from the nominal result per bin is taken as the systematic uncertainty. The overall uncertainty in the measurement is then derived by adding all contributions in quadrature, assuming the sources of systematic uncertainty to be fully uncorrelated.
 
Sources of systematic uncertainty relating only to the background prediction are evaluated by shifting the nominal distribution of the corresponding background process by its associated uncertainty. For the experimental uncertainties, the input is varied by the corresponding shift, which typically affects both the shape and normalisation of signal and background process distributions. The resulting distribution is unfolded and compared with the nominal unfolded distribution and the difference is assigned as an uncertainty. The systematic uncertainties due to signal modelling are evaluated by varying the signal corrections, i.e.\ the migration matrix $M_{kj}$, the efficiency $\epsilon_k$ and the fraction $f_{\mathrm{out},j}$, by the corresponding model parameter uncertainty and calculating the difference between the resulting unfolded distributions and the nominal ones.

\section{Fiducial inclusive cross-section measurement}
\label{sec:inclusive}
 
The number of signal events is extracted using a profile likelihood fit to the \ST distribution and is translated into the signal cross-section in the fiducial phase space given by the kinematic boundaries of the signal as described in Section~\ref{sec:strategy}.
 
The best-fit values of the nuisance parameters ranked highest in impact are shown in Figure~\ref{fig:ranking} along with their impact on the result.
Rate and shape uncertainties from the \tty PS model and \tty ISR variations are treated as separate nuisance parameters. This approach prevents pulls on the rate uncertainty due to differences in the shape of the \ST distribution between the data and simulation, in particular in the tail where the data overshoot the prediction and the fit compensates for this discrepancy by pulling the nuisance parameter of the \tty PS model shape uncertainty.
The impact of the individual nuisance parameters is evaluated as the difference between the reference best-fit value of the cross-section and the one obtained when fixing the corresponding nuisance parameter under scrutiny to its best-fit value and its $\pm$ one standard deviation $(\pm 1\sigma)$. Table~\ref{tab:syst_impact} shows the systematic uncertainties and their relative impact on the measurement of the fiducial inclusive cross-section. The effect of each category of uncertainties is calculated from the variance $(\sigma^2)$ difference between the total uncertainty in the measured fiducial cross-section and the uncertainty from the fit with the corresponding nuisance parameters fixed to their fitted values.
The uncertainties in the signal modelling, especially the rate uncertainties from the \tty PS model and the ISR variation, have the largest impact on the result.
 
\begin{figure}
\centering
\includegraphics[scale=0.55]{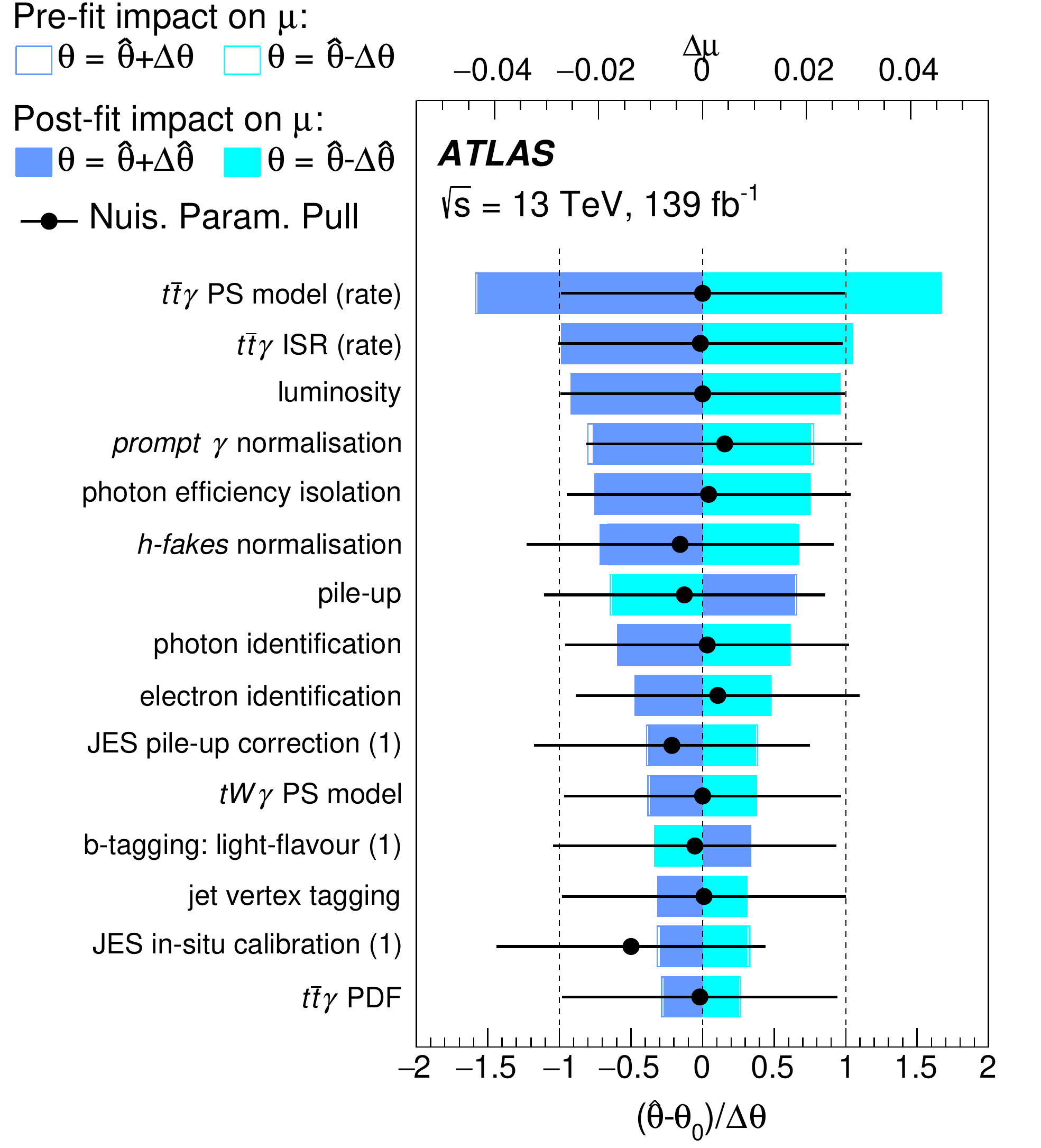}
\caption{
Ranking of the systematic uncertainties included in the profile likelihood fit used in the fiducial inclusive cross-section measurement.
The blue and turquoise bands indicate the post-fit impact on the fit result, whereas the outlined blue and turquoise rectangles show the pre-fit impact.
The difference between the two reflects the constraint of the nuisance parameter due to correlations in the fit.
Most nuisance parameters are not or only marginally constrained.
The impact is overlaid with the post-fit values of the nuisance parameters (pulls) shown by the black dots.
The black lines represent the post-fit uncertainties normalised to the pre-fit uncertainties.
For uncertainties parameterised with more than one nuisance parameter, the index (1) refers to the leading component.
}
\label{fig:ranking}
\end{figure}
 
\begin{table}
\centering
\caption{Illustrative summary of the systematic uncertainties on the fiducial inclusive cross-section measurement grouped into different categories and their relative impact on the measurement (symmetrised).
The categories `\tty/\tWy modelling' and `Background modelling' include all corresponding systematic uncertainties described in Section~\ref{sec:modellingsyst}.
The `\Wty parton definition' uncertainty is listed separately since it does not enter the profile likelihood fit directly as described in Section~\ref{sec:systtreatment}.
The category `Photons' corresponds to the uncertainties related to photon identification and isolation as well as photon energy scale and resolution.
`Jets' includes the total uncertainty from the JES, JER and JVT discriminant, while the $b$-tagging-related uncertainties are given in a separate category (`Flavour-tagging').
The category `Leptons' represents the uncertainties related to lepton identification, isolation and energy/momentum calibration. }
\label{tab:syst_impact}
\begin{tabular}{l S}
\toprule
Category                & {Uncertainty} \\
\midrule
\tty/\tWy modelling  & 3.8\% \\
Background modelling & 2.1\% \\
Photons              & 1.9\% \\
Luminosity           & 1.8\% \\
Jets                 & 1.6\% \\
Pile-up              & 1.3\% \\
Leptons              & 1.1\% \\
Flavour-tagging      & 1.1\% \\
MC statistics        & 0.4\% \\
Soft term \met       & 0.2\% \\
\midrule
\tWy parton definition & 2.8\% \\
\midrule
Total syst.           & 6.3\% \\
\bottomrule
\end{tabular}
\end{table}
 
The distribution of the fitted \ST variable is shown in Figure~\ref{fig:postfit}.
The dashed band represents the post-fit uncertainties.
The expected yields after the fit describe the data well.

\begin{figure}
\centering
\includegraphics[width=0.5\textwidth]{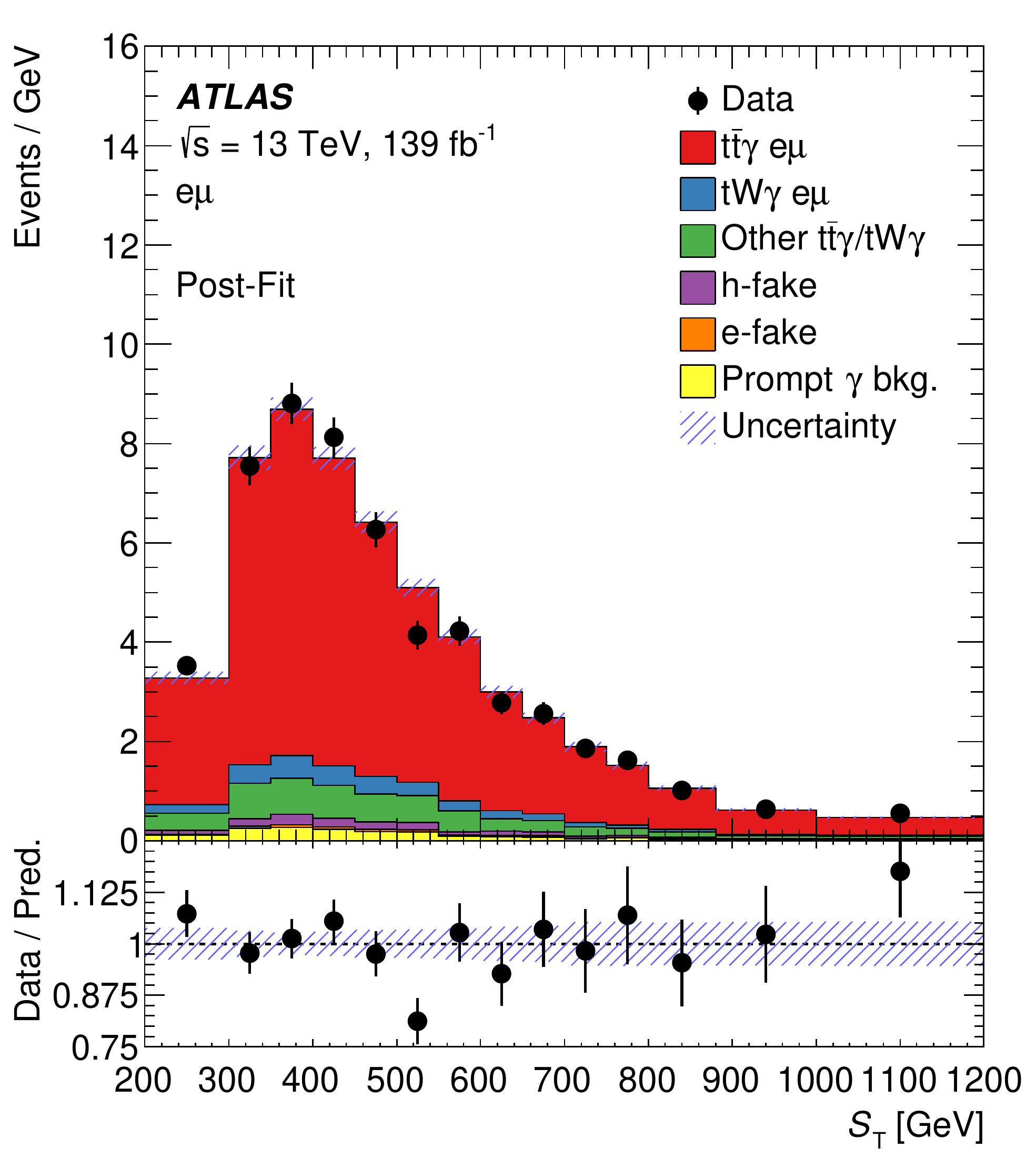}
\caption{
Post-fit distribution of the \ST variable. The uncertainty band represents the post-fit uncertainties. Underflow and overflow events are included in the first and last bins of the distribution, respectively. The lower part of the plot shows the ratio of the data to the prediction.
}
\label{fig:postfit}
\end{figure}
 
Extrapolated to the fiducial phase space using the correction factor $C$, the fit result corresponds to a fiducial inclusive cross-section for the combined \tty/\tWy process in the \chemu channel of
$\sigma_{\text{fid}}=39.6 \pm 0.8\,\textrm{(stat)} \,^{+2.6}_{-2.2}\,\textrm{(syst)}\,\textrm{fb} = 39.6\,^{+2.7}_{-2.3} \,\textrm{fb}$
. The measured cross-section is in good agreement with the dedicated theoretical calculation provided by the authors of Refs.~\cite{Bevilacqua:2018woc,Bevilacqua:2018dny}, which predicts a value of $\sigma_{\text{fid}}=38.50\,^{+0.56}_{-2.18}\,\textrm{(scale)}\,^{+1.04}_{-1.18}\,\textrm{(PDF)}\,\textrm{fb}$ for the chosen fiducial phase space using the CT14 PDF set~\cite{Dulat:2015mca}. The uncertainty in the theory prediction includes uncertainties owing to the scales and PDF. The PDF uncertainty is rescaled to the 68\% CL. In the theoretical calculation, the renormalisation and factorisation scales are chosen as 1/4 of the total transverse momentum of the system, defined as the scalar sum of the \pt of the leptons, $b$-jets, photon and the total missing \pt from the neutrinos. The mass of the top quark is set to 173.2~\GeV. The electroweak coupling in the calculation is derived from the Fermi constant $G_{\mu}$ and it is set to $\alpha_{G_{\mu}}\approx 1/132$, while it is 1/137 for the leading emission. Further details can be found in~Ref.~\cite{Bevilacqua:2018woc}.

\section{Differential cross-section measurements}
\label{sec:differential}
 
The absolute differential cross-sections are shown in Figure~\ref{fig:absolute_xsec} while the normalised measured differential cross-sections are presented in Figure~\ref{fig:normalised_xsec}. The cross-sections are compared with the NLO calculation in the same fiducial phase space and with the combination of the \tty and \Wty LO \madgraph simulations interfaced with  \PYTHIA8 and \herwig7, referred to as \madgraphshort+\pythia{}8 and \madgraphshort+\herwig{}7 in the following plots and tables. The calculated $\chi^2$/ndf values for the absolute and normalised cross-sections and their corresponding $p$-values are summarised in Tables~\ref{tab:chi2_absolute} and~\ref{tab:chi2_normalised}, quantifying the probability of compatibility between data
and each of the predictions. The $\chi^2$ values are calculated as:
\begin{equation*}
\chi^2 = \sum_{j,k} (\sigma_{j,\textrm{data}} - \sigma_{j,\textrm{pred.}})
\cdot C_{jk}^{-1} \cdot (\sigma_{k,\textrm{data}} - \sigma_{k,\textrm{pred.}})\,,
\end{equation*}
where $\sigma_{\textrm{data}}$ and $\sigma_{\textrm{pred.}}$ are the unfolded and predicted differential cross-sections, $C_{jk}$ is the covariance matrix of $\sigma_{\textrm{data}}$, calculated as the sum of the covariance matrix for the statistical uncertainty and the covariance matrices for the systematic uncertainties, and $j$ and $k$ are the binning indices of the distribution. The covariance matrix for each of the systematic uncertainties is estimated as $\sigma_j \times \sigma_k$, where $\sigma_j$ and $\sigma_k$ are the symmetrised uncertainties for bin $j$ and bin $k$ of the unfolded distribution. In the case of the normalised differential cross-sections, the last bin is removed from the $\chi^2$ calculation and the number of degrees of freedom is reduced by one.
 
The shape of the measured differential distributions is generally well described by both the LO MC predictions from \madgraph and the NLO theory prediction. The latter tends to describe the shape of the measured distribution slightly better. The shapes of $\Delta R (\gamma, \ell)_{\mathrm{min}}$ and $\Delta\phi(\ell,\ell)$ are not perfectly modelled by the \madgraph simulation, while the NLO prediction provides a better description of these distributions.
 
\begin{figure}[!htbp]
\centering
\includegraphics[width=0.45\textwidth]{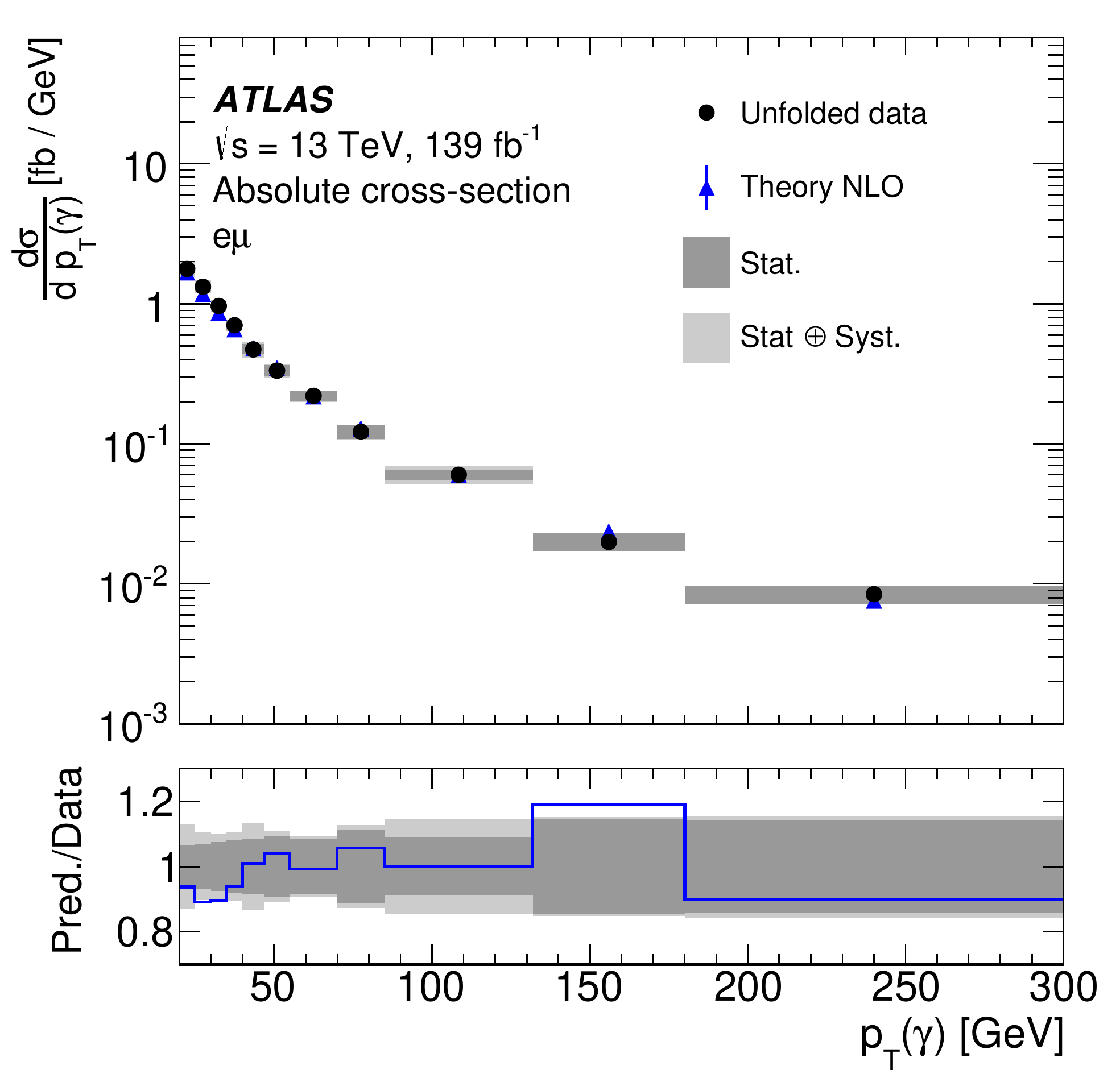}
\includegraphics[width=0.45\textwidth]{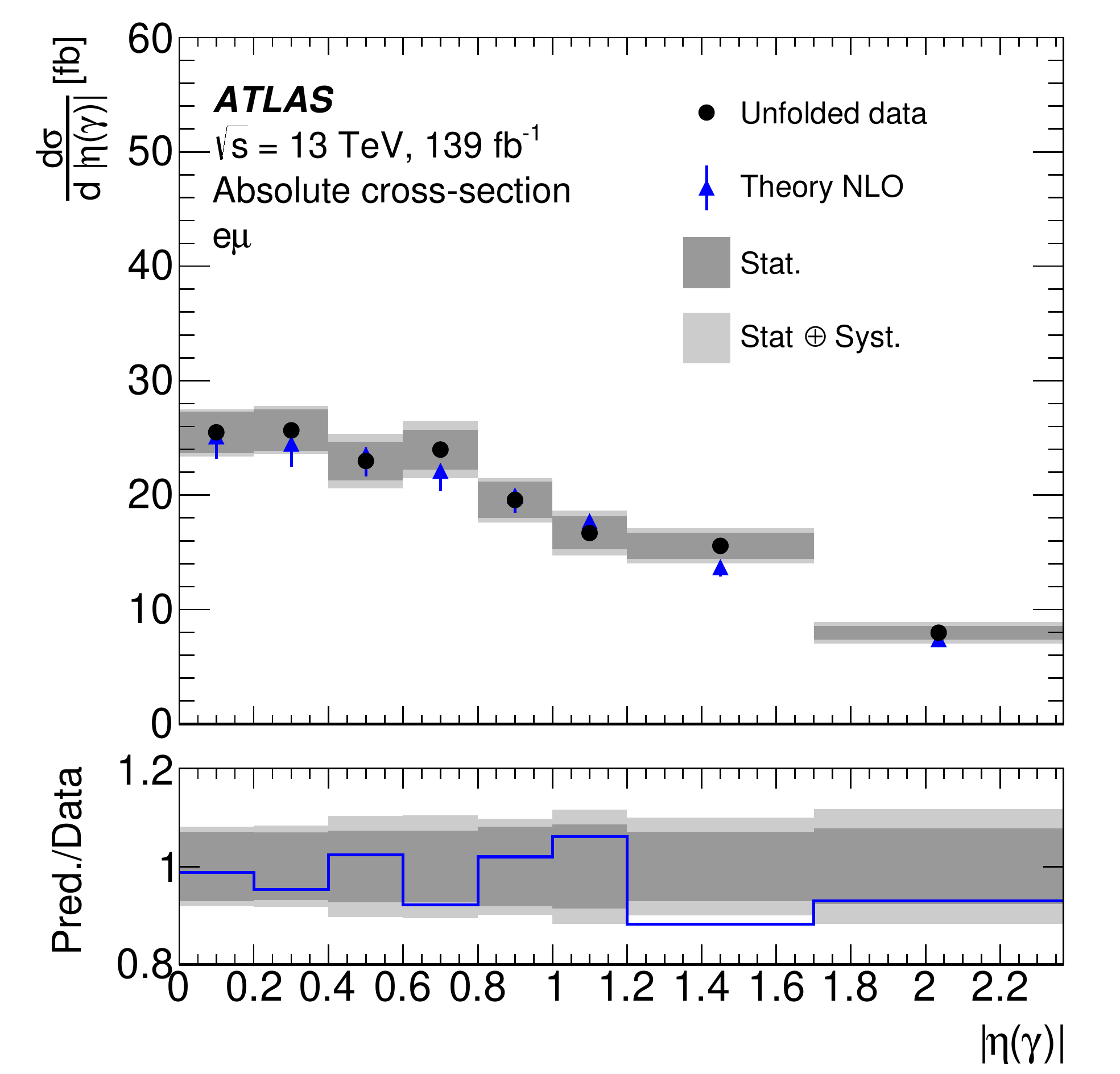}
\includegraphics[width=0.45\textwidth]{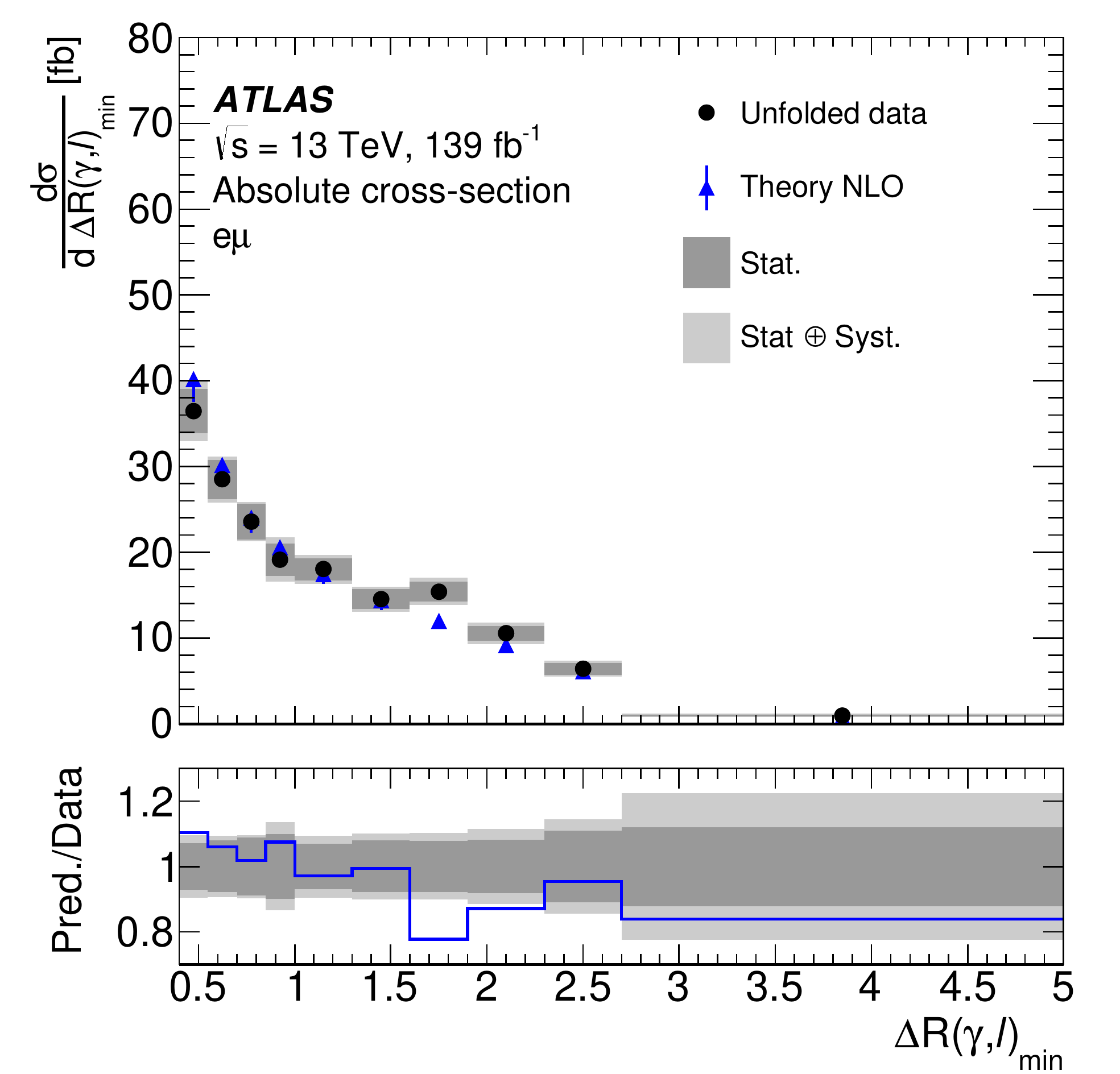}
\includegraphics[width=0.45\textwidth]{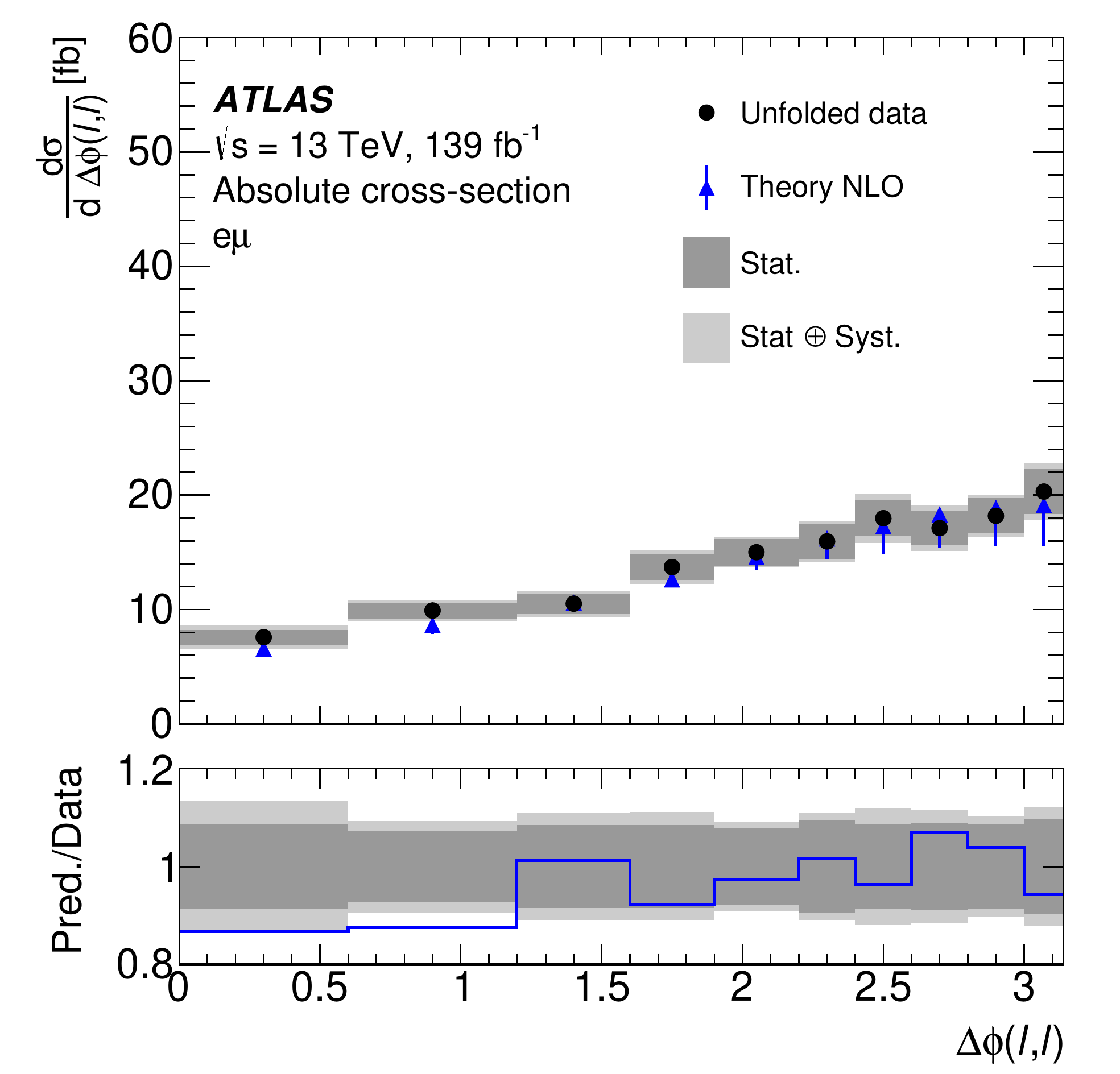}
\includegraphics[width=0.45\textwidth]{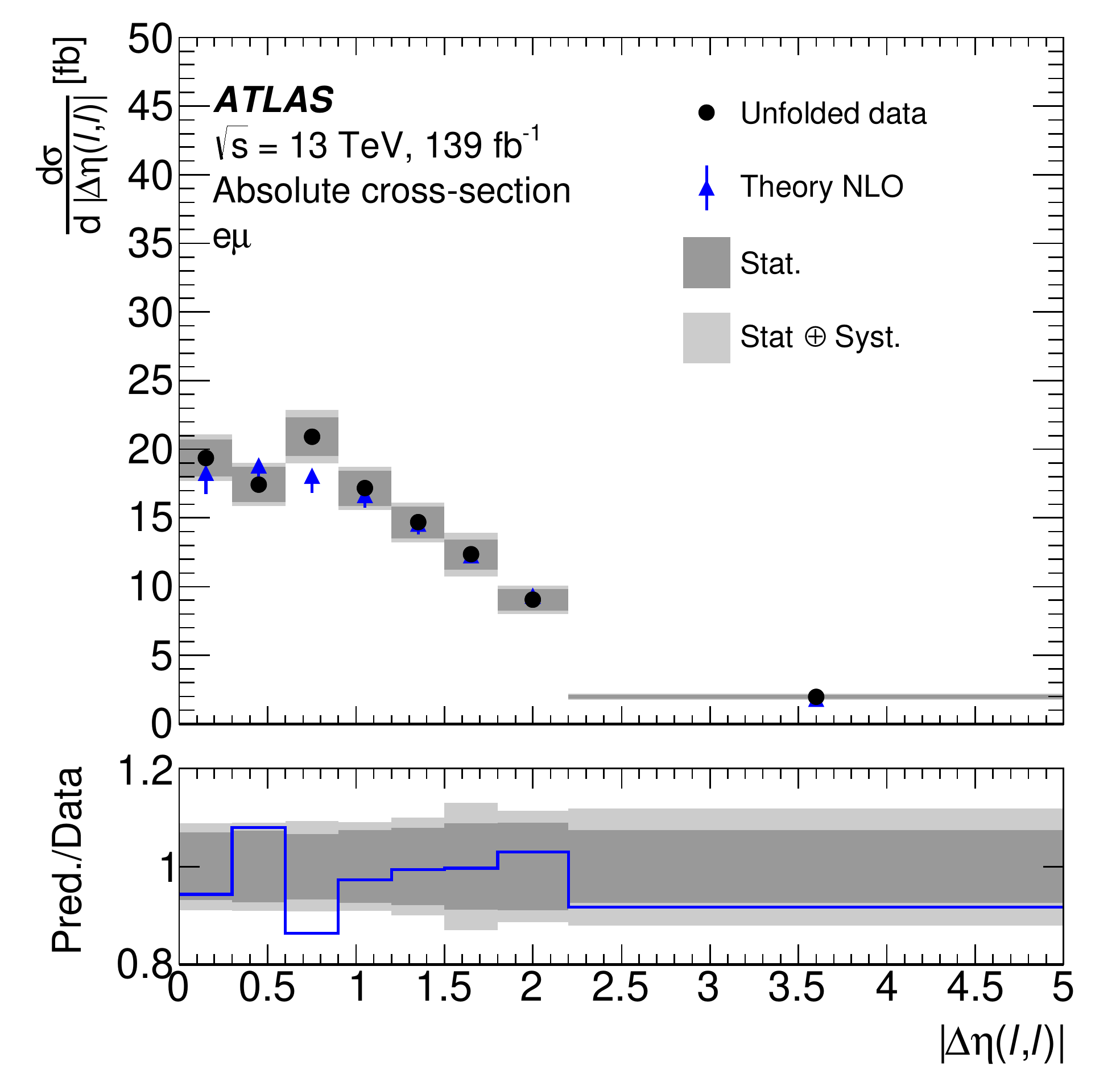}
\caption{Absolute differential cross-section measured in the fiducial phase space as a function of the photon \pt, photon $|\eta|$, $\Delta R (\gamma, \ell)_{\mathrm{min}}$, $\Delta\phi(\ell,\ell)$, and $|\Delta\eta(\ell,\ell)|$  (from left to right and top to bottom). Data are compared with the NLO calculation provided by the authors of Refs.~\cite{Bevilacqua:2018woc,Bevilacqua:2018dny}. The uncertainty in the calculation corresponds to the total scale and PDF uncertainties. The PDF uncertainty is rescaled to the 68\% CL. The lower part of each plot shows the ratio of the prediction to the data.}
\label{fig:absolute_xsec}
\end{figure}
 
\begin{figure}[!htbp]
\centering
\includegraphics[width=0.44\textwidth]{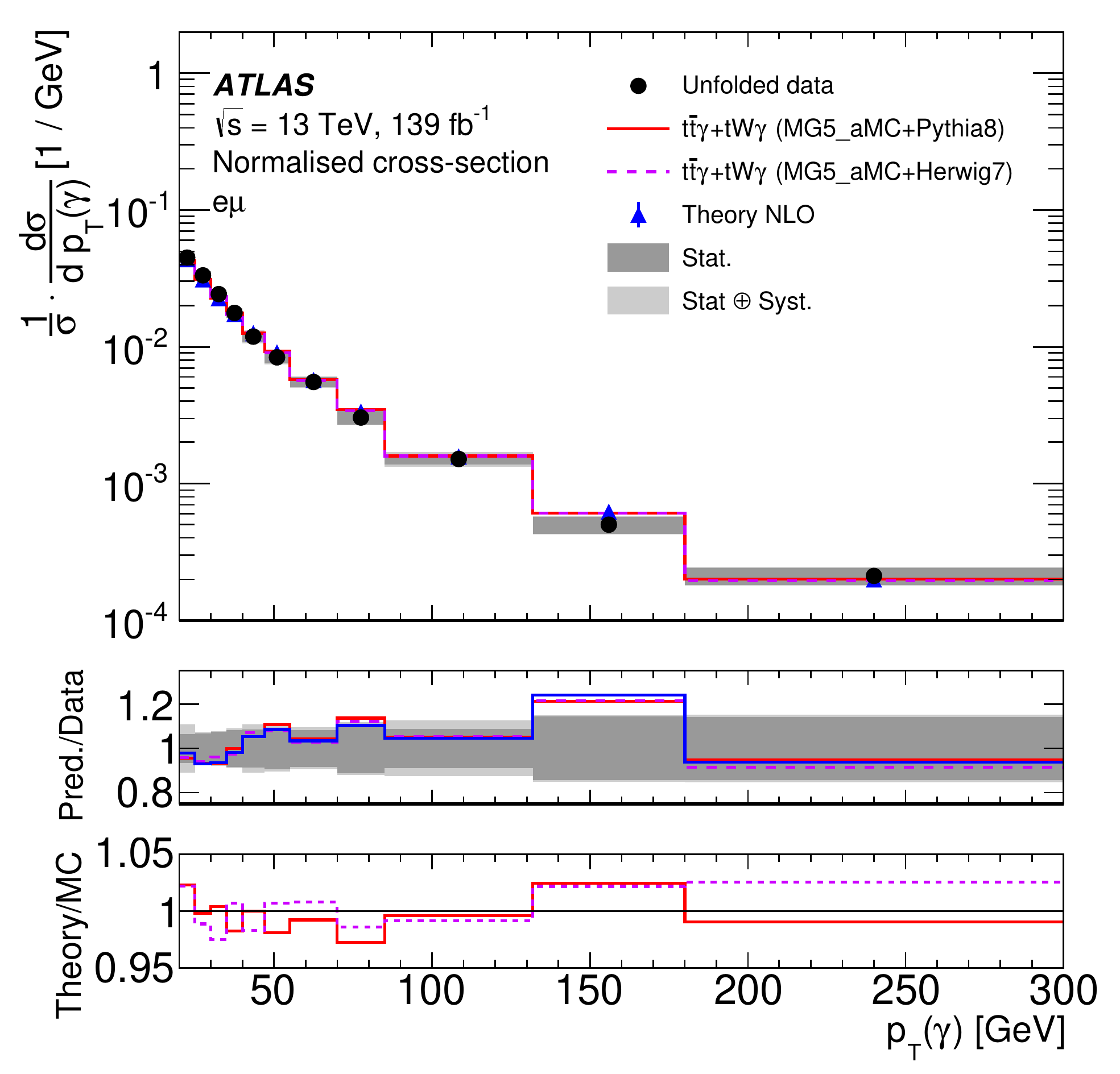}
\includegraphics[width=0.44\textwidth]{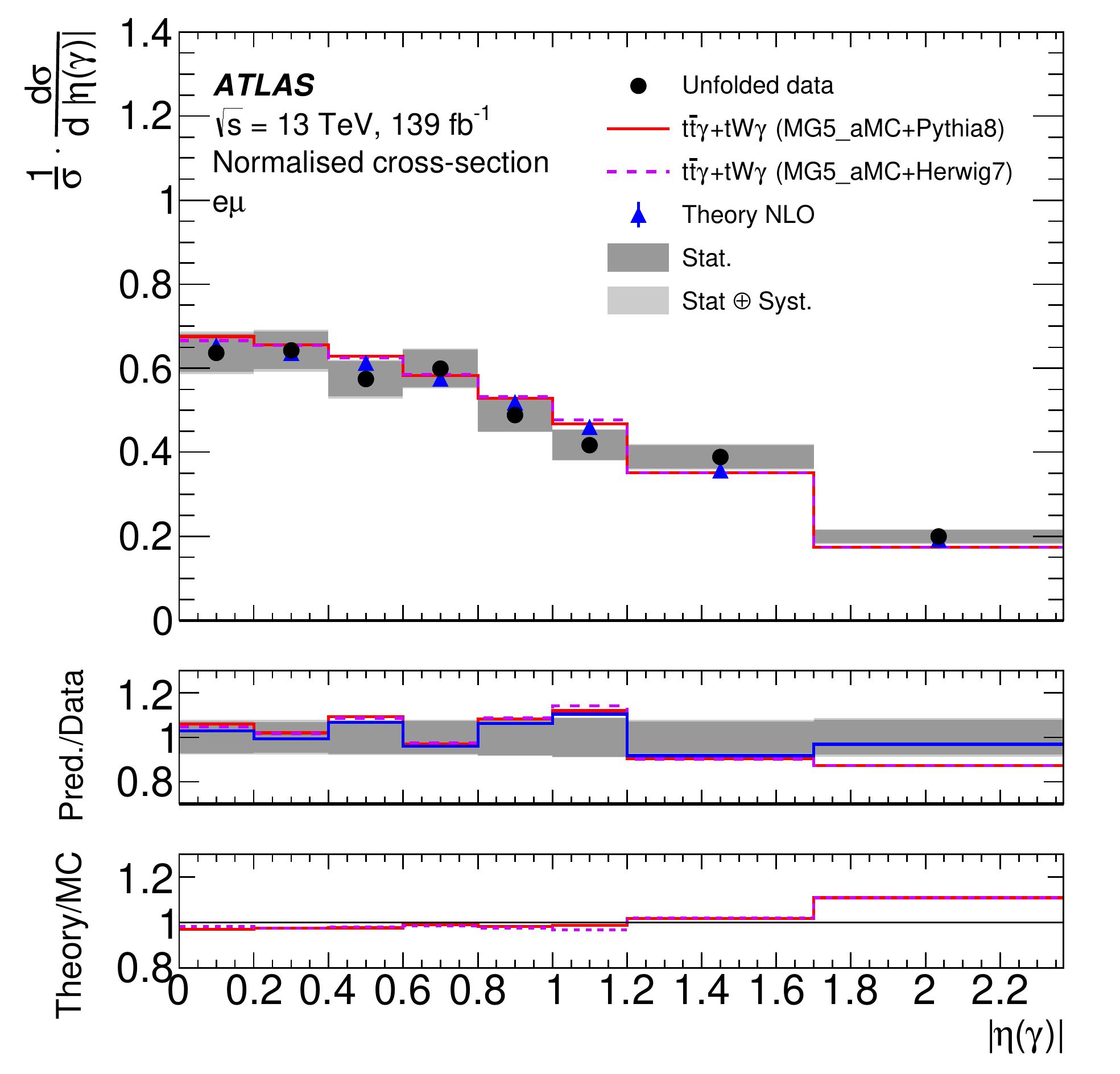}
\includegraphics[width=0.44\textwidth]{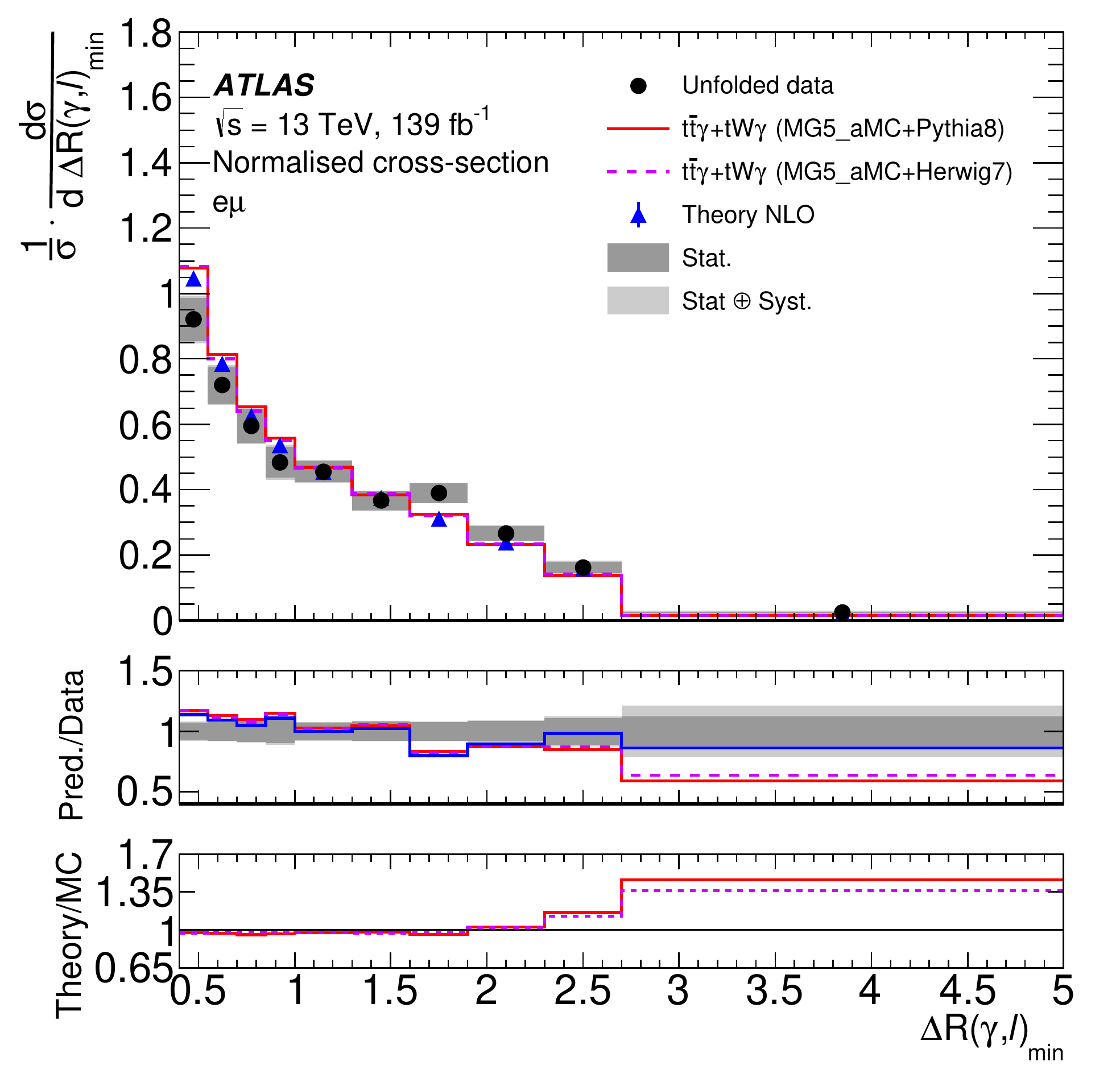}
\includegraphics[width=0.44\textwidth]{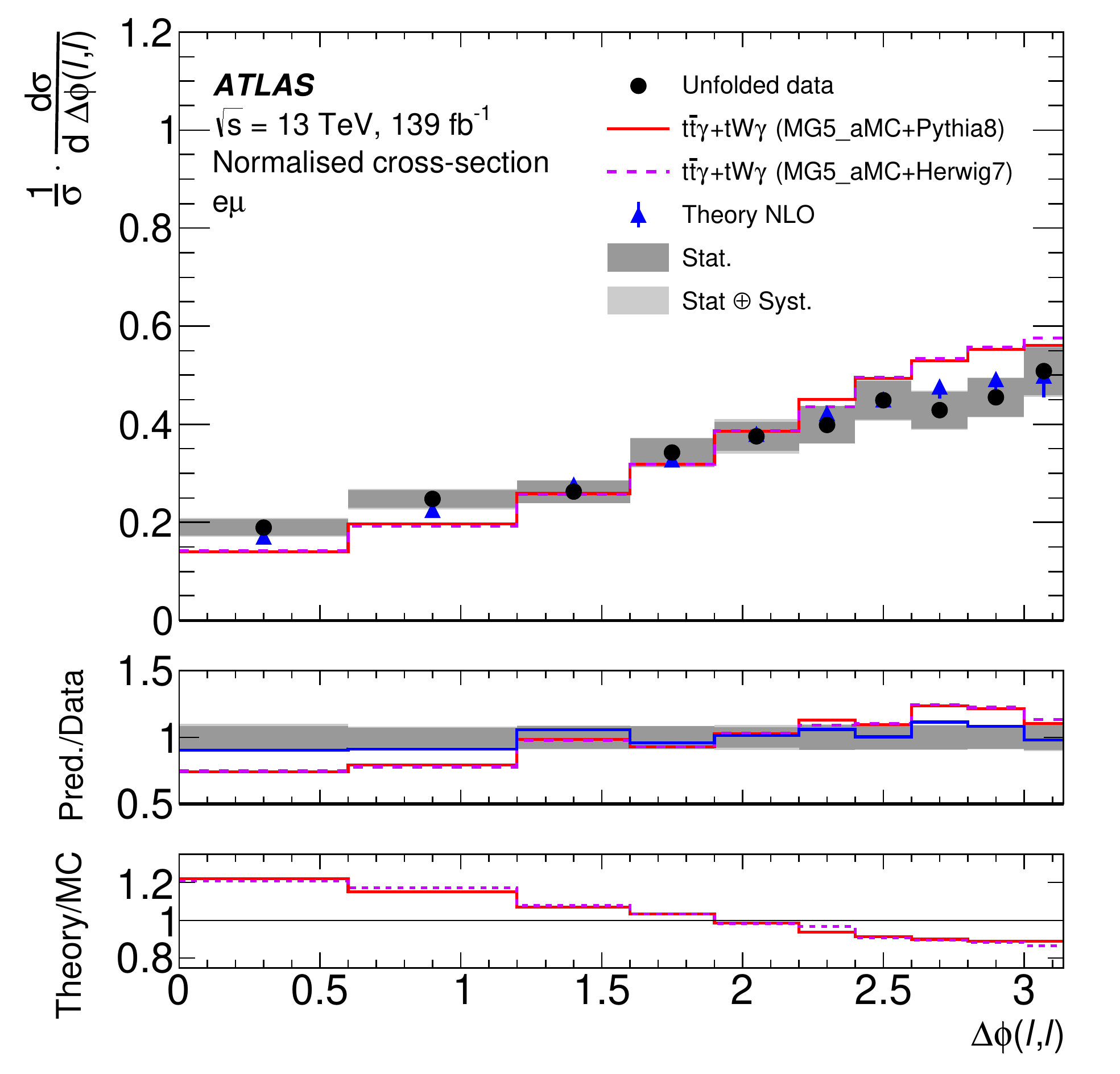}
\includegraphics[width=0.44\textwidth]{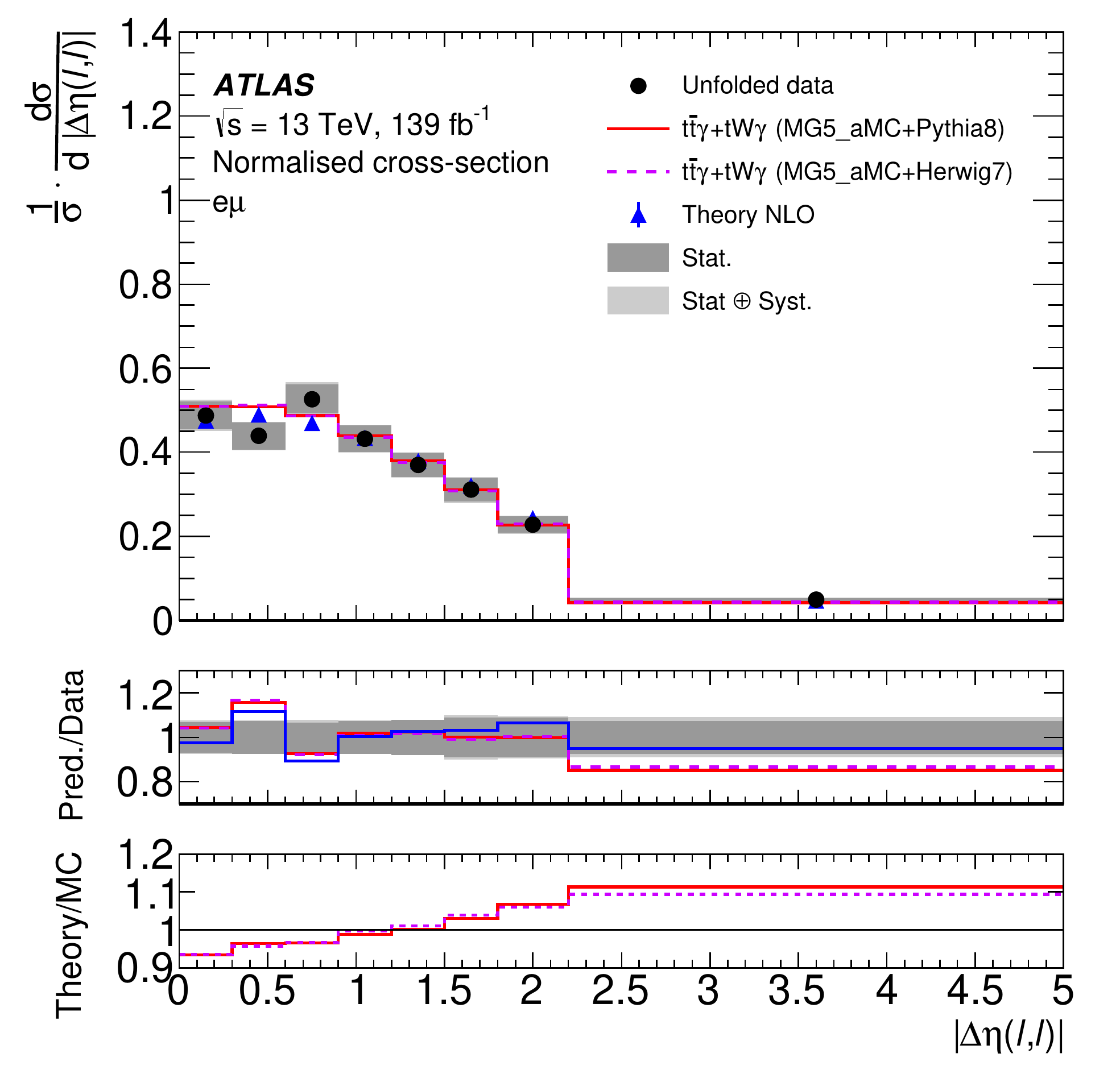}
\caption{Normalised differential cross-section measured in the fiducial phase space as a function of the photon \pt, photon $|\eta|$, $\Delta R (\gamma, \ell)_{\mathrm{min}}$, $\Delta\phi(\ell,\ell)$, and $|\Delta\eta(\ell,\ell)|$  (from left to right and top to bottom). Data are compared with the NLO calculation provided by the authors of Refs.~\cite{Bevilacqua:2018woc,Bevilacqua:2018dny} and the \madgraph simulation interfaced with \PYTHIA8 and \herwig7. The uncertainty in the calculation corresponds to the total scale and PDF uncertainties. The PDF uncertainty is rescaled to the 68\% CL. The lower parts of each plot show the ratio of the prediction to the data and the ratio of the NLO calculation to the MC simulations.}
\label{fig:normalised_xsec}
\end{figure}
 
\begin{table}[!htbp]
\scriptsize
\centering
\caption{$\chi^2$/ndf and $p$-values between the measured absolute cross-sections and the NLO calculation.}
\setlength{\tabcolsep}{4.9pt}
\begin{tabular}{l | c c | c c | c c | c c | c c }
\toprule
& \multicolumn{2}{c}{$p_\mathrm{T}(\gamma)$} & \multicolumn{2}{c}{$|\eta(\gamma)|$} & \multicolumn{2}{c}{$\Delta R (\gamma,\ell)_{\mathrm{min}}$} &
\multicolumn{2}{c}{$\Delta\phi(\ell,\ell)$} & \multicolumn{2}{c}{$|\Delta\eta(\ell,\ell)|$} \\
Predictions & $\chi^2$/ndf & $p$-value & $\chi^2$/ndf & $p$-value & $\chi^2$/ndf & $p$-value & $\chi^2$/ndf & $p$-value & $\chi^2$/ndf & $p$-value \\
\midrule
Theory NLO  & 6.1/11 & 0.87 & 4.5/8 & 0.81 & 11.7/10 & 0.31 & 5.8/10 & 0.83 & 6.2/8 & 0.62\\
\bottomrule
\end{tabular}
\label{tab:chi2_absolute}
\end{table}
 
\begin{table}[!htbp]
\scriptsize
\centering
\caption{$\chi^2$/ndf and $p$-values between the measured normalised cross-sections and various predictions from the MC simulation and the NLO calculation.}
\setlength{\tabcolsep}{4.9pt}
\begin{tabular}{l | c c | c c | c c | c c | c c }
\toprule
& \multicolumn{2}{c}{$p_\mathrm{T}(\gamma)$} & \multicolumn{2}{c}{$|\eta(\gamma)|$} & \multicolumn{2}{c}{$\Delta R (\gamma,\ell)_{\mathrm{min}}$} &
\multicolumn{2}{c}{$\Delta\phi(\ell,\ell)$} & \multicolumn{2}{c}{$|\Delta\eta(\ell,\ell)|$} \\
Predictions & $\chi^2$/ndf & $p$-value & $\chi^2$/ndf & $p$-value & $\chi^2$/ndf & $p$-value & $\chi^2$/ndf & $p$-value & $\chi^2$/ndf & $p$-value \\
\midrule
\tty + \Wty (\madgraphshort+\pythia{}8) & 6.3/10 & 0.79 & 7.3/7 & 0.40  & 20.1/9 & 0.02 & 30.8/9 & <0.01 & 6.5/7 & 0.48\\
\tty + \Wty (\madgraphshort+\herwig{}7) & 5.3/10 & 0.87 & 7.7/7 & 0.36 & 18.9/9 & 0.03 & 31.6/9 & <0.01 & 6.8/7 & 0.45\\
Theory NLO                         & 6.0/10 & 0.82 & 4.5/7 & 0.72& 13.5/9 & 0.14 & ~~5.8/9 & ~~0.76 & 5.6/7 & 0.59\\
\bottomrule
\end{tabular}
\label{tab:chi2_normalised}
\end{table}
 
The systematic uncertainties of the unfolded distributions are decomposed into signal modelling uncertainties, experimental uncertainties, and background modelling uncertainties. The breakdown of the categories of systematic uncertainties and the statistical one, which is the dominant source of uncertainty, is illustrated in Figures~\ref{fig:absolute_syst} and~\ref{fig:normalised_syst} for the absolute and normalised differential cross-sections, respectively. The systematic uncertainty is dominated by the background and signal modelling.
 
\begin{figure}[!htbp]
\centering
\includegraphics[width=0.48\textwidth]{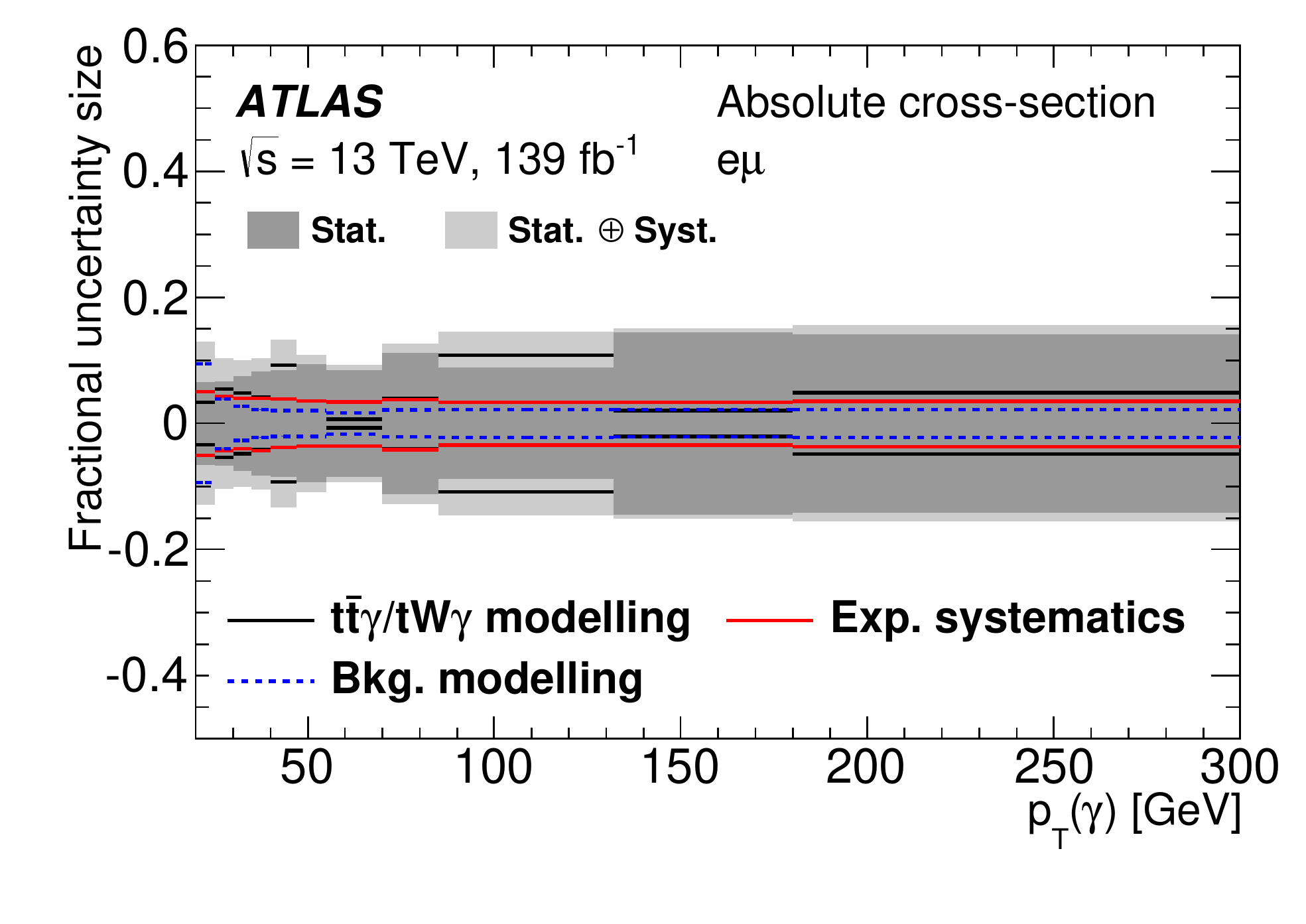}
\includegraphics[width=0.48\textwidth]{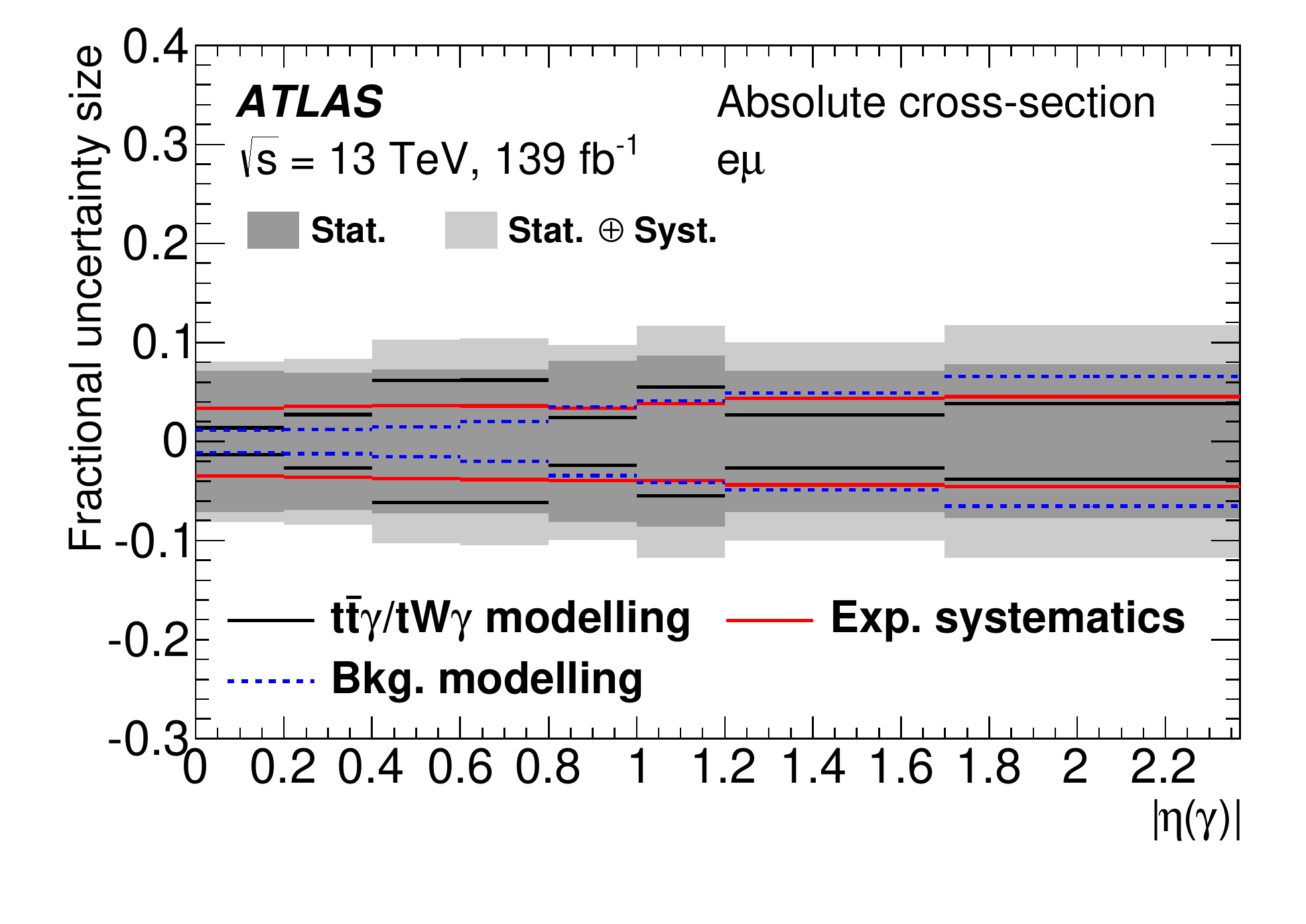}
\includegraphics[width=0.48\textwidth]{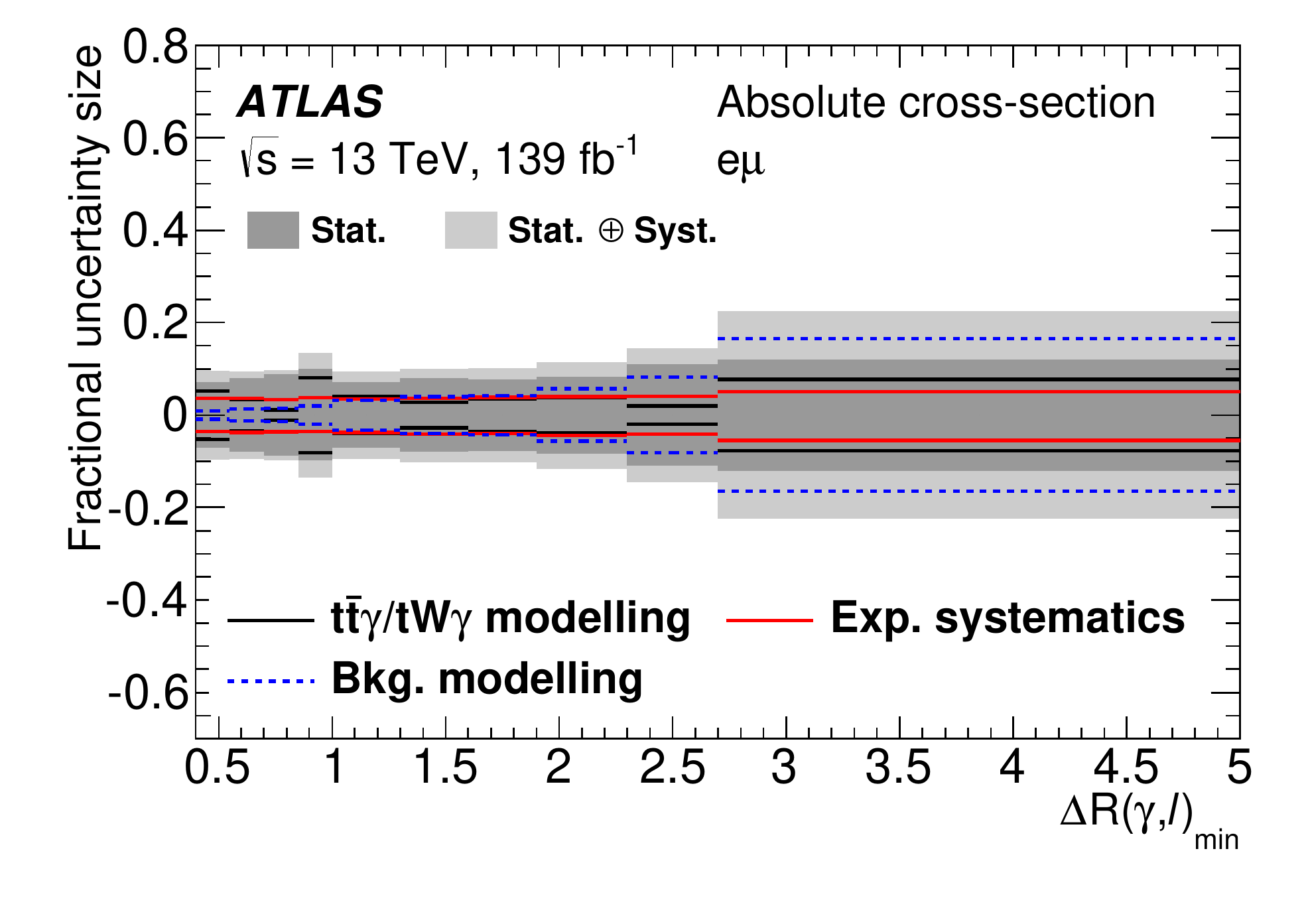}
\includegraphics[width=0.48\textwidth]{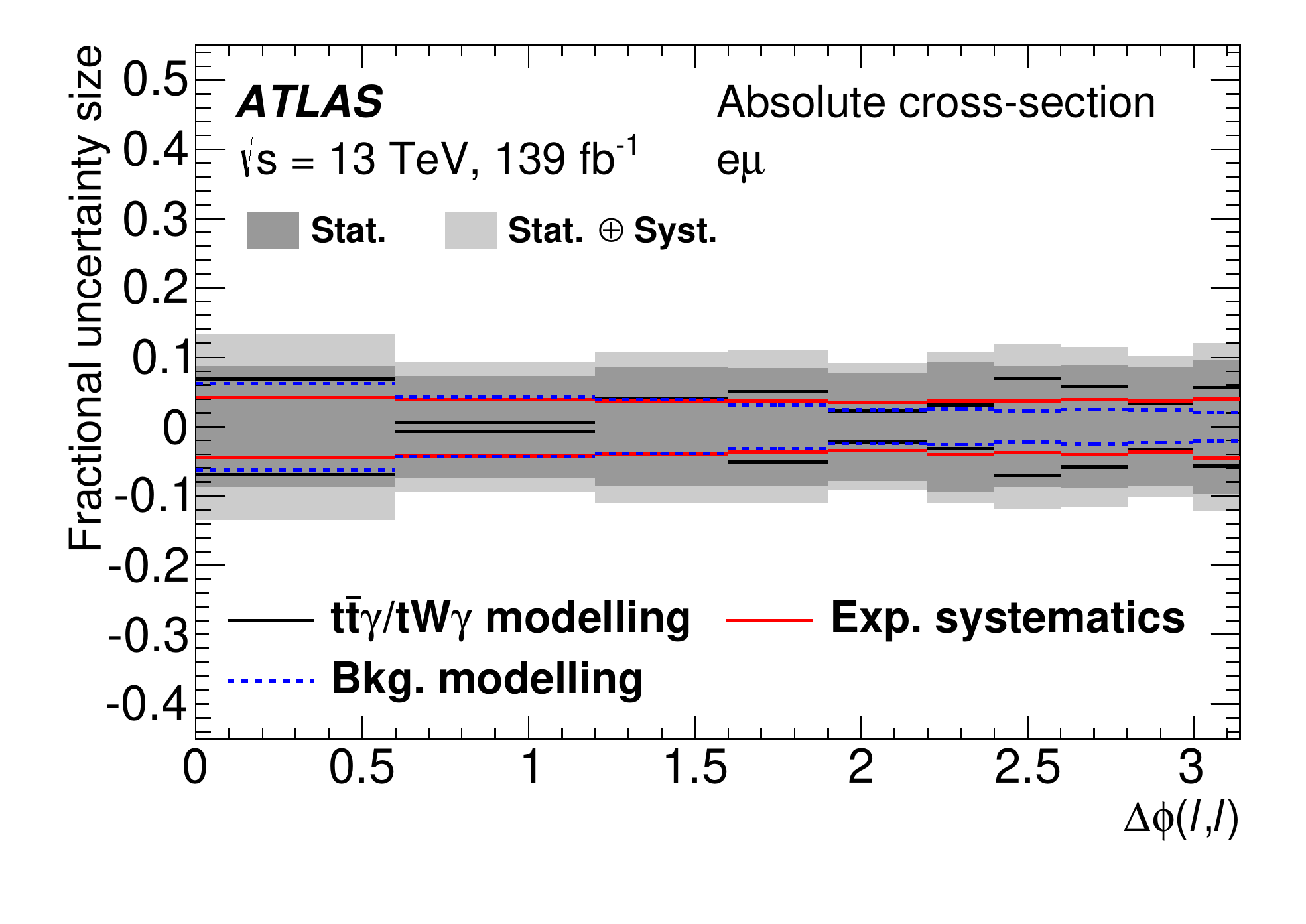}
\includegraphics[width=0.48\textwidth]{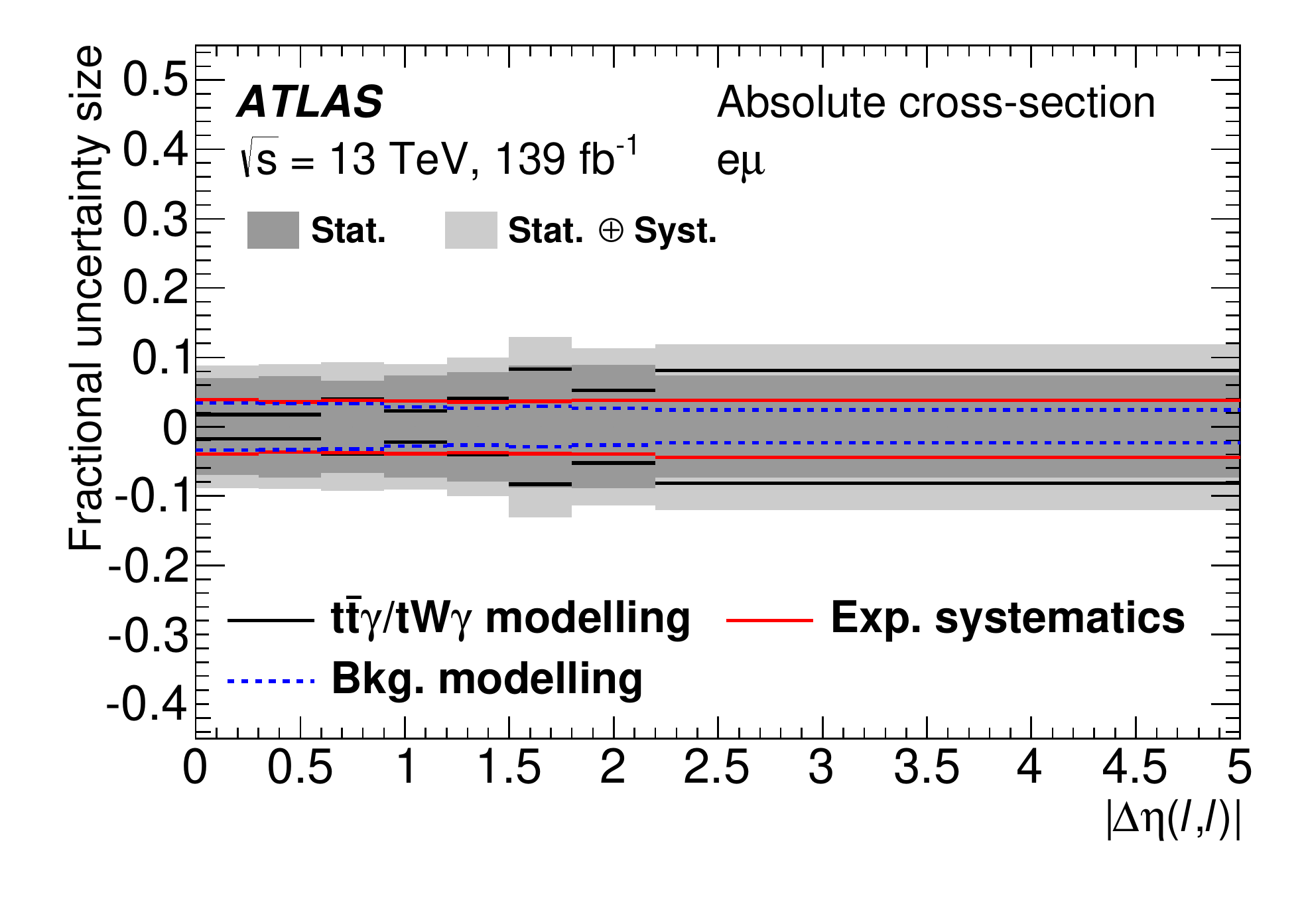}
\caption{Contribution of each category of systematic uncertainties in each bin of the measurement of the absolute cross-sections as functions of the photon \pt, photon $|\eta|$, $\Delta R (\gamma, \ell)_{\mathrm{min}}$, $\Delta\phi(\ell,\ell)$ and $|\Delta\eta(\ell,\ell)|$.}
\label{fig:absolute_syst}
\end{figure}
 
\begin{figure}[!htbp]
\centering
\includegraphics[width=0.48\textwidth]{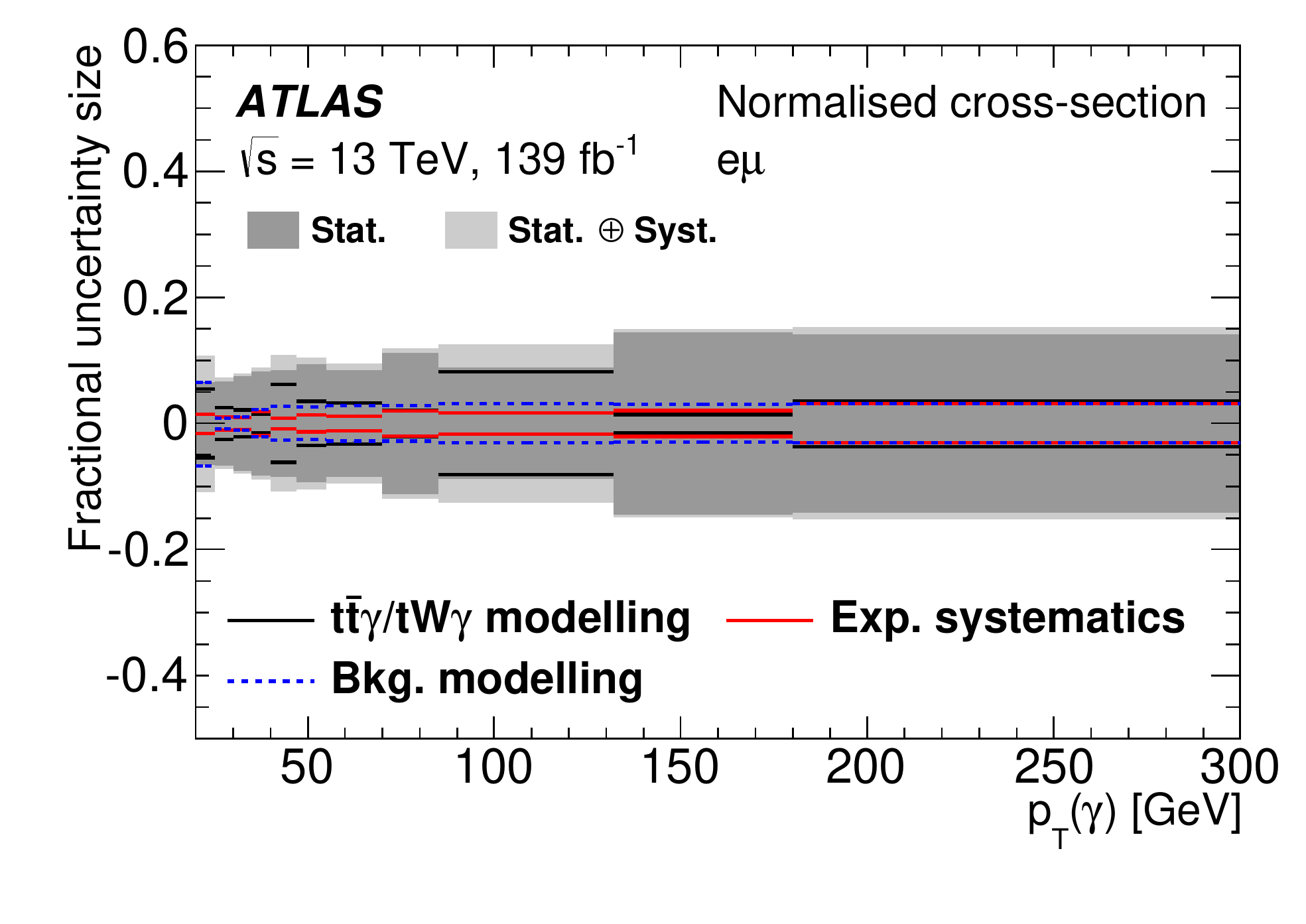}
\includegraphics[width=0.48\textwidth]{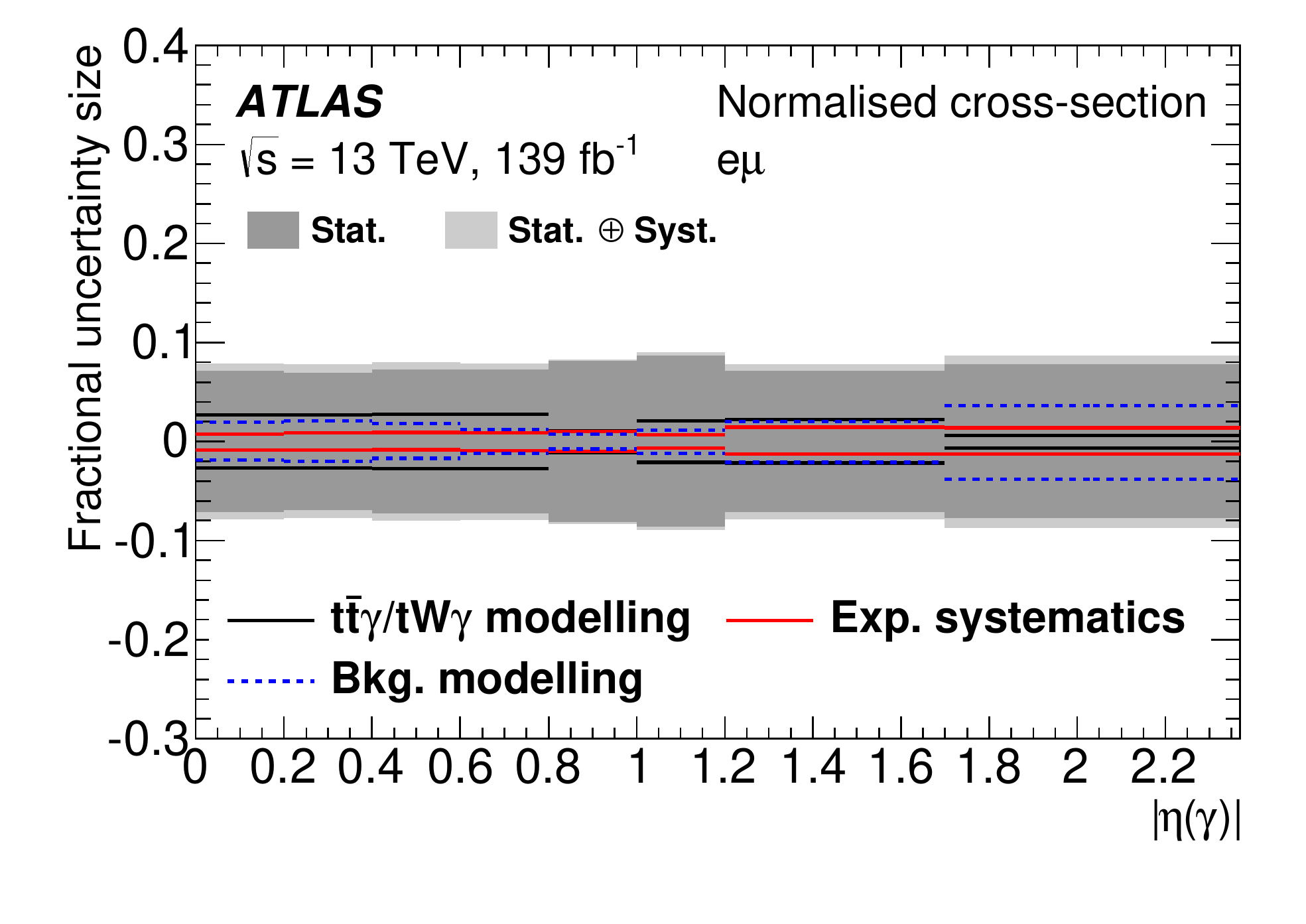}
\includegraphics[width=0.48\textwidth]{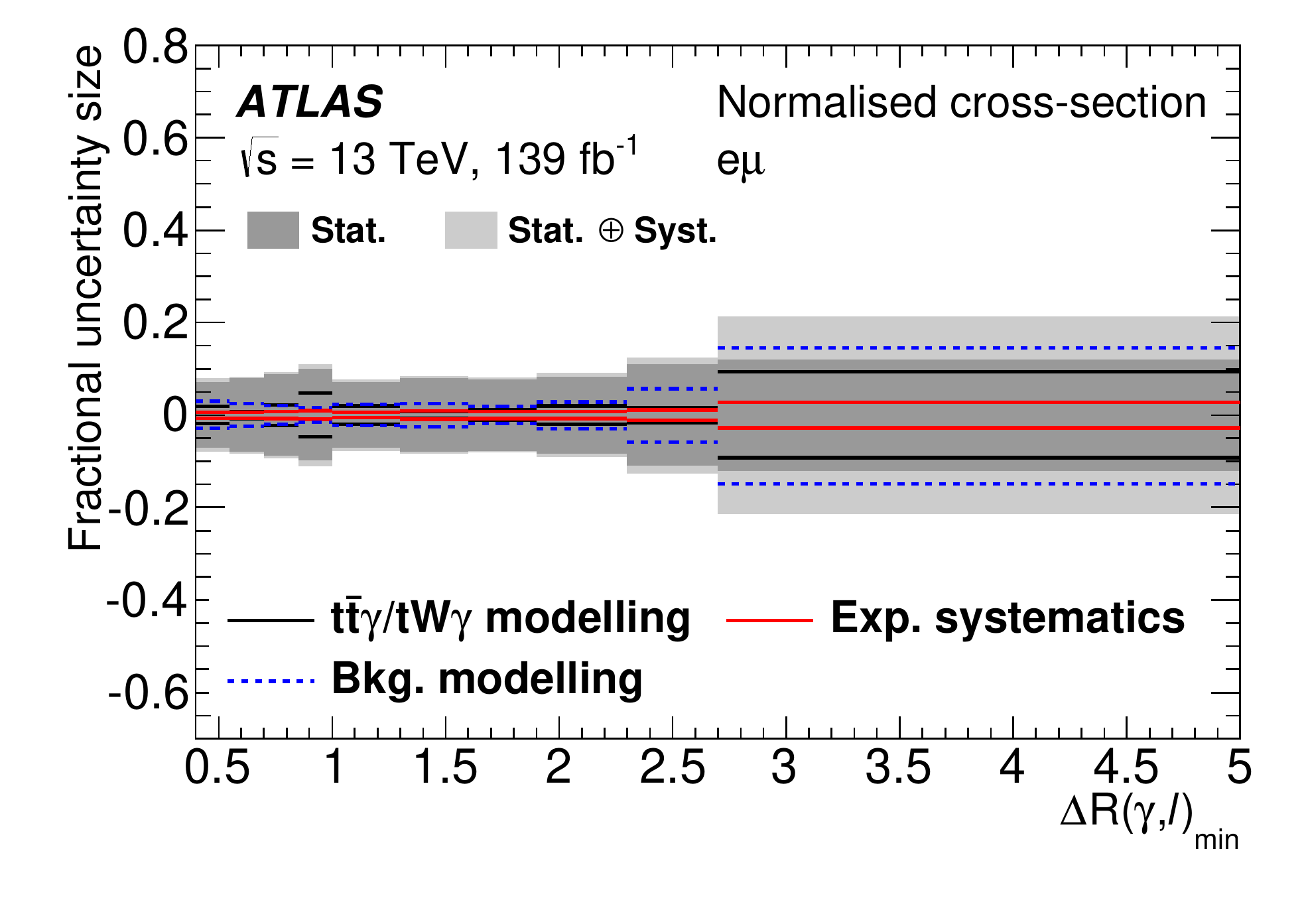}
\includegraphics[width=0.48\textwidth]{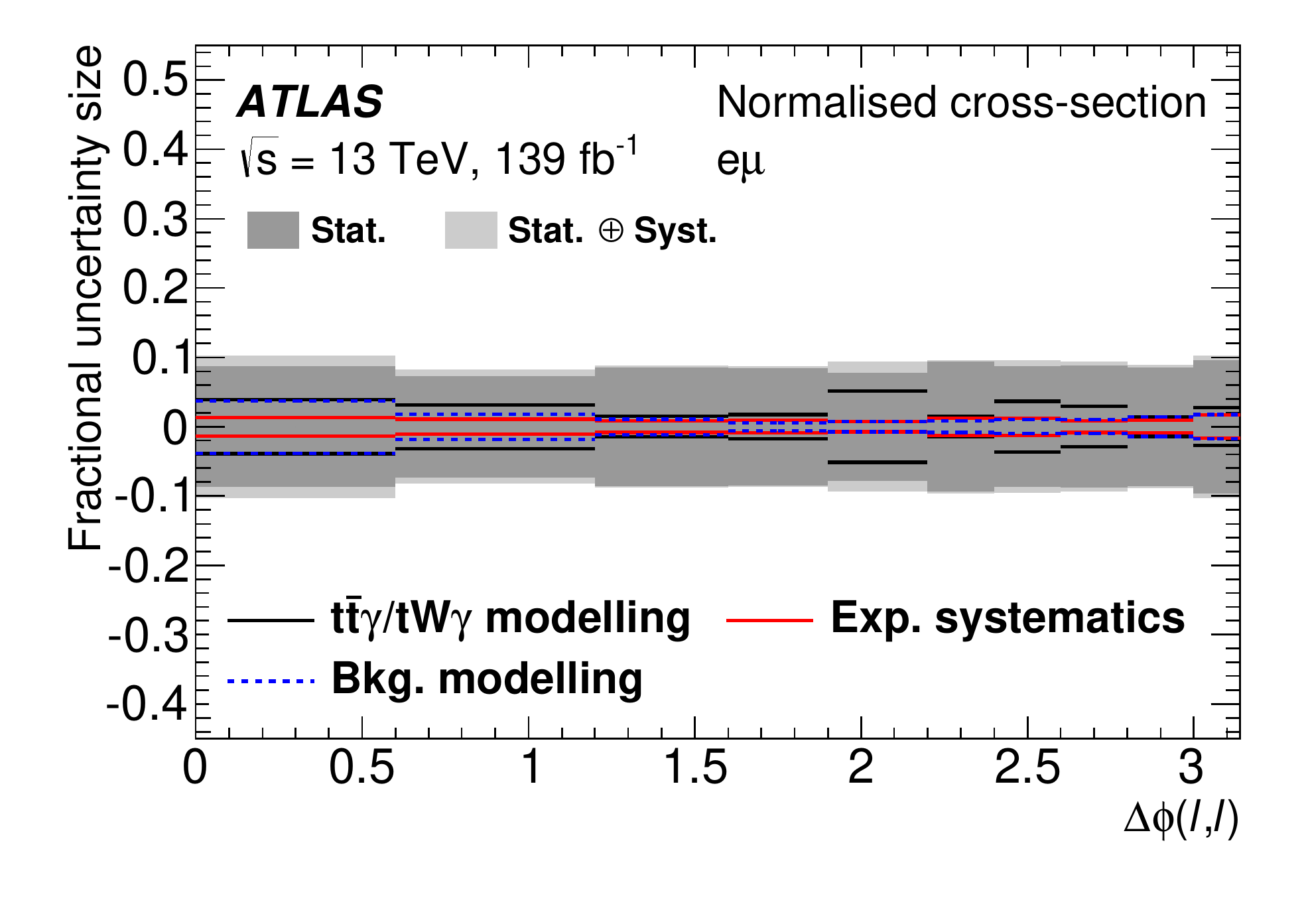}
\includegraphics[width=0.48\textwidth]{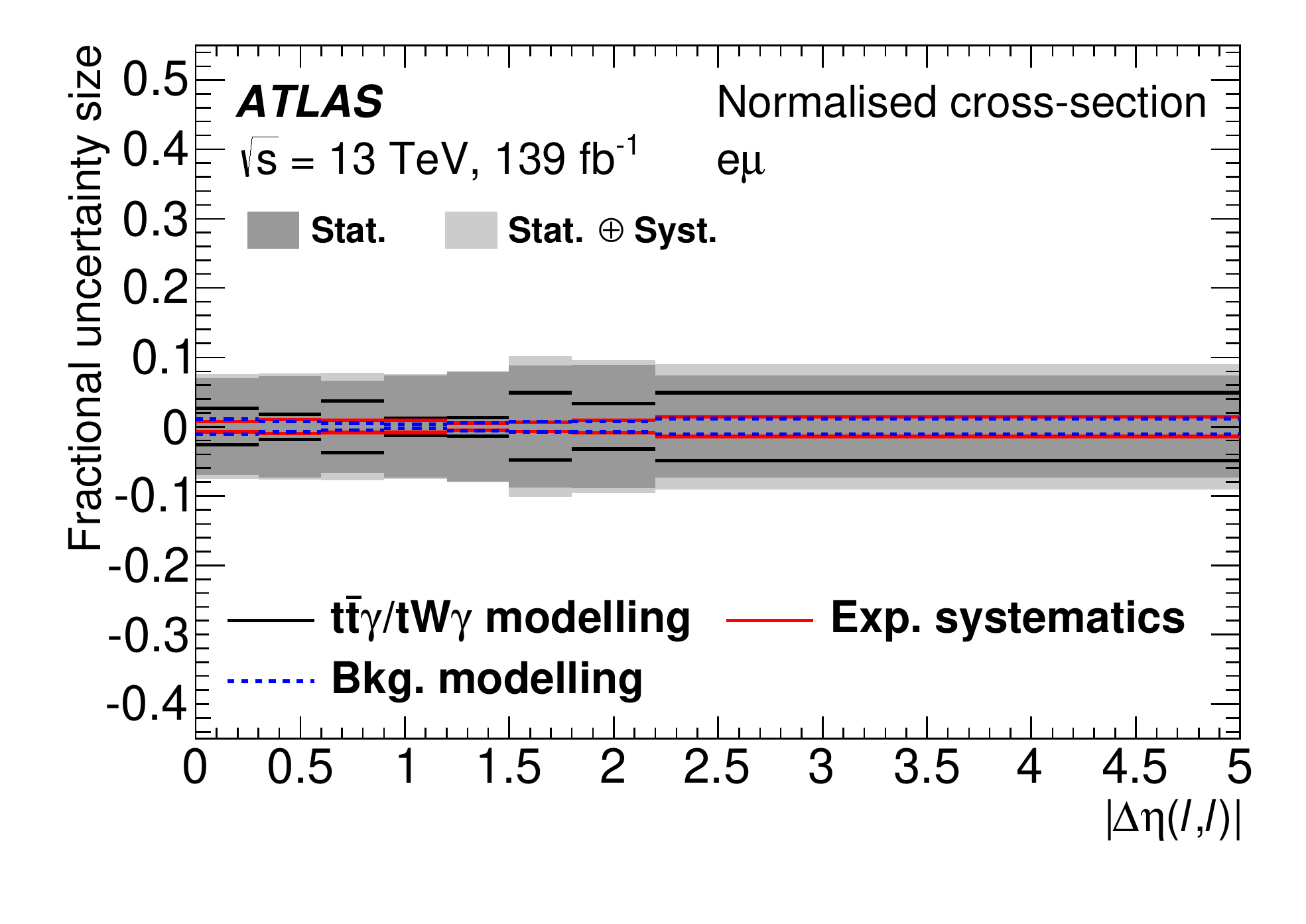}
\caption{Contribution of each category of systematic uncertainties in each bin of the measurement of the normalised cross-sections as functions of the photon \pt,  photon $|\eta|$, $\Delta R (\gamma, \ell)_{\mathrm{min}}$, $\Delta\phi(\ell,\ell)$ and $|\Delta\eta(\ell,\ell)|$ (from left to right and top to bottom). }
\label{fig:normalised_syst}
\end{figure}

\section{Conclusions}
\label{sec:conclusion}
 
Measurements of the fiducial inclusive production cross-section, as well as absolute and normalised differential production cross-sections, of the combined \tty/\tWy process in the \chemu decay channel are presented using $pp$ collisions at a centre-of-mass energy of 13~\TeV, corresponding to an integrated luminosity of 139~\ifb\ recorded by the ATLAS detector at the LHC.
For the estimation of efficiencies and acceptance corrections, a LO Monte Carlo simulation of the $2\rightarrow7$ process $pp\rightarrow e\nu\mu\nu bb\gamma$ was used for the \tty part of the signal.
The contribution from \tWy was estimated from a combination of LO Monte Carlo simulations for the $2\rightarrow3$ process  $pp\rightarrow tW\gamma$ and the $2\rightarrow6$ process $pp\rightarrow e\nu\mu\nu b\gamma$.
The simulations include initial- and final-state radiation of the photon from all involved objects in the matrix element.
The resonant top-quark production is taken into account in the simulation of \tty.
Possible singly resonant production leading to the same final state is included in the simulation of the \Wty process.
 
The results are compared with the prediction from the LO Monte Carlo simulations and also a dedicated NLO theory prediction which includes all off-shell contributions.
The measured fiducial inclusive cross-section of $\sigma=39.6\,^{+2.7}_{-2.3}\,\textrm{fb}$ is found to be in good agreement with the predicted NLO cross-section.
All considered differential distributions are also found to be well described by the NLO theory prediction.

\section*{Acknowledgements}
 
% The next lines are included from the .//acknowledgements/Acknowledgements.tex input file

We thank CERN for the very successful operation of the LHC, as well as the
support staff from our institutions without whom ATLAS could not be
operated efficiently.
 
We acknowledge the support of ANPCyT, Argentina; YerPhI, Armenia; ARC, Australia; BMWFW and FWF, Austria; ANAS, Azerbaijan; SSTC, Belarus; CNPq and FAPESP, Brazil; NSERC, NRC and CFI, Canada; CERN; CONICYT, Chile; CAS, MOST and NSFC, China; COLCIENCIAS, Colombia; MSMT CR, MPO CR and VSC CR, Czech Republic; DNRF and DNSRC, Denmark; IN2P3-CNRS and CEA-DRF/IRFU, France; SRNSFG, Georgia; BMBF, HGF and MPG, Germany; GSRT, Greece; RGC and Hong Kong SAR, China; ISF and Benoziyo Center, Israel; INFN, Italy; MEXT and JSPS, Japan; CNRST, Morocco; NWO, Netherlands; RCN, Norway; MNiSW and NCN, Poland; FCT, Portugal; MNE/IFA, Romania; MES of Russia and NRC KI, Russia Federation; JINR; MESTD, Serbia; MSSR, Slovakia; ARRS and MIZ\v{S}, Slovenia; DST/NRF, South Africa; MINECO, Spain; SRC and Wallenberg Foundation, Sweden; SERI, SNSF and Cantons of Bern and Geneva, Switzerland; MOST, Taiwan; TAEK, Turkey; STFC, United Kingdom; DOE and NSF, United States of America. In addition, individual groups and members have received support from BCKDF, CANARIE, Compute Canada and CRC, Canada; ERC, ERDF, Horizon 2020, Marie Sk{\l}odowska-Curie Actions and COST, European Union; Investissements d'Avenir Labex, Investissements d'Avenir Idex and ANR, France; DFG and AvH Foundation, Germany; Herakleitos, Thales and Aristeia programmes co-financed by EU-ESF and the Greek NSRF, Greece; BSF-NSF and GIF, Israel; CERCA Programme Generalitat de Catalunya and PROMETEO Programme Generalitat Valenciana, Spain; G\"{o}ran Gustafssons Stiftelse, Sweden; The Royal Society and Leverhulme Trust, United Kingdom.
 
The crucial computing support from all WLCG partners is acknowledged gratefully, in particular from CERN, the ATLAS Tier-1 facilities at TRIUMF (Canada), NDGF (Denmark, Norway, Sweden), CC-IN2P3 (France), KIT/GridKA (Germany), INFN-CNAF (Italy), NL-T1 (Netherlands), PIC (Spain), ASGC (Taiwan), RAL (UK) and BNL (USA), the Tier-2 facilities worldwide and large non-WLCG resource providers. Major contributors of computing resources are listed in Ref.~\cite{ATL-SOFT-PUB-2020-001}.
 
% End of text imported from the .//acknowledgements/Acknowledgements.tex input file

\printbibliography

@article{Denner:2016kdg,
      author         = "Denner, Ansgar and Dittmaier, Stefan and Hofer, Lars",
      title          = "{COLLIER: A fortran-based complex one-loop library in
                        extended regularizations}",
      journal        = "Comput. Phys. Commun.",
      volume         = "212",
      year           = "2017",
      pages          = "220-238",
      doi            = "10.1016/j.cpc.2016.10.013",
      eprint         = "1604.06792",
      archivePrefix  = "arXiv",
      primaryClass   = "hep-ph",
      reportNumber   = "FR-PHENO-2016-003, ICCUB-16-016",
      SLACcitation   = "%%CITATION = ARXIV:1604.06792;%%"
}

@article{Hoeche:2011fd,
      author         = "H{\"o}che, Stefan and Krauss, Frank and Sch{\"o}nherr, Marek and
                        Siegert, Frank",
      title          = "{A critical appraisal of NLO+PS matching methods}",
      journal        = "JHEP",
      volume         = "09",
      year           = "2012",
      pages          = "049",
      doi            = "10.1007/JHEP09(2012)049",
      eprint         = "1111.1220",
      archivePrefix  = "arXiv",
      primaryClass   = "hep-ph",
      reportNumber   = "SLAC-PUB-14661, IPPP-11-67, DCPT-11-134, LPN11-58,
                        FR-PHENO-2011-019, MCNET-11-24",
      SLACcitation   = "%%CITATION = ARXIV:1111.1220;%%"
}

@article{Catani:2001cc,
      author         = "Catani, S. and Krauss, F. and Webber, B. R. and Kuhn, R.",
      title          = "{QCD Matrix Elements + Parton Showers}",
      journal        = "JHEP",
      volume         = "11",
      year           = "2001",
      pages          = "063",
      doi            = "10.1088/1126-6708/2001/11/063",
      eprint         = "hep-ph/0109231",
      archivePrefix  = "arXiv",
      primaryClass   = "hep-ph",
      reportNumber   = "CERN-TH-2000-367, CAVENDISH-HEP-00-03",
      SLACcitation   = "%%CITATION = HEP-PH/0109231;%%"
}

@article{deFlorian:2016spz,
      author         = "de Florian, D. and others",
      title          = "{Handbook of LHC Higgs cross sections: 4. Deciphering the
                        nature of the Higgs sector}",
      collaboration  = "LHC Higgs Cross Section Working Group",
      doi            = "10.23731/CYRM-2017-002",
      volume         = "2/2017",
      month          = "10",
      year           = "2016",
      eprint         = "1610.07922",
      archivePrefix  = "arXiv",
      primaryClass   = "hep-ph",
      reportNumber   = "FERMILAB-FN-1025-T, CERN-2017-002-M",
      SLACcitation   = "%%CITATION = ARXIV:1610.07922;%%"
}

@article{Baur:2004uw,
      author         = "Baur, U. and Juste, A. and Orr, L. H. and Rainwater, D.",
      title          = "{Probing electroweak top quark couplings at hadron
                        colliders}",
      journal        = "Phys. Rev. D",
      volume         = "71",
      year           = "2005",
      pages          = "054013",
      doi            = "10.1103/PhysRevD.71.054013",
      eprint         = "hep-ph/0412021",
      archivePrefix  = "arXiv",
      primaryClass   = "hep-ph",
      reportNumber   = "UB-HET-04-03, FERMILAB-PUB-04-344-T",
      SLACcitation   = "%%CITATION = HEP-PH/0412021;%%"
}

@article{Bouzas:2012av,
      author         = "Bouzas, Antonio O. and Larios, F.",

      title          = "{Electromagnetic dipole moments of the top quark}",
      journal        = "Phys. Rev. D",
      volume         = "87",
      year           = "2013",
      number         = "7",
      pages          = "074015",
      doi            = "10.1103/PhysRevD.87.074015",
      eprint         = "1212.6575",
      archivePrefix  = "arXiv",
      primaryClass   = "hep-ph",

}

@article{Schulze:2016qas,
      author         = "Schulze, Markus and Soreq, Yotam",
      title          = "{Pinning down electroweak dipole operators of the top
                        quark}",
      journal        = "Eur. Phys. J. C",
      volume         = "76",
      year           = "2016",
      number         = "8",
      pages          = "466",
      doi            = "10.1140/epjc/s10052-016-4263-x",
      eprint         = "1603.08911",
      archivePrefix  = "arXiv",
      primaryClass   = "hep-ph",
      reportNumber   = "CERN-TH-2016-070, MIT-CTP-4790",
      SLACcitation   = "%%CITATION = ARXIV:1603.08911;%%"

}

@article{Bylund:2016phk,
      author         = "Bessidskaia Bylund, Olga and Maltoni, Fabio and Tsinikos,
                        Ioannis and Vryonidou, Eleni and Zhang, Cen",
      title          = "{Probing top quark neutral couplings in the Standard
                        Model Effective Field Theory at NLO in QCD}",
      journal        = "JHEP",
      volume         = "05",
      year           = "2016",
      pages          = "052",
      doi            = "10.1007/JHEP05(2016)052",
      eprint         = "1601.08193",
      archivePrefix  = "arXiv",
      primaryClass   = "hep-ph",

}

@article{Aaltonen:2011sp,
      author         = "{CDF Collaboration}",
      title          = "{Evidence for $t\bar{t}\gamma$ production and measurement
                        of $\sigma_{t\bar{t}\gamma} / \sigma_{t\bar{t}}$}",
      collaboration  = "CDF",
      journal        = "Phys. Rev. D",
      volume         = "84",
      year           = "2011",
      pages          = "031104",
      doi            = "10.1103/PhysRevD.84.031104",
      eprint         = "1106.3970",
      archivePrefix  = "arXiv",
      primaryClass   = "hep-ex",
}

@misc{RooUnfold,
title = "{RooUnfold package}",
url ="http://hepunx.rl.ac.uk/~adye/software/unfold/RooUnfold.html",
}

@article{Bevilacqua:2018woc,
      author         = "Bevilacqua, G. and Hartanto, H. B. and Kraus, M. and
                        Weber, T. and Worek, M.",
      title          = "{Hard photons in hadroproduction of top quarks with
                        realistic final states}",
      journal        = "JHEP",
      volume         = "10",
      year           = "2018",
      pages          = "158",
      doi            = "10.1007/JHEP10(2018)158",
      eprint         = "1803.09916",
      archivePrefix  = "arXiv",
      primaryClass   = "hep-ph",
      reportNumber   = "TTK-18-09, HU-EP-18/07, IPPP/18/17, HU-EP-18-07,
                        IPPP-18-17",
      SLACcitation   = "%%CITATION = ARXIV:1803.09916;%%",
      addendum      = "We thank the authors, in particular M. Worek, for fruitful discussions and for providing dedicated theory predictions for our measurements."
}

@article{Bevilacqua:2018dny,
    author = "Bevilacqua, G. and Hartanto, H.B. and Kraus, M. and Weber, T. and Worek, M.",
    archivePrefix = "arXiv",
    doi = "10.1007/JHEP01(2019)188",
    eprint = "1809.08562",
    journal = "JHEP",
    pages = "188",
    primaryClass = "hep-ph",
    reportNumber = "TTK-18-20, HU-EP-18/26, IPPP/18/77",
    title = "{Precise predictions for $t\bar{t}\gamma/t\bar{t}$ cross section ratios at the LHC}",
    volume = "01",
    year = "2019"
}

@Booklet{ATLAS-TDR-19,
      author       = "{ATLAS Collaboration}",
      title        = "ATLAS Insertable B-Layer Technical Design Report",
      howpublished = "ATLAS-TDR-19",
      year         = "2010",
      url          = "https://cds.cern.ch/record/1291633",
      addendum     = "{\textit{ATLAS Insertable B-Layer Technical Design Report Addendum}}, ATLAS-TDR-19-ADD-1, 2012, {\scriptsize{URL}}: \url{https://cds.cern.ch/record/1451888}",
}

@article{Abbott:2018ikt,
      author         = "Abbott, B. and others",
      title          = "{Production and integration of the ATLAS Insertable
                        B-Layer}",
      collaboration  = "ATLAS IBL",
      journal        = "JINST",
      volume         = "13",
      year           = "2018",
      number         = "05",
      pages          = "T05008",
      doi            = "10.1088/1748-0221/13/05/T05008",
      eprint         = "1803.00844",
      archivePrefix  = "arXiv",
      primaryClass   = "physics.ins-det",
      SLACcitation   = "%%CITATION = ARXIV:1803.00844;%%"
}

@article{DAgostini:1994fjx,
      author         = "D'Agostini, G.",
      title          = "{A multidimensional unfolding method based on Bayes'
                        theorem}",
      journal        = "Nucl. Instrum. Meth. A",
      volume         = "362",
      year           = "1995",
      pages          = "487",
      doi            = "10.1016/0168-9002(95)00274-X",
      reportNumber   = "DESY-94-099",
      SLACcitation   = "%%CITATION = NUIMA,A362,487;%%"
}

@article{Aaboud:2018xwy,
      author         = "{ATLAS Collaboration}",
      title          = "{Measurements of b-jet tagging efficiency with the ATLAS
                        detector using $ t\overline{t} $ events at $ \sqrt{s}=13 $
                        TeV}",
      collaboration  = "ATLAS",
      journal        = "JHEP",
      volume         = "08",
      year           = "2018",
      pages          = "089",
      doi            = "10.1007/JHEP08(2018)089",
      eprint         = "1805.01845",
      archivePrefix  = "arXiv",
      primaryClass   = "hep-ex",
      reportNumber   = "CERN-EP-2018-047",
      SLACcitation   = "%%CITATION = ARXIV:1805.01845;%%"
}

@article{LUCID2,
  author={G. Avoni and others},
  title={The new LUCID-2 detector for luminosity measurement and monitoring in ATLAS},
  journal={JINST},
  volume={13},
  number={07},
  pages={P07017},
  doi="10.1088/1748-0221/13/07/P07017",
  year={2018},
}

@article{Dulat:2015mca,
      author         = "Dulat, Sayipjamal and Hou, Tie-Jiun and Gao, Jun and
                        Guzzi, Marco and Huston, Joey and Nadolsky, Pavel and
                        Pumplin, Jon and Schmidt, Carl and Stump, Daniel and Yuan,
                        C. P.",
      title          = "{New parton distribution functions from a global analysis
                        of quantum chromodynamics}",
      journal        = "Phys. Rev. D",
      volume         = "93",
      year           = "2016",
      number         = "3",
      pages          = "033006",
      doi            = "10.1103/PhysRevD.93.033006",
      eprint         = "1506.07443",
      archivePrefix  = "arXiv",
      primaryClass   = "hep-ph",
      SLACcitation   = "%%CITATION = ARXIV:1506.07443;%%"
}

@Article{Sjostrand:2006za:mod,
     author    = "Sj{\"o}strand, Torbjorn and Mrenna, Stephen and Skands, Peter",
     title     = "{PYTHIA 6.4 physics and manual}",
     journal   = "JHEP",
     volume    = "05",
     year      = "2006",
     pages     = "026",
     eprint    = "hep-ph/0603175",
     archivePrefix = "arXiv",
     doi       = "10.1088/1126-6708/2006/05/026",
     SLACcitation  = "%%CITATION = HEP-PH/0603175;%%"
}

@Article{Gleisberg:2008ta:mod,
      author         = "Gleisberg, T. and others",
      title          = "{Event generation with SHERPA 1.1}",
      journal        = "JHEP",
      volume         = "02",
      pages          = "007",
      doi            = "10.1088/1126-6708/2009/02/007",
      year           = "2009",
      eprint         = "0811.4622",
      archivePrefix  = "arXiv",
      primaryClass   = "hep-ph",
}

@article{Hoeche:2012yf:mod,
      author         = "H{\"o}che, Stefan and Krauss, Frank and Sch{\"o}nherr, Marek and
                        Siegert, Frank",
      title          = "{QCD matrix elements + parton showers. The NLO case}",
      journal        = "JHEP",
      volume         = "04",
      year           = "2013",
      pages          = "027",
      doi            = "10.1007/JHEP04(2013)027",
      eprint         = "1207.5030",
      archivePrefix  = "arXiv",
      primaryClass   = "hep-ph",
      reportNumber   = "SLAC-PUB-15191, IPPP-12-52, DCPT-12-104, LPN12-081,
                        FR-PHENO-2012-017, MCNET-12-09, --FR-PHENO-2012-017",
      SLACcitation   = "%%CITATION = ARXIV:1207.5030;%%"
}

@article{Cascioli:2011va:mod,
      author         = "Cascioli, Fabio and Maierh{\"o}fer, Philipp and Pozzorini,
                        Stefano",
      title          = "{Scattering Amplitudes with Open Loops}",
      journal        = "Phys. Rev. Lett.",
      volume         = "108",
      year           = "2012",
      pages          = "111601",
      doi            = "10.1103/PhysRevLett.108.111601",
      eprint         = "1111.5206",
      archivePrefix  = "arXiv",
      primaryClass   = "hep-ph",
      reportNumber   = "ZU-TH-23-11, LPN11-66",
      SLACcitation   = "%%CITATION = ARXIV:1111.5206;%%"
}

@article{Czakon:2011xx:mod,
  author         = "Czakon, Michal and Mitov, Alexander",
  title          = "{Top++: A program for the calculation of the top-pair
                    cross-section at hadron colliders}",
  journal        = "Comput. Phys. Commun.",
  volume         = "185",
  year           = "2014",
  pages          = "2930",
  doi            = "10.1016/j.cpc.2014.06.021",
  eprint         = "1112.5675",
  archivePrefix  = "arXiv",
  primaryClass   = "hep-ph",
  reportNumber   = "CERN-PH-TH-2011-303, TTK-11-58",
}

@article{Frixione:1998jh,
      author         = "Frixione, Stefano",
      title          = "{Isolated photons in perturbative QCD}",
      journal        = "Phys. Lett. B",
      volume         = "429",
      year           = "1998",
      pages          = "369-374",
      doi            = "10.1016/S0370-2693(98)00454-7",
      eprint         = "hep-ph/9801442",
      archivePrefix  = "arXiv",
      primaryClass   = "hep-ph",
      reportNumber   = "ETH-TH-97-40",
      SLACcitation   = "%%CITATION = HEP-PH/9801442;%%"
}

@article{Aad:2020uyd,
    author = "{ATLAS Collaboration}",
    collaboration = "ATLAS",
    title = "{Performance of the ATLAS muon triggers in Run 2}",
    eprint = "2004.13447",
    archivePrefix = "arXiv",
    primaryClass = "hep-ex",
    reportNumber = "CERN-EP-2020-031",
    month = "4",
    year = "2020"
}

@article{Cowan:2010js:fixed,
    author = "Cowan, Glen and Cranmer, Kyle and Gross, Eilam and Vitells, Ofer",
    title = "{Asymptotic formulae for likelihood-based tests of new physics}",
    eprint = "1007.1727",
    archivePrefix = "arXiv",
    primaryClass = "physics.data-an",
    doi = "10.1140/epjc/s10052-011-1554-0",
    journal = "Eur. Phys. J. C",
    volume = "71",
    pages = "1554",
    year = "2011",
    note = "[Erratum: \href{https://link.springer.com/article/10.1140\%2Fepjc\%2Fs10052-013-2501-z}{Eur. Phys. J. C \textbf{73} (2013) 2501}]"
}

@article{Aad:2019wsl,
    author = "{ATLAS Collaboration}",
    collaboration = "ATLAS",
    title = "{Performance of electron and photon triggers in ATLAS during LHC Run 2}",
    eprint = "1909.00761",
    archivePrefix = "arXiv",
    primaryClass = "hep-ex",
    reportNumber = "CERN-EP-2019-169",
    doi = "10.1140/epjc/s10052-019-7500-2",
    journal = "Eur. Phys. J. C",
    volume = "80",
    number = "1",
    pages = "47",
    year = "2020"
}

@Article{Cacciari:2008gp,
     author    = "Cacciari, Matteo and Salam, Gavin P. and Soyez, Gregory",
     title     = "{The anti-\(k_t\) jet clustering algorithm}",
     journal   = "JHEP",
     volume    = "04",
     year      = "2008",
     pages     = "063",
     eprint    = "0802.1189",
     archivePrefix = "arXiv",
     primaryClass  =  "hep-ph",
     doi       = "10.1088/1126-6708/2008/04/063",
     SLACcitation  = "%%CITATION = 0802.1189;%%"
}

@Article{Fastjet,
      author         = "Cacciari, Matteo and Salam, Gavin P. and Soyez, Gregory",
      title          = "{FastJet user manual}",
      journal        = "Eur. Phys. J. C",
      volume         = "72",
      year           = "2012",
      pages          = "1896",
      doi            = "10.1140/epjc/s10052-012-1896-2",
      eprint         = "1111.6097",
      archivePrefix  = "arXiv",
      primaryClass   = "hep-ph",
      reportNumber   = "CERN-PH-TH-2011-297",
      SLACcitation   = "%%CITATION = ARXIV:1111.6097;%%"
}

@Article{Pumplin:2002vw,
     author    = "Pumplin, J. and others",
     title     = "{New Generation of Parton Distributions with Uncertainties from Global QCD Analysis}",
     journal   = "JHEP",
     volume    = "07",
     year      = "2002",
     pages     = "012",
     doi       = "10.1088/1126-6708/2002/07/012",
     eprint    = "hep-ph/0201195",
     archivePrefix = "arXiv",
     SLACcitation  = "%%CITATION = HEP-PH/0201195;%%"
}

@article{Ball:2014uwa,
      author         = "Ball, Richard D. and others",
      title          = "{Parton distributions for the LHC Run II}",
      collaboration  = "NNPDF",
      journal        = "JHEP",
      volume         = "04",
      year           = "2015",
      pages          = "040",
      doi            = "10.1007/JHEP04(2015)040",
      eprint         = "1410.8849",
      archivePrefix  = "arXiv",
      primaryClass   = "hep-ph",
      reportNumber   = "EDINBURGH-2014-15, IFUM-1034-FT, CERN-PH-TH-2013-253,
                        OUTP-14-11P, CAVENDISH-HEP-14-11",
      SLACcitation   = "%%CITATION = ARXIV:1410.8849;%%"
}

@article{Sjostrand:2014zea,
      author         = "Sj{\"o}strand, Torbj{\"o}rn and Ask, Stefan and Christiansen,
                        Jesper R. and Corke, Richard and Desai, Nishita and Ilten,
                        Philip and Mrenna, Stephen and Prestel, Stefan and
                        Rasmussen, Christine O. and Skands, Peter Z.",
      title          = "{An introduction to PYTHIA 8.2}",
      journal        = "Comput. Phys. Commun.",
      volume         = "191",
      year           = "2015",
      pages          = "159",
      doi            = "10.1016/j.cpc.2015.01.024",
      eprint         = "1410.3012",
      archivePrefix  = "arXiv",
      primaryClass   = "hep-ph",
      reportNumber   = "LU-TP-14-36, MCNET-14-22, CERN-PH-TH-2014-190,
                        FERMILAB-PUB-14-316-CD, DESY-14-178, SLAC-PUB-16122,
                        --FERMILAB-PUB-14-316-CD",
      SLACcitation   = "%%CITATION = ARXIV:1410.3012;%%"
}

@Article{Lange:2001uf,
      author         = "Lange, D. J.",
      title          = "{The EvtGen particle decay simulation package}",
      booktitle      = "{Proceedings, 7th International Conference on B physics
                        at hadron machines (BEAUTY 2000)}",
      journal        = "Nucl. Instrum. Meth. A",
      volume         = "462",
      year           = "2001",
      pages          = "152",
      doi            = "10.1016/S0168-9002(01)00089-4",
      SLACcitation   = "%%CITATION = NUIMA,A462,152;%%"
}

@Article{Alwall:2014hca,
      author         = "Alwall, J. and Frederix, R. and Frixione, S. and Hirschi,
                        V. and Maltoni, F. and Mattelaer, O. and Shao, H. -S. and
                        Stelzer, T. and Torrielli, P. and Zaro, M.",
      title          = "{The automated computation of tree-level and
                        next-to-leading order differential cross sections, and
                        their matching to parton shower simulations}",
      journal        = "JHEP",
      volume         = "07",
      year           = "2014",
      pages          = "079",
      doi            = "10.1007/JHEP07(2014)079",
      eprint         = "1405.0301",
      archivePrefix  = "arXiv",
      primaryClass   = "hep-ph",
      reportNumber   = "CERN-PH-TH-2014-064, CP3-14-18, LPN14-066, MCNET-14-09,
                        ZU-TH-14-14",
      SLACcitation   = "%%CITATION = ARXIV:1405.0301;%%"
}

@Article{Bahr:2008pv,
      author         = "Bahr, M. and others",
      title          = "{Herwig++ physics and manual}",
      journal        = "Eur. Phys. J. C",
      volume         = "58",
      year           = "2008",
      pages          = "639",
      doi            = "10.1140/epjc/s10052-008-0798-9",
      eprint         = "0803.0883",
      archivePrefix  = "arXiv",
      primaryClass   = "hep-ph",
      reportNumber   = "CERN-PH-TH-2008-038, CAVENDISH-HEP-08-03, KA-TP-05-2008,
                        DCPT-08-22, IPPP-08-11, CP3-08-05",
      SLACcitation   = "%%CITATION = ARXIV:0803.0883;%%"
}

@Article{Bellm:2015jjp,
      author         = "Bellm, Johannes and others",
      title          = "{Herwig 7.0/Herwig++ 3.0 release note}",
      journal        = "Eur. Phys. J. C",
      volume         = "76",
      year           = "2016",
      number         = "4",
      pages          = "196",
      doi            = "10.1140/epjc/s10052-016-4018-8",
      eprint         = "1512.01178",
      archivePrefix  = "arXiv",
      primaryClass   = "hep-ph",
      reportNumber   = "CERN-PH-TH-2015-289, MAN-HEP-2015-15, IFJPAN-IV-2015-13,
                        HERWIG-2015-01, KA-TP-18-2015, DCPT-15-142, MCNET-15-28,
                        IPPP-15-71, --HERWIG-2015-01",
      SLACcitation   = "%%CITATION = ARXIV:1512.01178;%%"
}

@Article{Campbell:1999ah,
      author         = "Campbell, John M. and Ellis, R. Keith",
      title          = "{Update on vector boson pair production at hadron colliders}",
      journal        = "Phys. Rev. D",
      volume         = "60",
      pages          = "113006",
      doi            = "10.1103/PhysRevD.60.113006",
      year           = "1999",
      eprint         = "hep-ph/9905386",
      archivePrefix  = "arXiv",
      reportNumber   = "FERMILAB-PUB-99-146-T",
      SLACcitation   = "%%CITATION = HEP-PH/9905386;%%",
}

@Article{Nason:2004rx,
      author         = "Nason, Paolo",
      title          = "{A new method for combining NLO QCD with shower Monte Carlo algorithms}",
      journal        = "JHEP",
      volume         = "11",
      pages          = "040",
      doi            = "10.1088/1126-6708/2004/11/040",
      year           = "2004",
      eprint         = "hep-ph/0409146",
      archivePrefix  = "arXiv",
}

@Article{Frixione:2007vw,
      author         = "Frixione, Stefano and Nason, Paolo and Oleari, Carlo",
      title          = "{Matching NLO QCD computations with parton shower
                        simulations: the POWHEG method}",
      journal        = "JHEP",
      volume         = "11",
      pages          = "070",
      doi            = "10.1088/1126-6708/2007/11/070",
      year           = "2007",
      eprint         = "0709.2092",
      archivePrefix  = "arXiv",
      primaryClass   = "hep-ph",
}

@Article{Alioli:2010xd,
      author         = "Alioli, Simone and Nason, Paolo and Oleari, Carlo and Re,
                        Emanuele",
      title          = "{A general framework for implementing NLO calculations in
                        shower Monte Carlo programs: the POWHEG BOX}",
      journal        = "JHEP",
      volume         = "06",
      pages          = "043",
      doi            = "10.1007/JHEP06(2010)043",
      year           = "2010",
      eprint         = "1002.2581",
      archivePrefix  = "arXiv",
      primaryClass   = "hep-ph",
}

@Article{Hoeche:2009rj,
      author         = "H{\"o}che, Stefan and Krauss, Frank and Schumann, Steffen and Siegert, Frank",
      title          = "{QCD matrix elements and truncated showers}",
      journal        = "JHEP",
      volume         = "05",
      pages          = "053",
      doi            = "10.1088/1126-6708/2009/05/053",
      year           = "2009",
      eprint         = "0903.1219",
      archivePrefix  = "arXiv",
      primaryClass   = "hep-ph",
}

@Article{Gleisberg:2008fv,
      author         = "Gleisberg, Tanju and H{\"o}che, Stefan",
      title          = "{Comix, a new matrix element generator}",
      journal        = "JHEP",
      volume         = "12",
      pages          = "039",
      doi            = "10.1088/1126-6708/2008/12/039",
      year           = "2008",
      eprint         = "0808.3674",
      archivePrefix  = "arXiv",
      primaryClass   = "hep-ph",
}

@Article{Schumann:2007mg,
      author         = "Schumann, Steffen and Krauss, Frank",
      title          = "{A parton shower algorithm based on Catani--Seymour dipole factorisation}",
      journal        = "JHEP",
      volume         = "03",
      pages          = "038",
      doi            = "10.1088/1126-6708/2008/03/038",
      year           = "2008",
      eprint         = "0709.1027",
      archivePrefix  = "arXiv",
      primaryClass   = "hep-ph",
}

@article{Agostinelli:2002hh,
      author         = "Agostinelli, S. and others",
      title          = "{\textsc{Geant4} -- a simulation toolkit}",
      journal        = "Nucl. Instrum. Meth. A",
      volume         = "506",
      year           = "2003",
      pages          = "250",
      doi            = "10.1016/S0168-9002(03)01368-8",
      reportNumber   = "SLAC-PUB-9350, FERMILAB-PUB-03-339",
      SLACcitation   = "%%CITATION = NUIMA,A506,250;%%"
}

@Booklet{Adye:2011gm,
      author         = "Adye, Tim",
      title          = "{Unfolding algorithms and tests using RooUnfold}",
      year           = "2011",
      eprint         = "1105.1160",
      archivePrefix  = "arXiv",
      primaryClass   = "physics.data-an"
}

@Article{PERF-2007-01,
    author         = "{ATLAS Collaboration}",
    title          = "{The ATLAS Experiment at the CERN Large Hadron Collider}",
    journal        = "JINST",
    volume         = "3",
    year           = "2008",
    pages          = "S08003",
    doi            = "10.1088/1748-0221/3/08/S08003",
    primaryClass   = "hep-ex",
}

@Article{SOFT-2010-01,
    author         = "{ATLAS Collaboration}",
    title          = "{The ATLAS Simulation Infrastructure}",
    journal        = "Eur. Phys. J. C",
    volume         = "70",
    year           = "2010",
    pages          = "823",
    doi            = "10.1140/epjc/s10052-010-1429-9",
    eprint         = "1005.4568",
    archivePrefix  = "arXiv",
    primaryClass   = "physics.ins-det",
}

@Article{TOPQ-2012-07,
    author         = "{ATLAS Collaboration}",
    title          = "{Observation of top-quark pair production in association with a photon and measurement of the \(t\bar{t}\gamma\) production cross section in \(pp\) collisions at \(\sqrt{s} = 7\,\text{TeV}\) using the ATLAS detector}",
    journal        = "Phys. Rev. D",
    volume         = "91",
    year           = "2015",
    pages          = "072007",
    doi            = "10.1103/PhysRevD.91.072007",
    reportNumber   = "CERN-PH-EP-2014-284",
    eprint         = "1502.00586",
    archivePrefix  = "arXiv",
    primaryClass   = "hep-ex",
}

@Article{PERF-2015-10,
    author         = "{ATLAS Collaboration}",
    title          = "{Muon reconstruction performance of the ATLAS detector in proton--proton collision data at \(\sqrt{s} = 13\,\text{TeV}\)}",
    journal        = "Eur. Phys. J. C",
    volume         = "76",
    year           = "2016",
    pages          = "292",
    doi            = "10.1140/epjc/s10052-016-4120-y",
    reportNumber   = "CERN-EP-2016-033",
    eprint         = "1603.05598",
    archivePrefix  = "arXiv",
    primaryClass   = "hep-ex",
}

@Article{STDM-2015-03,
    author         = "{ATLAS Collaboration}",
    title          = "{Measurement of \(W^{\pm}\) and \(Z\)-boson production cross sections in \(pp\) collisions at \(\sqrt{s} = 13\,\text{TeV}\) with the ATLAS detector}",
    journal        = "Phys. Lett. B",
    volume         = "759",
    year           = "2016",
    pages          = "601",
    doi            = "10.1016/j.physletb.2016.06.023",
    reportNumber   = "CERN-EP-2016-069",
    eprint         = "1603.09222",
    archivePrefix  = "arXiv",
    primaryClass   = "hep-ex",
}

@Article{TOPQ-2015-21,
    author         = "{ATLAS Collaboration}",
    title          = "{Measurement of the \(t\bar{t}\gamma\) production cross section in proton--proton collisions at \(\sqrt{s} = 8\,\text{TeV}\) with the ATLAS detector}",
    journal        = "JHEP",
    volume         = "11",
    year           = "2017",
    pages          = "086",
    doi            = "10.1007/JHEP11(2017)086",
    reportNumber   = "CERN-EP-2017-086",
    eprint         = "1706.03046",
    archivePrefix  = "arXiv",
    primaryClass   = "hep-ex",
}

@Article{PERF-2016-04,
    author         = "{ATLAS Collaboration}",
    title          = "{Jet energy scale measurements and their systematic uncertainties in proton--proton collisions at \(\sqrt{s} = 13\,\text{TeV}\) with the ATLAS detector}",
    journal        = "Phys. Rev. D",
    volume         = "96",
    year           = "2017",
    pages          = "072002",
    doi            = "10.1103/PhysRevD.96.072002",
    reportNumber   = "CERN-EP-2017-038",
    eprint         = "1703.09665",
    archivePrefix  = "arXiv",
    primaryClass   = "hep-ex",
}

@Article{PERF-2016-05,
    author         = "{ATLAS Collaboration}",
    title          = "{Measurements of \(b\)-jet tagging efficiency with the ATLAS detector using \(t\bar{t}\) events at \(\sqrt{s} = 13\,\text{TeV}\)}",
    journal        = "JHEP",
    volume         = "08",
    year           = "2018",
    pages          = "089",
    doi            = "10.1007/JHEP08(2018)089",
    reportNumber   = "CERN-EP-2018-047",
    eprint         = "1805.01845",
    archivePrefix  = "arXiv",
    primaryClass   = "hep-ex",
}

@Article{PERF-2016-07,
    author         = "{ATLAS Collaboration}",
    title          = "{Performance of missing transverse momentum reconstruction with the ATLAS detector using proton--proton collisions at \(\sqrt{s} = 13\,\text{TeV}\)}",
    journal        = "Eur. Phys. J. C",
    volume         = "78",
    year           = "2018",
    pages          = "903",
    doi            = "10.1140/epjc/s10052-018-6288-9",
    reportNumber   = "CERN-EP-2017-274",
    eprint         = "1802.08168",
    archivePrefix  = "arXiv",
    primaryClass   = "hep-ex",
}

@Article{TRIG-2016-01,
    author         = "{ATLAS Collaboration}",
    title          = "{Performance of the ATLAS trigger system in 2015}",
    journal        = "Eur. Phys. J. C",
    volume         = "77",
    year           = "2017",
    pages          = "317",
    doi            = "10.1140/epjc/s10052-017-4852-3",
    reportNumber   = "CERN-EP-2016-241",
    eprint         = "1611.09661",
    archivePrefix  = "arXiv",
    primaryClass   = "hep-ex",
}

@Article{PERF-2017-02,
    author         = "{ATLAS Collaboration}",
    title          = "{Measurement of the photon identification efficiencies with the ATLAS detector using LHC Run~2 data collected in 2015 and 2016}",
    journal        = "Eur. Phys. J. C",
    volume         = "79",
    year           = "2019",
    pages          = "205",
    doi            = "10.1140/epjc/s10052-019-6650-6",
    reportNumber   = "CERN-EP-2018-216",
    eprint         = "1810.05087",
    archivePrefix  = "arXiv",
    primaryClass   = "hep-ex",
}

@Article{PERF-2017-03,
    author         = "{ATLAS Collaboration}",
    title          = "{Electron and photon energy calibration with the ATLAS detector using 2015--2016 LHC proton--proton collision data}",
    journal        = "JINST",
    volume         = "14",
    year           = "2019",
    pages          = "P03017",
    doi            = "10.1088/1748-0221/14/03/P03017",
    reportNumber   = "CERN-EP-2018-296",
    eprint         = "1812.03848",
    archivePrefix  = "arXiv",
    primaryClass   = "hep-ex",
}

@Article{TOPQ-2017-14,
    author         = "{ATLAS Collaboration}",
    title          = "{Measurements of inclusive and differential fiducial cross-sections of \(t\bar{t}\gamma\) production in leptonic final states at \(\sqrt{s} =  13\,\text{TeV}\) in ATLAS}",
    journal        = "Eur. Phys. J. C",
    volume         = "79",
    year           = "2019",
    pages          = "382",
    doi            = "10.1140/epjc/s10052-019-6849-6",
    reportNumber   = "CERN-EP-2018-302",
    eprint         = "1812.01697",
    archivePrefix  = "arXiv",
    primaryClass   = "hep-ex",
}

@Article{EGAM-2018-01,
    author         = "{ATLAS Collaboration}",
    title          = "{Electron and photon performance measurements with the ATLAS detector using the 2015--2017 LHC proton--proton collision data}",
    journal        = "JINST",
    volume         = "14",
    year           = "2019",
    pages          = "P12006",
    doi            = "10.1088/1748-0221/14/12/P12006",
    reportNumber   = "CERN-EP-2019-145",
    eprint         = "1908.00005",
    archivePrefix  = "arXiv",
    primaryClass   = "hep-ex",
}

@Booklet{ATL-SOFT-PUB-2020-001,
    author         = "{ATLAS Collaboration}",
    title          = "{ATLAS Computing Acknowledgements}",
    howpublished   = "{ATL-SOFT-PUB-2020-001}",
    url            = "https://cds.cern.ch/record/2717821",
    year           = "20XX",
}

@Article{CMS-TOP-14-008,
    author         = "{CMS Collaboration}",
    title          = "{Measurement of the semileptonic \(t\bar{t} + \gamma\) production cross section in \(pp\) collisions at \(\sqrt{s} = 8\,\text{TeV}\)}",
    journal        = "JHEP",
    volume         = "10",
    year           = "2017",
    pages          = "006",
    doi            = "10.1007/JHEP10(2017)006",
    reportNumber   = "CERN-EP-2017-112",
    eprint         = "1706.08128",
    archivePrefix  = "arXiv",
    primaryClass   = "hep-ex",
}

@Booklet{ATLAS-CONF-2013-082,
    author         = "{ATLAS Collaboration}",
    title          = "{Performance of Missing Transverse Momentum Reconstruction in ATLAS studied in Proton--Proton Collisions recorded in 2012 at \(\sqrt{s} = 8~\text{TeV}\)}",
    howpublished   = "{ATLAS-CONF-2013-082}",
    url            = "https://cds.cern.ch/record/1570993",
    year           = "2013",
}

@Booklet{ATLAS-CONF-2014-018,
    author         = "{ATLAS Collaboration}",
    title          = "{Tagging and suppression of pileup jets with the ATLAS detector}",
    howpublished   = "{ATLAS-CONF-2014-018}",
    url            = "https://cds.cern.ch/record/1700870",
    year           = "2014",
}

@Booklet{ATLAS-CONF-2018-001,
    author         = "{ATLAS Collaboration}",
    title          = "{Measurement of \(b\)-tagging efficiency of \(c\)-jets in \(t\bar{t}\) events using a likelihood approach with the ATLAS detector}",
    howpublished   = "{ATLAS-CONF-2018-001}",
    url            = "https://cds.cern.ch/record/2306649",
    year           = "2018",
}

@Booklet{ATLAS-CONF-2018-006,
    author         = "{ATLAS Collaboration}",
    title          = "{Calibration of light-flavour \(b\)-jet mistagging rates using ATLAS proton--proton collision data at \(\sqrt{s} = 13~\text{TeV}\)}",
    howpublished   = "{ATLAS-CONF-2018-006}",
    url            = "https://cds.cern.ch/record/2314418",
    year           = "2018",
}

@Booklet{ATLAS-CONF-2018-023,
    author         = "{ATLAS Collaboration}",
    title          = "{\(E_{\text{T}}^{\text{miss}}\) performance in the ATLAS detector using 2015--2016 LHC \(pp\) collisions}",
    howpublished   = "{ATLAS-CONF-2018-023}",
    url            = "https://cds.cern.ch/record/2625233",
    year           = "2018",
}

@Booklet{ATLAS-CONF-2019-021,
    author         = "{ATLAS Collaboration}",
    title          = "{Luminosity determination in \(pp\) collisions at \(\sqrt{s} = 13\,\text{TeV}\) using the ATLAS detector at the LHC}",
    howpublished   = "{ATLAS-CONF-2019-021}",
    url            = "https://cds.cern.ch/record/2677054",
    year           = "2019",
}

@Booklet{ATL-PHYS-PUB-2012-003,
    author         = "{ATLAS Collaboration}",
    title          = "{Summary of ATLAS Pythia 8 tunes}",
    howpublished   = "{ATL-PHYS-PUB-2012-003}",
    url            = "https://cds.cern.ch/record/1474107",
    year           = "2012",
}

@Booklet{ATL-PHYS-PUB-2014-021,
    author         = "{ATLAS Collaboration}",
    title          = "{ATLAS Pythia~8 tunes to \(7~\text{TeV}\) data}",
    howpublished   = "{ATL-PHYS-PUB-2014-021}",
    url            = "https://cds.cern.ch/record/1966419",
    year           = "2014",
}

@Booklet{ATL-PHYS-PUB-2015-015,
    author         = "{ATLAS Collaboration}",
    title          = "{Jet Calibration and Systematic Uncertainties for Jets Reconstructed in the ATLAS Detector at \(\sqrt{s} = 13~\text{TeV}\)}",
    howpublished   = "{ATL-PHYS-PUB-2015-015}",
    url            = "https://cds.cern.ch/record/2037613",
    year           = "2015",
}

@Booklet{ATL-PHYS-PUB-2015-036,
    author         = "{ATLAS Collaboration}",
    title          = "{Properties of jets and inputs to jet reconstruction and calibration with the ATLAS detector using proton--proton collisions at \(\sqrt{s} = 13~\text{TeV}\)}",
    howpublished   = "{ATL-PHYS-PUB-2015-036}",
    url            = "https://cds.cern.ch/record/2044564",
    year           = "2015",
}

@Booklet{ATL-PHYS-PUB-2016-017,
    author         = "{ATLAS Collaboration}",
    title          = "{The Pythia~8 A3 tune description of ATLAS minimum bias and inelastic measurements incorporating the Donnachie--Landshoff diffractive model}",
    howpublished   = "{ATL-PHYS-PUB-2016-017}",
    url            = "https://cds.cern.ch/record/2206965",
    year           = "2016",
}

@Booklet{ATL-PHYS-PUB-2017-013,
    author         = "{ATLAS Collaboration}",
    title          = "{Optimisation and performance studies of the ATLAS \(b\)-tagging algorithms for the 2017-18 LHC run}",
    howpublished   = "{ATL-PHYS-PUB-2017-013}",
    url            = "https://cds.cern.ch/record/2273281",
    year           = "2017",
}

@Booklet{ATL-PHYS-PUB-2018-009,
    author         = "{ATLAS Collaboration}",
    title          = "{Improvements in \(t\bar{t}\) modelling using NLO+PS Monte Carlo generators for Run~2}",
    howpublished   = "{ATL-PHYS-PUB-2018-009}",
    url            = "https://cds.cern.ch/record/2630327",
    year           = "2018",
}

\clearpage
% ATLAS Collaboration author list
% Reference date of TOPQ-2020-03 is 2020-04-27
% Author list last updated on date 14-JUL-20
% Data extracted on 14-Jul-2020 for paper reference TOPQ-2020-03
% at 11:43am
 
\begin{flushleft}
\hypersetup{urlcolor=black}
{\Large The ATLAS Collaboration}

\bigskip

\AtlasOrcid[0000-0002-6665-4934]{G.~Aad}$^\textrm{\scriptsize 102}$,    
\AtlasOrcid[0000-0002-5888-2734]{B.~Abbott}$^\textrm{\scriptsize 128}$,    
\AtlasOrcid{D.C.~Abbott}$^\textrm{\scriptsize 103}$,    
\AtlasOrcid[0000-0002-2788-3822]{A.~Abed~Abud}$^\textrm{\scriptsize 36}$,    
\AtlasOrcid[0000-0002-1002-1652]{K.~Abeling}$^\textrm{\scriptsize 53}$,    
\AtlasOrcid[0000-0002-2987-4006]{D.K.~Abhayasinghe}$^\textrm{\scriptsize 94}$,    
\AtlasOrcid[0000-0002-8496-9294]{S.H.~Abidi}$^\textrm{\scriptsize 166}$,    
\AtlasOrcid[0000-0002-8279-9324]{O.S.~AbouZeid}$^\textrm{\scriptsize 40}$,    
\AtlasOrcid{N.L.~Abraham}$^\textrm{\scriptsize 155}$,    
\AtlasOrcid[0000-0001-5329-6640]{H.~Abramowicz}$^\textrm{\scriptsize 160}$
% ...
\bigskip
\\

% ...
$^{36}$CERN, Geneva; Switzerland.\\
% ...
$^{40}$Niels Bohr Institute, University of Copenhagen, Copenhagen; Denmark.\\
% ...
$^{53}$II. Physikalisches Institut, Georg-August-Universit\"{a}t G\"ottingen, G\"ottingen; Germany.\\
% ...
$^{94}$Department of Physics, Royal Holloway University of London, Egham; United Kingdom.\\
% ...
$^{102}$CPPM, Aix-Marseille Universit\'e, CNRS/IN2P3, Marseille; France.\\
$^{103}$Department of Physics, University of Massachusetts, Amherst MA; United States of America.\\
% ...
$^{128}$Homer L. Dodge Department of Physics and Astronomy, University of Oklahoma, Norman OK; United States of America.\\
% ...
$^{155}$Department of Physics and Astronomy, University of Sussex, Brighton; United Kingdom.\\
% ...
$^{160}$Raymond and Beverly Sackler School of Physics and Astronomy, Tel Aviv University, Tel Aviv; Israel.\\
% ...
$^{166}$Department of Physics, University of Toronto, Toronto ON; Canada.\\
% ...

\end{flushleft}

% Created with Glance <Atlas.Glance@cern.ch>
 
\end{document}